\documentstyle[12pt]{article}
\catcode`@=11
\def\@cite#1{$^{\scriptsize{#1}}$}
\def\@refe#1{#1}
\def\@biblabel#1{{\normalsize\bf{#1}}}
\def\refe{\@ifnextchar
[{\@tempswatrue\@citexr}{\@tempswafalse\@citexr[]}}
\def\@citexr[#1]#2{\if@filesw\immediate\write\@auxout{\string\citation
{#2}}\fi
  \def\@citea{}\@refe{\@for\@citeb:=#2\do
    {\@citea\def\@citea{,}\@ifundefined
       {b@\@citeb}{{\bf ?}\@warning
       {Citation `\@citeb' on page \thepage \space undefined}}%
\hbox{\csname b@\@citeb\endcsname}}}{#1}}
 \catcode`@=12
\title{Critical phenomena for Riemannian manifolds:\\
Simple Homotopy\\
 and\\
simplicial quantum gravity}
\author{M.Carfora*, A.Marzuoli**\\
International School for Advanced Studies, SISSA-ISAS*\\
Via Beirut 2-4, 34013 Trieste, Italy\\
Istituto Nazionale di Fisica Nucleare, Sezione di Pavia,Italy\\
Dipartimento di Fisica Nucleare e Teorica dell'Universit\`a di
Pavia**\\
Via Bassi 6, I-27100 Pavia, Italy \\
 Istituto Nazionale di Fisica Nucleare, Sezione di Pavia,Italy}
\date{}
\def\Ricco{{\cal{R}}(r,D,V)} 
\begin{document}
\maketitle
\begin{abstract}
Simplicial quantum gravity is an approach to quantizing gravity
by using as a regularization scheme dynamical triangulations of
riemannian manifolds. In dimension two, such an approach has
led to considerable progress in our understanding of $2D$-gravity.
In dimension three and four, the current state of affairs is
less favourable, and dynamically triangulated models of quantum
gravity suffer from serious difficulties which can be only
partially removed by the use of computer simulations.
In this paper, which to a large extent reviews some of our recent
work, we show how Gromov's spaces of bounded geometries provide
a general mathematical framework for addressing and solving many
of the issues of $3D$-simplicial quantum gravity.
In particular, we establish entropy estimates characterizing the
asymptotic distribution of combinatorially inequivalent triangulated
$3$-manifolds, as the number of tetrahedra diverges.
Moreover, we offer a rather detailed presentation of
how spaces of three-dimensional riemannian manifolds with
natural bounds on curvatures, diameter, and volume can be
used to prove that three-dimensional simplicial quantum gravity
is connected to a Gaussian model determined by the simple
homotopy types of the underlying manifolds. This connection is
determined by  a Gaussian measure defined over the
general linear group $GL({\bf R},\infty)$.\par
By exploiting these results it is shown that the partition
function of three-dimensional simplicial quantum gravity
is well-defined, in the thermodynamic limit, for a suitable range of
values of the gravitational and cosmological coupling constants.
Such values are determined by the Reidemeister-Franz torsion
invariants
associated with an orthogonal representation of the fundamental
groups of the set of manifolds considered. The geometrical system
considered
shows also critical behavior, and in such a case the partition
function is exactly evaluated and shown to be equal to the
Reidemeister-Franz torsion. The phase structure in the
thermodynamical limit is also discussed. In particular, there are
either phase
transitions describing the passage from a simple homotopy type
to another, and (first order) phase transitions within a
given simple homotopy type which seem to confirm, on an analytical
ground, the  picture of three-dimensional
quantum gravity recently  suggested by numerical simulations.
\end{abstract}
\vfill\eject
\tableofcontents
\vfill\eject

\section{Introduction}

Random surface theory\cite{Pfister} and the strictly related notion of
random triangulations\cite{{Garrido},{David},{Dur},{Kazakov1}}
have played a key role in our understanding of models of quantum
gravity. Advances have been particularly significant in
dimension two\cite{{Brez},{Doug},{Gros},{Ambj}}
where dynamically triangulated models of two-dimensional
quantum gravity, (either pure or coupled to suitable
matter fields\cite{{Kazakov2},{Kazakov3}}),
can be put in  correspondence with an appropriate matrix field
theory\cite{{Kostov1},{Kostov2},{Kostov3}}. The relation between two-
dimensional random triangulations
${\cal T}$ and matrix models arises from the observation\cite{Pari}
that each vertex of the graph associated with the dual triangulation,
${\cal T}^*$, (the graph the vertices of which are the centers of the
triangles and the edges of which arise from pairs of adjacent
triangles), is of order three, (see\cite{{Ambj},{Fernandez}}
for a thorough account of such topics).
 This remark implies that every
Feynman diagram in the formal diagrammatic expansion of a
${\phi}^3$ field theory can be labelled by  dual
triangulations ${\cal T}^*$. It is well known that similar mappings
into a matrix model, mappings
which are instrumental to the success of the two-dimensional theory,
are
not so well-behaved in higher dimensions.
The techniques of matrix field theory  are not  helpful
in such a case, at least apparently, and one must resort to more
direct combinatorial
constructions, (often computer-assisted), in the theory of
triangulated
manifolds in order to investigate the properties
of three and four-dimensional simplicial quantum gravity models.\par
\vskip 0.5 cm
At the basis of the difficulties in dealing with these higher
dimensional models of simplicial quantum gravity, (difficulties which
are
present also in the  more standard Regge calculus
approach\cite{Regge}),
there  is a lack of control of the interplay between manifolds
topologies and the riemannian invariants which, in their discretized
form, provide the action of the theory. To what extent such
control is necessary follows by noticing that in order to
construct a sensible statistical
theory of riemannian manifolds out of random triangulations one needs
{\it entropy bounds} assuring that the number of
triangulated manifolds of given dimension,
volume and fixed topology grows with the volume at most at an
exponential rate, possibly
with a non-trivial subleading asymptotics.\par
In the case of surfaces the required
entropy bounds  can be proved
either by direct counting
arguments, or by quantum field theory techniques as applied
to graphs enumeration\cite{Pari}, a technique, this latter, that
has found its way in a number of far reaching applications in
surface theory\cite{{Witten},{Konse},{Penner}}.
In higher dimension, the natural generalization of
such entropy bounds has, up to now, defied analytical proofs,
even if it is known that they hold under some suitable restriction on
the
way one constructs triangulations, and their validity
is  confirmed by
numerical simulation\cite{{Jana1},{Jana2},{Jana3}}. It must also be
stressed
that without any restriction on topology, the situation gets
considerably more involved since it can be shown for instance
\cite{Jana1} that the number of distinct three-
manifolds,
with given
volume $V$ and arbitrary topological type, grows at least factorially
with $V$. Thus a sistematic
method for understanding and enforcing entropy bounds relating
topology to riemannian invariants appears as
a major issue
to deal with in order to extend to higher
dimensions\cite{{Jana1},{Jana2},{Jana3}}, (in particular
to $n=3$ and $n=4$, see for instance\cite{{Jana4},{Jana5}} and
also\cite{{Jana6},{Jana7}}), the results of the two-
dimensional theory. \par
\vskip 0.5 cm
Of particular interest in this line of research is the study of the
three-dimensional case,
either as a necessary step before moving on to the physical relevant
case of dimension four\cite{{Jana8},{Jana9},{Jana10},{Jana11}},
or because one hopes in some
explicit solvability, as hinted by the Chern-Simons approach fostered
by Witten\cite{Witt} in the field theoretic setting.\par
In order to describe in barest essential such three-dimensional
models, let us recall that dynamical triangulations ${\cal T}$ for
three-dimensional manifolds
are constructed through regular tetrahedra glued together along their
two-dimensional faces in such a way so as to  form a
piecewise linear manifold. The {\it dynamical variable} of
the resulting theory is shifted from the length of the links,
used in the usual Regge calculus approach, and here set to one,
to the connectivity
properties of the manifolds. This has the effect that the discretized
version of the continuus Einstein-Hilbert action (with cosmological
term), takes on the
form
\begin{eqnarray}
S[{\cal T}]\equiv k_3N^{(3)}({\cal T})-k_1N^{(1)}({\cal T})
\label{EH}
\end{eqnarray}
where $N^{(3)}({\cal T})$ and $N^{(1)}({\cal T})$ respectively
denote the number of tetrahedra and the number of links in the
given triangulation ${\cal T}$, and where $k_3$ is a bare cosmological
constant while $k_1$  can be related to a gravitational coupling
constant.\par
A simplicial quantum gravity model is realized by considering
as statistical sum associated with the action $S[{\cal T}]$ the
expression
\begin{eqnarray}
\sum_{{\cal T}}{\rho}({\cal T})exp [- S[{\cal T}]]
\label{funzione}
\end{eqnarray}
where the sum is over a suitable class of triangulations, and
where the factor ${\rho}({\cal T})$ is a weight for each
triangulation,
weight which is usually  assumed equal to one.\par
\vskip 0.5 cm
A clear account
of the geometry and the physics underlying such construction
can be found in the quoted papers by Ambj\o rn\cite{Jana1},
suffices here to say
that in order to make sense of the partition function
(\ref{funzione}), one has to impose strong restrictions of topological
nature on the class of allowable triangulations ${\cal T}$ to be
summed over. For instance, a reasonable model
\cite{{Jana1},{Jana2},{Jana3},{Jana4}} can be constructed
by restricting
the topology of the manifolds, simplicially
approximated by ${\cal T}$, to the three-sphere topology.
\vskip 0.5 cm
Under such assumptions, it is possible to
discuss the general characteristics of the critical behavior
of the model, so as to conclude that the partition
function (\ref{funzione}) is well defined in a convex
region  ${\cal D}$ in the parameters space $(k_1,k_3)$.
Moreover, numerical simulations suggest that  a phase transition,
presumably of first order, occurs. Such phase transition
separates two phases, one, ({\it hot}), in which {\it crumpled}
geometries dominate, while the other, ({\it cold}), is characterized
by geometries of a more regular nature (at least in the sense that
their Hausdorff dimension is three).\par
By reference
to the structure of the action in (\ref{funzione}), these
results hint toward a mechanism which control, in terms of the
gravitational coupling the onset of a {\it regular}
geometry out of crumbled {\it quantum} universes.
It is not yet clear if one can take
any significant continuum limit of the theory in the critical regime,
however it seems that a sensible vacuum exists in the cold phase
of the model developed by Boulatov and
Krzywicki\cite{Jana5}. Such
vacuum is a reasonable candidate for a physically acceptable
continuum limit of the theory.\par
\vskip 0.5 cm
The starting point of this Paper, which
reviews and extends some recent
work\cite{{Carf},{popa},{IJMPA},{Marz}} of
ours, is the observation
 that the ideas and techniques introduced in
Riemannian geometry  by Gromov's
\cite{Grom} coarse grained description of the space of $n$-
dimensional riemannian structures provide a natural framework
for addressing and, to a large extent, solving the above issues.
In particular, we show that the partial results on
three-dimensional simplicial quantum gravity recalled above, most
of which are based on computer simulations, can analytically recovered
and put in a systematic geometrical perspective within a general
regularization scheme which exploits the coarse-graining description
of riemannian manifolds introduced by Gromov. The rationale which
motivates such an approach follows by noticing that the
problems encountered in evaluating partitions functions in
simplicial quantum gravity are counting
problems, typically the numbering of the different topological types
of manifolds contributing with a given volume or average curvature to
the
state of the system. Similar questions appear in disguise in
riemannian geometry in the form of finiteness theorems concerning the
topology of riemannian manifolds, and the basic observation underlying
Gromov's analysis is that such finiteness theorems do not depend
 on a detailed local description of the
geometry of the manifold, but they are rather connected to controlling
geometry in the large by imposing some natural constraints on the size
of the manifold. Such constraints concern the diameter, the
 volume and the sectional curvatures of the manifold, and it is not
too
 difficult to realize, already at this point,
that they are the geometric conditions under which (a weak form of)
the entropy bounds
mentioned above hold true. These entropy bounds, which we prove here
as a by-product of a detailed analysis of three-dimensional
simplicial gravity, are
basically a consequence of Gromov's compactness theorem (see below).
They are one of the main reasons that can  make Gromov's spaces
 of bounded geometries an
appropriate mathematical setting for discussing models of $n$-
dimensional quantum
gravity.\par
 \vskip 0.5 cm
Other important motivations for adopting Gromov's coarse grained point
of view can be seen at work when we exploit the structure of Gromov's
spaces to the
effect of
proving the basic fact that three-dimensional
simplicial
quantum gravity is connected with a gaussian theory describing
the random insertion of three-dimensional cells onto the
two-dimensional skeleton of  suitable triangulations.
It is interesting to remark that, from a geometrical point of view,
such role of the two-skeleton  does not surprise,
 since it is known that
triangulated $n$-manifolds are determined by their
$[n/2]+1$-skeletons\cite{Dancis}. In particular, for $n=3$, a
triangulated
three-manifold is determined by the (simplicial) isomorphism
class of its two-skeleton. Obviously is to be stressed that the
two-skeleton of a triangulation is not, in general, the triangulation
of a two-dimensional surface, but it is possible
to take care of this difficulty.
In this connection, rather than generic triangulations,
 we use as a regularization framework
the simplicial approximations generated by
minimal geodesic balls coverings of three-dimensional
riemannian manifolds\cite{Grov}. The riemannian manifolds in question
are thought of as points in the infinite dimensional
compact space of bounded geometries generated by
Gromov-Hausdorff completion of the set of all riemannian
three-manifolds with suitable bounds on curvatures, volume and
diameter\cite{Grom}. As already remarked\cite{{Carf},{popa}} the
choice of such configurational space for the theory, allows
for a natural control of the topology of the manifolds
 involved in the statistical sum, and yields naturally for
those entropy bounds which, otherwise are to be imposed
{\it ad hoc}. \par
\vskip 0.5 cm
It must be noticed that the constraints on curvatures, volume and
diameter,
(sectional  curvatures bounded
below by a real constant, volume bounded below and diameter bounded
above), can be
removed if one carefully perform a direct limit of such Gromov
spaces. This is a most delicate issue, (its counterpart in dimension
two is connected to summation over all surface genera), however we
defer the analysis of this further limiting
procedure to a forthcoming paper: even if our treatment is
so biased, we trust that the reader will be
convinced of the usefulness of Gromov's spaces of bounded geometries
in yielding a unified framework within which to study dynamically
triangulated quantum gravity models.\par
\vskip 0.5 cm
Now we briefly delineate the strategy adopted and the main results
obtained.\par
\vskip 0.5 cm
For the convenience of the reader, we begin by providing a brief
review of some elementary aspects of Gromov's theory of spaces of
bounded geometry. We apologize to the expert reader who may skip
section 2. \par
The starting point of our
analysis is the observation that geodesic
balls coverings of manifolds of bounded geometry naturally yields
for simplicial approximations which are strictly related to the
topology of the underlying manifold. In particular
the homotopy type of the manifold can be labelled by the
one-skeleton  of the covering\cite{Grov}. This is  a
rather non-trivial fact
which allows to discuss fluctuations of the homotopy types
of manifolds of bounded geometry, in any dimension.
Such matters have already
been discussed at length in the papers\cite{{Carf},{popa}}, where
they have been applied to the generic $n$-dimensional manifold
of bounded geometry. Here, by combining reviews of our former work and
new results, we specialize to dimension three, and
exploit the theorem of Dancis\cite{Dancis} recalled above,
theorem which relates the combinatorial isomorphism class of a
simplicial
three-manifold to its two-skeleton. In this way,
we can  put into work  the above homotopy classification  by
writing down, rather than the discrete gravitational action
(\ref{EH}),
the simplicial quantum gravity partition function for the
two-skeleton
of the geodesic ball coverings of the three-manifolds in question,
and by studing the thermodynamical limit of such partition function
as the covering becomes finer and finer.\par
\vskip 0.5 cm
A key point in this part of the
paper, is the connection which can be established between the
statistical
model so defined and the classical statistical mechanics of a
{\it lattice gas} on a suitable (denumerable) infinite graph.
The interplay between lattice gas theory and the compactness
properties
of Gromov's space of bounded geometries allow to establish the
existence of the thermodynamical limit of the partition function,
to understand its phase structure, and in particular to prove the
existence
of a critical regime of the theory where the role of the topology of
the manifolds come to the fore.\par
The use of a partition function associated with the two-skeleton of
a geodesic ball covering for describing
a three-dimensional manifold, ({\it i.e.}, of a two-dimensional
statistical
theory for describing a three-dimensional object), has some
advantages and some drawbacks. The advantages concern a qualitative
similarity between the theory described here and the general features
of two-dimensional quantum gravity models based on random
triangulation of surfaces. The drawbacks come about by noticing that
the resulting statistical
approach is rather
combinatorial and measure-theoretic in spirit, to the effect that
geometry plays only a relatively auxiliary role in the theory.
In particular, it is difficult to provide a sharp estimate of the
entropic contribution to the partition function.
Moreover, it is difficult to get information on the existence and
on the properties of a sensible continuum limit of the theory, (if any
such limit exists). This circumstance does not surprise, since
the fact that we are dealing with three-dimensional objects is only
indirectly exploited (through the role of the first order intersection
pattern of the covering), and topology only enters in a relatively
elementary way in the theory.\par
A more fundamental approach, based on first principles which from
the very onset call into play geometry and topology is thus called
for. Such an approach is dealt with in the second part of the paper,
and
its implementation relies on the use of few elentary facts of
{\it simple homotopy theory}.
\vskip 0.5 cm
Roughly speaking,
simple homotopy can be related to the way the tetrahedra of the nerve
of the covering are attached to  the two-dimensional faces of the
two-skeleton. Inequivalent ways of attaching, up to a suitable
action of the fundamental group, provide an obstruction to lifting
an homotopy equivalence to a topological equivalence, and
are of natural interest for simplicial gravity.\par
In general, the simple homotopy description of the insertion of
simplices into a lower dimensional complex is a difficult
matter to handle. However, in dimension three,
such insertion can be
equivalently described by attaching a finite number of
two-dimensional and three-dimensional cells to a vertex
of the two-dimensional skeleton, (the boundary of the
three-cells being non-trivially pasted onto the two-cells via
the action of a suitable incidence matrix).\par
If one is willing
to blend a statistical formalism with such geometrical framework,
it is natural to attach such cells according to a Gaussian
distribution, (in the sense that, given
an orthogonal representation of the fundamental group,
 the characteristic maps describing
the insertion of the cells in the two-skeleton, are distributed
according to a Gaussian probability law with variance related to
the representation of the incidence matrix).
Not surprisingly, the outcome of this
construction is that a Reidemeister-Franz representation
torsion\cite{DeRham}
makes its appearence in the statistical sum; what is less obvious
is that such Gaussian pasting reproduces the statistical sum
of simplicial three-dimensional quantum gravity, (\ref{funzione}).\par
The Reidemeister-Franz representation torsion is a combinatorial
invariant that, in its analytical counterpart, namely
as Ray-Singer\cite{RayS}
torsion, also naturally appears in Schartz-Witten's approach to
three-dimensional gravity\cite{Witt}. Thus the combinatorial
result we get indicates that one is on the right track for
connecting the simplicial theory to the continuus formalism.\par
\vskip 0.5 cm
As the geodesic ball covering becomes
finer and finer, we prove that the three-dimensional
partition function  obtained by the above construction
is well defined if the couplings $(k_1,k_2)$ vary in a
convex region whose boundary is determided by the
Reidemeister torsions sampled.
The rationale underlying this result is an intriguing connection
between the thermodynamical limit of the partition function and
a group-theoretic construction of a Gaussian measure on the
general linear group $GL({\bf R})$.\par
The critical regime of the system is also discussed
to the effect of
conferming the existence of a non-trivial phase structure of
the theory. In the thermodynamical limit the system exhibits
either a phase transition describing
the passage from a simple homotopy type to another, or,
within a given simple homotopy type, a first order phase
transition. In such a critical regime, the partition function is
evaluated and shown to be equal to the corresponding value of the
Reidemeister-Franz torsion.\par
\vskip 0.5 cm
{}From a physical point of view, since the representation torsions
are combinatorial invariant, it follows that at this latter critical
point the partition function has lost memory of the details of the
simplicial approximation used, ({\it viz}, that the given
combinatorial approximation was obtained through
geodesic balls coverings), and a sort of scaling limit is active in
such a
case, (obviously, a full scaling limit would require also the
removal of the cut-offs on curvatures, diameter and volume which
specify the space of bounded geometry in which the statistical
sum has been computed. In this way, as already remarked, one can
sample all topologies). Such critical point
can be rather naturally identified as the transition numerically
found by Boulatov and Krzywicki\cite{Jana5}, and which appears as a
suitable candidate for the {\it vacuum} of the theory.
As in their case,
the critical regime we discuss holds in a {\it cold} phase, namely for
geometries of finite (Hausdorff) dimension. This follows by noticing
that  spaces
of bounded geometry are compact\cite{Grom}, and simplicial
approximations associated with minimal geodesic balls coverings
yield, as the covering becomes finer and finer, objects which
are no more singular than three-dimensional homology
manifolds\cite{{Pete},{Pans}}. Moreover, the phase transition
in question is of first order, a fact which completely consistent
with the numerical analysis quoted above. This first order nature
also seems to indicate that there is no reasonable continuum limit
of the theory.  This lack of existence
of a continuum limit is to be seen as  statistical counterpart of
the fact that in dimension three homology manifolds are
the (Gromov-Hausdorff) closure of the space of riemannian
manifolds of bounded geometry. The Gromov-Hausdorff
topology, which appears as the natural topology
to put on the space of riemannian structures in order to implement
a statistical formalism of use in (euclidean) simplicial quantum
gravity,
is so weak that in the thermodynamical limit one gets
a large entropic contribution from
geometrical
object which are so convoluted as to have no control on their
topological structure.\par
In higher dimensions ({\it e.g.},
in dimension four), the (Gromov-Hausdorff) closure of spaces of
bounded geometries is nicer, being generated by
topological manifolds, and on such grounds we would
expect the existence of a more reasonable continuum limit
of the corresponding statistical theory. Here we do not address
four-dimensional simplicial quantum gravity, even if some of the
results presented in this paper appear to be readily generalizable
to dimension four. A rough explanation of this restriction comes about
by noticing that our construction based on simple homotopy theory
is intrinsically three-dimensional. To be more precise,
given an incidence ${\bf Z}({\pi}_1)$-matrix, (see section 5, for
definitions), representing the simple homotopy type of
a four-dimensional manifold, (or  more in general, of a $n$-
dimensional
manifold), then,  there always exists a three-dimensional $CW$-complex
realizing such incidence matrix. Thus, in order to address the case
of four dimension, we must put into work something finer than simple
homotopy theory. In the concluding remarks of the paper, we provide
few, rather non-conclusive, indications in this direction.\par
\vskip 0.5 cm

\section{Geodesic balls coverings of manifolds of bounded geometries}
  Our starting point is a  classical  result  of M.Gromov \cite{Grom}
  according to which  the space of all riemannian  structures
  of a given dimension, not necessarily of a given topology,  with
natural
  constraints on their diameter and Ricci curvature
  is characterized  by some remarkable compactness
  properties  which make the global  characteristics of geometric
functionals,
  (typically their boundedness), very trasparent and natural. The
compactness we are referring to
  can be very effectively brought to surface by introducing a
suitable notion of
  distance, (actually, a metrizable uniform structure), on the space
of all riemannian manifolds of dimension $n$,
(possibly with different underlying topologies)
considered as elements of the more general category of all
compact metric spaces . This
distance, first
  introduced by M.Gromov \cite{Grom}, generalizes the familiar notion
of
  Hausdorff distance between compact subsets of a given metric space,
and even if it looks rather formal, it provides a sort of
{\it visual} distance between manifolds, a remark, this latter, that
will be made more precise below by discussing geodesic balls
coverings of manifolds of bounded geometry.\par
\vskip 0.5 cm
Consider two riemannian
   manifolds $M_1$   and $M_2$, and let
  $i_1(M_1)$ and $i_2(M_2)$ denote respectively isometric embeddings
of
$M_1$  and $M_2$ in some metric space  $(A,d)$, (this may be a
Euclidean
space of
  sufficiently high dimension, but in full generality we must allow
for curved
  and possibly infinite-dimensional embedding spaces). A
  Hausdorff distance in $(A,d)$ between $i_1(M_1)$ and $i_2(M_2)$ can
be
introduced according to
\begin{eqnarray}
 d_H^A[i_1(M_1),i_2(M_2)]  & = \inf \{\epsilon >0,
U_{\epsilon}(i_1(M_1)) \supset
i_2(M_2) \nonumber \\
&   U_{\epsilon}(i_2(M_2)) \supset i_1(M_1)\}
\end{eqnarray}
where
$U_{\epsilon}(i_1(M_1))=\{z \in A \colon d(z,i_1(M_1))\leq \epsilon
\}$
 and similarly for $U_{\epsilon}(i_2(M_2))$.  In other words
$d_H^A[i_1(M_1),i_2(M_2)]$ is
the lower bound of the  $\epsilon$ such that $i_1(M_1)$  is contained
in
the $\epsilon$ -
neighborhood of $i_2(M_2)$ and vice versa.  The  {\it Gromov distance}
between the two
riemannian manifolds $M_1$ and  $M_2$,  $d_G(M_1,M_2)$, is then
defined
according
to the following\cite{Grom} \par
\newtheorem{Gaiad}{Definition}
\begin{Gaiad}
  $d_G(M_1,M_2)$ is the lower bound of   the  Hausdorff distances
\begin{eqnarray}
  d_H^A[i_1(M_1),i_2(M_2)]
\end{eqnarray}
as  $A$ varies in the set of
metric spaces and $i_{1}$, $i_{2}$ vary in the set of all isometric
embeddings
of $M_1$ and $M_2$ in $(A,d)$.
\label{miauno}
\end{Gaiad}
\vskip 0.5 cm
Notice that $d_G$ is not, propoerly speaking,  a distance since it
does not satisfy triangle inequality, it rather gives rise to a
metrizable uniform structure. For our purposes, we can use it as if it
were a distance function. In any case,
in working with Gromov distance one gets a sense
of a geometric
 nearness among riemannian structures having to do with a
classification of riemannian
manifolds according to how they can be covered by small  geodesic
balls.  In
particular two riemannian manifolds can be considered  near to each
other
in the Gromov distance if the can  both be covered with a collection
of
small geodesic  balls arranged in a similar packing configuration.
In order to explain this remark, perhaps the main source of
geometrical
intuition in working with the Gromov distance, we are naturally
led to introduce the following class of
riemannian structures\cite{Grom} \par
 \begin{Gaiad}
For $k$ a real number and $D$ a positive real number let
$Ric[n,k,D]$  denote the space of isometry classes of closed connected
 n-dimensional riemannian manifolds $(M,g)$ with
$Ric_{M} \geq -(n-1)k$ and $diam_{M}\leq D$.
\label{miadue}
\end{Gaiad}
(Recall that, for a closed ($n$-dimensional) riemannian manifold $M$
we define the diameter of $M$ as $\sup_{(p,q)\in M\times M}d(p,q)$
where
$d(\cdot , \cdot)$ denotes the distance function of $M$. And
if $Ric(M)$ denotes the Ricci curvature of $M$, we let
\begin{eqnarray}
k(x)\equiv \inf \{ Ric(u,u) \colon u \in T_xM, |u_x|=1\}\nonumber
\end{eqnarray}
thus, the lower bound of the Ricci tensor of a riemannian manifold $M$
is
defined as the lower bound of $k(x)$ as $x$ varies in $M$).\par
The point of the introduction of $R[n,k,D]$ is that for any
manifold $M$ in such class
it is possible to introduce suitable coverings by geodesic balls
providing
a coarse classification of the riemannian structures occurring in
$Ric[n,k,D]$. In
particular \cite{Grov}  for any manifold M in such class and for any
given  $\epsilon >0$ it is always
possible to find an ordered set of points $\{p_1,\ldots,p_N\}$ in $M$,
 so that\par
\noindent  {\it (i)} the open balls
$B_{M}(p_{i},\epsilon) = \{x \in M \vert d(x, p_{i})\leq
\epsilon\}$, $i=1,\ldots,N$, cover $M$; in
other words the collection
\begin{eqnarray}
{\{p_1,\ldots,p_N\}}
\end{eqnarray}
  is an  $\epsilon$-net in $M$.\par
\noindent  {\it(ii)} the open balls $B_{M}(p_{i},{\epsilon\over 2})$,
$i=1,\ldots,N$, are
disjoint, {\it i.e.}, $\{p_1,\ldots,p_N\}$
is a {\it minimal} $\epsilon$-net in $M$.\par
  Any such
minimal net is also characterized by its {\it intersection pattern},
namely by
the set
\begin{eqnarray}
 I_{\epsilon}(M) \equiv  \{(i,j)\vert i,j= 1,\ldots,N
\vert B(p_{i},\epsilon)\cap B(p_{j},\epsilon)\not=\emptyset\}
\end{eqnarray}
  Any two manifolds $ M_{1}$ and $ M_{2}$ in $Ric[n,k,D]$ endowed
with  $\epsilon$-minimal nets
$\{p_1,\ldots,p_N\}$ and $\{q_1,\ldots,q_L\}$ , respectively,
 are said to be equivalent if and
only if $ N=L$ and if they
have the same intersection pattern. Namely if, up to combinatorial
isomorphisms, we can write
\begin{eqnarray}
N_{(\epsilon)}^{(0)}(M_1) & = N_{(\epsilon)}^{(0)}(M_2) \nonumber \\
I_{(\epsilon)}(M_1) & = I_{(\epsilon)}(M_2)
\label{equivalence}
\end{eqnarray}
where we have introduced the  {\it filling function}
$N_{(\epsilon)}^{(0)}(M)$
of the covering, {\it i.e.}, the function which associates with $M$
the maximum
number of geodesic balls realizing a minimal $\epsilon$-net on
$M$.\par
The filling function and the (first) intesection pattern
characterize  the   $1$-skeleton,
${\Gamma}_{(\epsilon)}(M)$, of the geodesic  balls covering.
Similarly, upon
considering   the higher  order intersection patterns of the set
of balls $\{B_i(\epsilon)\}$, we can define
the  two-skeleton ${\Gamma}_2(M)$, and eventually  the
nerve ${\cal N}\{B_i\}$ of the geodesic balls covering of   the
manifold   $M$, this is a symplicial complex realized according to the
following rules : \par
{\it (i)} the vertices $p_i^{(0)}$ of ${\cal N}$ correspond to
the balls $B_i(\epsilon)$, \par
{\it (ii)} the edges $p_{ij}^{(1)}$ correspond to pairs of
geodesic balls $\{B_i(\epsilon),B_j(\epsilon)\}$ having a not-empty
intersection $B_i(\epsilon) \cap B_j(\epsilon) \not= \emptyset$, \par
{\it (iii)} the faces $p_{ijk}^{(2)}$ correspond to triples of
geodesic balls with not-empty intersection $B_i(\epsilon)\cap
{B}_j(\epsilon)\cap {B}_k(\epsilon) \not= \emptyset$, \par
{\it (iv)} the $k$-symplices $p_{i_1i_2 \ldots i_{k+1}}^{(k)}$
correspond
to collections of $k+1$ geodesic balls such that
$B_1 \cap {B}_2 \cap \ldots \cap {B}_{k+1} \not= \emptyset$. \par
 If  $\epsilon$ is
sufficiently  small  this nerve gives rise to a
polytope  which
approximate   the  manifold  $M$. In  particular,  the
$1$-
skeleton is just the graph providing the vertex-edge structure  of
this approximation.  It is to be stressed that since for manifolds in
$Ric[n,k,D]$ arbitrarily small metric balls need not be contractible,
this approximation is rather rough. In any case,
what happens is that the inclusion of sufficiently small geodesic
balls into suitably larger balls is homotopically trivial. Thus
one gets a polytope which is homotopically dominating the underlying
manifold. This homotopical approximation yielding for
the homotopy finiteness theorem recalled below is all we need for
the analysis of simplicial gravity that follows.\par
\vskip 0.5 cm
The equivalence relation (\ref{equivalence})  partitions
$Ric[n,k,D]$  into disjoint equivalence classes whose
finite number can be estimated in terms of the parameters
$n$, $k$, $D$.
 In particular, if  we set
\begin{eqnarray}
{\cal  O}^{(\epsilon)}_{(\lambda, \Gamma)} \equiv \{M \in Ric[n,k,D]
 \colon N_{(\epsilon)}^{(0)}(M) = \lambda, I_{(\epsilon)}(M)= \Gamma
\}
\end{eqnarray}
  where $\lambda$ is a given positive integer and $\Gamma$ is
a given
 graph defined by a collection of $\lambda$ vertices
$\{p_1, \ldots, p_{\lambda}\}$, and  by a
collection of edges $\{p_i, p_j\}$ joining the vertices $p_i$ and
$p_j$. Then
for each given $\epsilon$ we can write $Ric[n,k,D]$ as a disjoint
finite
union
\begin{eqnarray}
Ric[n,k,D] = \bigcup_{(\lambda, \Gamma)} {\cal
O}^{(\epsilon)}_{(\lambda,
\Gamma)}
\label{disjoint}
\end{eqnarray}
going over  all possible choices of $\lambda$ and $\Gamma$
accessible to
the manifolds in $Ric[n,k,D]$. \par
Thus each equivalence
  class of manifolds is
characterized by the abstract (unlabelled)    graph
${\Gamma}_{(\epsilon)}$
defined  by the  $1$-skeleton of the $L(\epsilon)$-covering. The
order of
any
 such graph ({\it i.e.}, the number of
vertices) is  provided by  the filling  function
$N_{(\epsilon)}^{(0)}$,
while
the structure  of the  edge set  of ${\Gamma}_{(\epsilon)}$  is
defined
by  the intersection pattern $I_{(\epsilon)}(M)$. It is important
to remark that  on $Ric[n,k,D]$ either the filling function or the
intersection pattern cannot  be arbitrary. The former is always
bounded
 above for each  given $\epsilon$, and  the best  filling of
a riemannian manifolds  with geodesic  balls  of  radius  $\epsilon$
is realized on (portions of)  spaces  of  constant  curvature
\cite{Grom}.  The latter is  similarly  controlled  through  the
geometry
 of the  manifold to  the effect  that the  average degree,
$d(\Gamma)$,  of  the  graph ${\Gamma}_{(\epsilon)}$, ({\it
i.e.}, the  average number of edges incident on a
vertex of  the graph), is bounded above by  a constant  as the
radius of the balls defining  the covering  tend to  zero, ({\it
i.e.},
as $\epsilon \to 0$ ).  Such constant  is independent  from
$\epsilon$,
and can be estimated in terms of the parameters $n$, $k$, $D$
\cite{Grov}.
These remarks are the content of three basic propositions which obtain
for manifolds  in $Ric[n,k,D]$.
 The first proposition
is an immediate consequence of the Bishop-Gromov volume comparision
theorem, \cite{Grom}. It provides the quoted geometrical bound on the
maximum
number of geodesic balls realizing a minimal
$\epsilon$-net\cite{Grom}.\par
\newtheorem{Gaiap}{Proposition}
\begin{Gaiap}
Let $Ric(g)$ and  $d(M)$ respectively  denote  the
  Ricci   tensor  and the diameter of a manifold  $M \in  Ric[n,k,D]$.
Assume
  that  $Ric(g) \geq (n-1)kg$ for some real number
  $k$,  and let us consider a geodesic ball, ${\bar  B}(d(M))$, of
  radius  $d(M)$  in  the model space ${\bar M}_k$ of constant
sectional curvature
  equal to $k$. Denote by   ${\bar N}_{(\epsilon)}(k)$  the
value that the  filling function
  $N_{(\epsilon)}^{(0)}$   attains on   ${\bar  B}(d(M))$,  {\it
viz.}
\begin{eqnarray}
  {\bar  N}_{(\epsilon)}(k) =\frac {\int_0^{d(M)} {\bar J}(t)^{n-1}dt}
{\int_0^{\epsilon} {\bar J}(t)^{n-1}dt}
\label{barra}
\end{eqnarray}
  where
${\bar J}(t) = t$, $\frac{\sin \sqrt {kt}}{\sqrt {k}}$,
$\frac{\sinh \sqrt {-kt}}{\sqrt  {-k}}$
  for $k=0$, $>0$, $<0$ respectively.  Then
 \begin{eqnarray}
  N_{(\epsilon)}^{(0)}(M)  \leq  {\bar N}_{(\epsilon)}(k)
 \end{eqnarray}
\label{miatre}
\end{Gaiap}
The next  proposition\cite{Grov} provides an  a priori estimate on the
average degree of the graph ${\Gamma}_{(\epsilon)}$ describing
the intersection pattern for minimal nets \par
\begin{Gaiap}
Let $\{p_1,\ldots,p_N\}$ be any $\epsilon$-net in
$M \in Ric[n,k,D]$. Then there is a constant
$N_2$ depending  on $k$, and $D$, such that
for any $x\in M$, the geodesic ball $B_{x}(\epsilon)$ intersects at
most
$N_2$ of the balls $B(p_1,\epsilon),\ldots,B(p_N,\epsilon)$, and as
$\epsilon \to 0$ we get
\begin{eqnarray}
\limsup_{\epsilon \to 0} N_2(\epsilon) \leq  (C_{n,k,D})^{n-1}
\end{eqnarray}
where $C_{n,k,D}$ is a constant depending only on the
parameters $n$, $k$, $D$, (and not from
$\epsilon$),  which characterize $Ric[n,k,D]$.
\label{miaquattro}
\end{Gaiap}
 The bound on
$d({\Gamma}_{(\epsilon)})$ follows by applying the above theorem
in particular to each of the balls of the covering. Then
$N_2(\epsilon)$ has the meaning of an upper bound to the
degree of ${\Gamma}_{(\epsilon)}$ at the vertex $x=p_i$ considered.
Since this bound holds for any vertex $p_i$,
$i=1,\ldots,N_{(\epsilon)}^{(0)}$,we get as well
\begin{eqnarray}
d({\Gamma}_{(\epsilon)}) \leq  (C_{n,k,D})^{n-1}
\end{eqnarray}
Finally the third proposition\cite{Grov} allows us to compare
distances in
minimal
$\epsilon$-nets with the some intersection pattern.\par
 \begin{Gaiap}
Let $\{p_1,\ldots,p_N\}$ and $\{q_1,\ldots,q_N\}$
 respectively denote minimal  $\epsilon$-
nets with the
same intersection pattern in  $M_1$ and $M_2 \in Ric[n,k,D]$. Then for
any $K$ such that
\begin{eqnarray}
d_{{M}_1}(p_i,p_j) < K \cdot \epsilon
\end{eqnarray}
there is a constant  $N_3(K)$, depending on $K$, and $r,D,n$ for
which
\begin{eqnarray}
d_{{M}_2}(q_i,q_j) < N_3(K) \cdot \epsilon
\end{eqnarray}
\label{miacinque}
\end{Gaiap}
In other words, equivalent  $\epsilon$-nets on manifolds in
$Ric[n,k,D]$, are metrically
equivalent up to a dilatation. This remark can be made more precise if
we recall the notion  of Lipschitz distance\cite{Grom} between two
metric
spaces.\par
\begin{Gaiad}
Let  $X$ and $Y$ be two metric spaces and let
$f\colon  X \rightarrow Y$  be a  (bi\-Lipschitz) homeomorphism. The
 Lipschitz distance between $X$ and $Y$, $d_{L}\{X,Y\}$,
is the lower bound,  as $f$ varies in the collection of the above
homeomorphisms,
of the $L$ such that
\begin{eqnarray}
e^{-L}\leq \frac{d_{Y}(f(p),f(q))}{d_{X}(p,q)}\leq e^{L}
\end{eqnarray}
for  all  $ p\not= q $ in  $ X$.
\label{miasei}
\end{Gaiad}
According to this latter definition, minimal $\epsilon$-nets with the
same
intersection patterns  in any given two manifolds $M_1$ and
$M_2 \in  Ric[n,k,D]$, are at finite Lipschitz distance from each
other;
{\it i.e.},
\begin{eqnarray}
d_{L}(\{p_1,\ldots,p_N\},\{q_1,\ldots,q_N\}) < C
\end{eqnarray}
 where  the
constant $C$  is related to  $N_3$. It can be also shown that
the two riemannian manifolds  $M_1$ and $M_2$ are
at finite Gromov distance from each other. This circumstance is
connected
to the following result\cite{Grom} which relates
the notion of Gromov distance between
riemannian manifolds  to the Lipschitz distance between  nets
of points on such manifolds.\par
\begin{Gaiap}
If a sequence $\{M_{i}\}$ of riemannian manifolds
converges to a space $M$, (not necessarily a riemannian manifold, as
we shall
see momentarily), for the Gromov distance, then for every positive
$\epsilon$ and $\bar {\epsilon} > \epsilon$, every $\epsilon$ -
net of $M$
 is the limit for the Lipschitz distance of a sequence $N_{i}$ where
$N_{i}$ is
 an $\bar {\epsilon}$-net of $M_{i}$. Conversely,  if $M$ and
$M_{i}$ are
 riemannian manifolds in $Ric[n,k,D]$, and if  for  every $\epsilon
>0$
there exists
  an $\epsilon$-net of $M$ which is the limit   for  the Lipschitz
distance of a
  sequence of $\epsilon$-nets $N_{i}$ of $M_{i}$, then $M_{i}$
converges to
  $M$  for the Gromov distance.
\label{miasette}
\end{Gaiap}
In other words,two riemannian manifolds $M_1$ and $M_2 \in Ric[n,k,D]$
are nearby  to each other in
the Gromov distance if, given a suitably small $\epsilon >0$,  they
admit
$\epsilon$-nets, $\{p_1,\ldots,p_N\}$ and $\{q_1,\ldots,q_N\}$
respectively, such that their
Lipschitz distance is small.\par
  Thus  two riemannian
manifolds in  $Ric[n,k,D]$ get closer and
closer in the Gromov distance if we can introduce on them finer and
finer
minimal $\epsilon$-nets of geodesic balls with the some intersection
patterns. \par
As an immediate consequence of proposition \ref{miasette}
one can easily show that any
two riemannian manifolds $M_1$ and $M_2$, such that $d_G(M_1,M_2)=0$
are necessarily isometric. More in general, the preceding results
imply\cite{Grom} the
following {\it approximation}\par
 \newtheorem{Gaiat}{Theorem}
\begin{Gaiat}
For any $\epsilon > 0$ there is a {\it finite} collection
$\{M^*_a\}$, ($a=1,\ldots, k(\epsilon)$), of riemannian manifolds in
$Ric[n,k,D]$, such that
\begin{eqnarray}
 d_{G}(M, \{M^{*}_{a} \}) < \epsilon
\end{eqnarray}
 for any riemannian manifold $M$ in $Ric[n,k,D]$.
\label{miaotto}
\end{Gaiat}
  In  other words, the set $Ric[n,k,D]$ of closed riemannian
manifolds $(M,g)$
 of dimension $n$,  with Ricci curvature bounded below and diameter
 bounded above, is {\it precompact} in the set of all compact metric
spaces  endowed with the Gromov
 distance.\par
This last theorem implies that we can approximate any $M$ in
$Ric[n,k,D]$ by one of
 the $\{M^*_a\}$ in the sense that, for a given cut-off length scale
$\epsilon$,
 we can choose a finite number of {\it model} riemannian manifolds
$M^{*}_a$
 such that any given manifold in $Ric[n,k,D]$ is, on length scales
sufficiently larger
 than the cut-off $\epsilon$, metrically indistinguishable from one of
the
 manifolds $M^{*}_a$.\par
\vskip 0.5 cm
It must be stressed that for manifolds in $Ric[n,k,D]$ we have no a
priori bounds on the
injectivity radius,
and  small geodesic balls can be topologically quite complicated. This
situation is made manifest by the fact that a sequence of riemannian
manifold
in $Ric[n,k,D]$ , $\{M_i\}$,  may converge, under the Gromov distance,
to an object
which is no longer a manifold. Namely, the space $(Ric[n,k,D], d_G)$
is not
complete , (this lack of completeness explains why the approximation
theorem above  is a {\it precompactness} result). What happens is that
the
manifolds in the sequence $\{M_i\}$ may degenerate in the sense that
the
constraints on the elements of $Ric[n,k,D]$ are so weak that {\it
pathological}
metrics with  sectional curvatures uniformly bounded, diameter
uniformly
bounded above, but with injectivity radius converging uniformly toward
zero
everywhere are in $Ric[n,k,D]$. It follows that
  metric spaces of a very complex structure may result from the
Gromov-
  convergence of a sequence of riemannian manifolds in $Ric[n,k,D]$.
In general such singular spaces
can be
  characterized as {\it length spaces} \cite{Grom}, that is metric
spaces
where the
  distance between any two points is the lower bound of the length of
the
  curves joining such two points. This is a rather weak
characterization and
  the explicit structure of the boundary points of $Ric[n,k,D]$ is
largely unknown.
  The reader interested to non-trivial examples and further analysis
of these matters may profitably consult the papers of
Fukaya\cite{Grom} and Pansu\cite{Pans}.  \par
 The situation described above gets considerably simpler if we
consider the
  subspace of $Ric[n,k,D]$ generated by the class $R[n,r,D,V]$ of
closed
  riemannian n-dimensional manifolds in $Ric[n,k,D]$
with sectional curvatures, (rather than Ricci curvature), bounded
below by $r$,  and
 with volume
bounded
  below by a constant $V$. For such subspace the following results
obtain\cite{Pete} \par
\begin{Gaiat}
Let ${\cal R}(n,r,D,V)$ denote the Gromov-Hausdorff closure of the set
of
  closed riemannian n-manifolds with sectional curvatures bounded
below by
  $r$, diameter bounded above by $D$, and volume bounded below by
  $V$. Then each length space in ${\cal R}(n,r,D,V)$ is either a
riemannian $n$-manifold
  or a $n$-dimensional homology manifold. Moreover, if $n \geq 5$, a
$d_G$-limit
  of a sequence of riemannian manifolds in ${\cal R}(n,r,D,V)$ is a
topological
  $n$-manifold.
\label{mianove}
\end{Gaiat}
\vskip 0.5 cm
\noindent {\it (For notational convenience in what follows, we shall
indifferently use the word manifold and the symbol $M$ either to mean
a regular  riemannian manifold or a metric homology manifold arising
as
$d_G$-limit of a sequence of riemannian manifolds. When a more
explicit characterization is necessary, the specifications
 riemannian or  homology are  used)}.
\vskip 0.5 cm
We conclude this section by recalling the following basic
homotopy finiteness result\cite{Grov}. It provides the
topological rationale underlying the use of spaces
of bounded geometries in simplicial quantum gravity
\begin{Gaiat}
For any dimension $n \geq 2$, and  for $m$
sufficiently large, manifolds in ${\cal R}(n,r,D,V)$ with the same
$1$-skeleton ${\Gamma}_{(m)}$ are homotopically equivalent, and
the number of different homotopy-types of manifolds
realized in $\Ricco$ is finite.
\end{Gaiat}
(we recall
that two manifolds $M_1$ and $M_2$ are said to have the same
 homotopy type if there exists a continuous
 application $\phi$ of $M_1$ into $M_2$ and $f$ of $M_2$ into $M_1$,
such that both $f \cdot \phi$ and $\phi \cdot f$ are homotopic
to the respective identity mappings, $I_{M_1}$ and $I_{M_2}$.
 Obviously, two homeomorphic manifolds are of
the same homotopy type, but the converse is not true).\par
Further details and many useful examples of Gromov-Hausdorff
convergence of manifolds of bounded geometry can be found in the
quoted paper of K.Fukaya\cite{Grom}.\par

\section{Homotopy types of three-manifolds of bounded geometry
and lattice gas statistical mechanics}

In the following sections we shall explicitly assume that $n=3$, and
henceforth we shall denote by $\Ricco$ the generic
compact metric space of three-dimensional (homology) manifolds
of bounded geometry, characterized by Gromov-Hausdorff completion
of closed riemannian three-manifolds with sectional curvatures
bounded below by $k$, diameter bounded above by $D$, and volume
bounded below by $V$. As alredy stressed, many of the statements
which follows hold true regardless of the dimension, ($n\geq 2$), and
as a matter of fact one can develop a rather general theory\cite{popa}
along the line discussed below.\par
 \vskip 0.5 cm
Let us begin by recalling a theorem of Dancis\cite{Dancis}, according
to which the
simplicial isomorphism class of a simplicial three-manifold is
determined by its underlying two-skeleton. In its more general
setting,
the theorem reads (see\cite{Dancis}, th.1)
\begin{Gaiat}
Let $M$ and $W$ be compact, triangulated homology $n$-manifolds.
Let $k\geq n/2 +\frac{1}{2}$. Given a simplicial isomorphism
$f\colon {Skel}^{(k)}(M)\to {Skel}^{(k)}(W)$, there is a
simplicial isomorphism $f_k$ of $M$ onto $W$ which is an
extension of $f$.
\end{Gaiat}
(where ${Skel}^{(k)}(M)$ and ${Skel}^{(k)}(W)$ denote the
$k$-skeletons of the triangulated homology manifolds $M$ and $W$,
respectively).\par
This result suggests that
three-dimensional simplicial quantum gravity may, to some extent,
be described by the (dynamical triangulation) partition function
associated with the two-skeleton of the underlying simplicial
approximation scheme adopted.\par
\vskip 0.5 cm
\subsection{A two-dimensional partition function for three-manifolds}
In order to implement the point of view implied by
Dancis' theorem, let
$L(m)\equiv 1/m$, $0<m<\infty$ denote a cut-off
parameter to be interpreted as the radius $\epsilon /2$
of minimal geodesic balls coverings on manifolds of
bounded geometry. With this notational remark and
within the  geometrical framework delineated above,
the most suitable form, (at least for our purposes),
of a simplicial quantum gravity partition function to be
associated with a $L(m)$-geodesic ball two-skeleton is
\begin{eqnarray}
{\Xi}(m,z)  =   \sum_{\Gamma}z^{N_{(m)}^{(0)}(\Gamma)}\exp
{[\beta  N^{(1)}_{(m)}(\Gamma)]}
\label{dueuno}
\end{eqnarray}
\noindent   where $-\ln z  \equiv c$, and $\beta$  are
 constants, and where the (finite) sum  is over
 all inequivalent  $1$-skeletons ${\Gamma}_{(m)}$, with order
$N_{(m)}^{(0)}(\Gamma)$ and
size $N^{(1)}_{(m)}(\Gamma)$, realized by the  possible $L(m)$-
geodesic balls covering  over manifolds  in $\Ricco$,
(the size $N^{(1)}_{(m)}(\Gamma)$ is the number of edges in the
graph ${\Gamma}_{(m)}$). Notice that the more general form
of such partition function would be
\begin{eqnarray}
  \sum_{{\Gamma}^{(2)}_{(m)}} \exp \{-
[c_0N^{(0)}_{(m)}(\Gamma)-c_1N^{(1)}_{(m)}(\Gamma)+
c_2N^{(2)}_{(m)}(\Gamma)]\}
\label{Euler}
\end{eqnarray}
where $N_{(m)}^{(2)}(\Gamma)$ is the number of two-faces in the
two-skeleton ${\Gamma}^{(2)}_{(m)}$, and $c_0$, $c_1$, $c_2$ are
suitable constants. By exploiting Euler's relation relating the number
of faces, edges and vertices to the Euler number of the two-skeleton,
one can arrange the terms in (\ref{Euler}) so as to riproduce
(\ref{dueuno}) or other equivalent expressions.
However, it must be stressed
that the Euler number of the two-skeleton has no obvious topological
meaning for the underlying three-manifold, (contrary to what happens
for a surface), and thus there is some ambiguity in the choice
of the two-dimensional partition function to be associated with
the two-skeleton. Our choice (\ref{dueuno}) is the simplest
partition function yielding for non-trivial results. It
directly evidentiates the role of the one-skeleton
${\Gamma}^{(1)}_{(m)}$ of the geodesic balls covering, rather
then of the two-skeleton ${\Gamma}^{(2)}_{(m)}$. As a
consequence of the homotopy finiteness theorem recalled in the
previous paragraph, it follows that, at least for manifolds
in $\Ricco$, there is no loss in
generality in doing so. We shall came back to the homotopical
interplay between the one-skeleton and the two-skeleton later on,
when estimating, in terms of the  presentation of the
fundamental group, the configurational entropy associated with
(\ref{dueuno}).\par
\vskip 0.5 cm
A rigorous analysis of the
properties of (\ref{dueuno}) follows by noticing that
from a thermodynamical point of view, ({\it i.e.}, when
$m \to \infty$),  ${\Xi}(m,z)$ has the
structure of  a grand-partition  function computed  for a lattice
gas  with   negative  pair   interactions  evaluated  at inverse
temperature $\beta$  and for  a fugacity $z$. In order to
avoid any misunderstanding, it must be stressed
that the lattice gas in
question evolves on a denumerable graph $\Omega$, (to be defined
below), rather than on a regular, say hypercubic, lattice.\par
To discuss this equivalence more in details
we explicitly identify the geodesic balls of a minimal $L(m)$-net on a
manifold  $M$ of bounded geometry with the vertices
 $\{p_1^{(0)},\ldots p_N^{(0)}\}$,
 of an abstract graph ${\Gamma}(M)$. Such ${\Gamma}(M)$ is
defined by connecting the $\{p_i^{(0)} \}$ among themselves with
undirected
edges $p_{ij}^{(1)}=\{p_i^{(0)}, p_j^{(0)} \}$ if and only if the
geodesic balls
labelled $p_i^{(0)}$ and
$p_j^{(0)}$ have a non-empty intersection when we double their radius.
Any
two such graph are considered equivalent if, up to the labelling
$\{p_i^{(0)} \}$ of the balls, they have the same vertex-edge scheme,
{\it viz.} we are considering unlabelled graphs associated
with the possible one-skeletons of geodesic balls coverings of
manifolds of bounded geometries.
For any given $m$, any such graph can be topologically
imbedded in a larger regular,({\it
i.e.},
whose vertices have all the same degree), graph
${\Omega}_{(m)}$, ($\Omega$ for short), having as its
connected subgraphs all possible $1$-
skeletons ${\Gamma}_{(m)}$ which are realized by $L(m)$-geodesic
balls coverings as $M$ varies in $\Ricco$. One may think of such
$\Omega$ as the configuration space for the finite set of
possible $1$-skeletons that can
be realized by minimal nets on the manifolds in $\Ricco$.\par
For a given $m$, a graph realizing $\Omega$ can be constructively
defined  first by labelling the graphs ${\Gamma}_{(m)}$ and
order them  by inclusion,
{\it i.e.}, ${\Gamma}^{(i)}_{(m)} \subset {\Gamma}^{(j)}_{(m)}$ if
${\Gamma}^{(i)}_{(m)}$ is a subgraph of ${\Gamma}^{(j)}_{(m)}$,
and then by considering the union $H_{(m)}$ over the
{\it prime} graphs ${\Gamma}^{(k)}_{(m)}$, ({\it i.e.}, over
those graphs which do not appear as subgraphs of other
${\Gamma}_{(m)}$):
\begin{eqnarray}
H_{(m)} \equiv {\cup}^*_j {\Gamma}_{(m)}^{(j)}
\label{acca}
\end{eqnarray}
Let $d({\Omega}_{(m)})$ the maximum degree of the vertices of
$H$, (or which is the same, the maximum degree of the vertices among
those occurring in the various ${\Gamma}_{(m)}^{(j)}$). Notice that,
for every $m$, we have $d(\Omega) \leq (C_{n,r,D})^{n-1}$,
where the constant $C_{n,r,D}$ is the upper bound to the average
degree of
the graphs ${\Gamma}_{(m)}$ associated with the
 $L(m)$-geodesic balls realized in $\Ricco$, (see
proposition \ref{miaquattro}). With these
preliminary remarks, we characterize
the graph ${\Omega}_{(m)}$ according to the following \par
\newtheorem{Gaial}{Lemma}
\begin{Gaial}
The graph ${\Omega}_{(m)}$ is
 the regular graph, of degree
$d({\Omega}_{(m)})$, of the least possible order having as its
connected subgraphs all possible one-skeletons
${\Gamma}_{(m)}$ which are realized by $L(m)$-geodesic
balls coverings as $M$ varies in $\Ricco$
\label{miadieci}
\end{Gaial}
{\it PROOF}. The existence of
such graph is assured by a theorem of P.Erdos and P.Kelly\cite{Erdo},
and  in order to construct it,  we proceed as follows. First add
to $H_{(m)}$ a set $I$ of $b_{(m)}$ isolated points,(the number of
which, as indicated, is a fuction of the given $m$), a new graph is
formed from $H_{(m)}$
and $I$ by adding edges between pairs of points in $I$ and $H_{(m)}$,
(notice that no edges are added between vertices in $H_{(m)}$). The
strategy is to give rise, in this way, to a new graph which is
regular of degree $d({\Omega}_{(m)})$ while keeping the
number $b_{(m)}$ of vertices to be added to $H_{(m)}$ as small as
possible.
As expected, the number $b_{(m)}$ of added vertices depends only on
the
degree sequence of the graph $H_{(m)}$. In particular if $d_i$
denote the various degrees at the vertices of $H_{(m)}$,
then $b_{(m)}$ is the least integer satisfying: {\it (i)}
$b_{(m)}d({\Omega}_{(m)})\geq \sum_i(d(\Omega)-d_i)$, {\it (ii)}
$b_{(m)}^2-(d(\Omega)+1)b_{(m)}+\sum_i(d(\Omega)-d_i) \geq 0$, {\it
(iii)}
$b_{(m)} \geq \max (d(\Omega)-d_i)$, and {\it (iv)}
$(b_{(m)}+h_{(m)})d(\Omega)$ is even, where $h_{(m)}$ is the order
of the union graph $H_{(m)}$\cite{Erdo}.\par
It should be noted that in the large $m$ limit, ${\Omega}_{(m)}$ is
a regular graph whose degree,
$d({\Omega}_{(m)})$, is  independent
from $m$, being bounded above in terms of $n$,$r$, $D$, by
$(C_{n,r,D})^{n-1}$. $\clubsuit$ \par
\vskip 1 cm
With these preliminary remarks we can work out the correspondence
between the statistical system described by (\ref{dueuno}) and a
lattice
gas by identifying
 the collection of
 geodesic balls,   providing the dense packing of $M$, with a gas  of
indistinguishable particles.
 Each particle can
  occupy at most one  site of  ${\Omega}_{(m)}$,
    (the geodesic balls of radius $L(m)=1/m$ are
  disjoint), and  the configuration of occupied sites corresponding to
the net of
  balls covering $M$ is  defined by the vertices of the graph
  ${\Gamma}(M) \subset {\Omega}_{(m)}$.  \par
  \noindent  The interaction energy between two occupied sites is
supposed
  to be different from zero only if the sites in question  are
connected by an edge $\{p_i^{(0)}, p_j^{(0)}\}$  of ${\Gamma}(M)$,
 namely  if  the points of
the minimal $L(m)$-net
corresponding to the  lattice sites in question are in the
intersection pattern
 of the manifold  according to the definition recalled above, (the
 geodesic balls $p_i^{(0)}$ and $p_j^{(0)}$ have a non-empty
 intersection when their radius is doubled). More in general,
 $p_i^{(0)}$ and $p_j^{(0)}$ will be said to be
neighbors if $(p_i^{(0)},p_j^{(0)})$ is an edge of the graph
${\Omega}_{(m)}$. Obviously, ${\Omega}_{(m)}$ will contain as
possible configurations of occupied sites, ({\it i.e.}, as possible
subgraphs), not only those graphs ${\Gamma}(M)$ representing
one-skeleton of manifolds in $\Ricco$, but also graphs
associated to the $b(m)$ added vertices (and to the corresponding
edges), needed in order to construct ${\Omega}_{(m)}$.
In any case,
given a configuration of occupied sites in ${\Omega}_{(m)}$
represented
by a graph
 ${\Gamma} = {\Gamma}(M)$ with $N_{(m)}^{(0)}(M)$ vertices,
  the total interaction energy corresponding to such
configuration is  given   by
\begin{eqnarray}
E({\Gamma}(M)) =  - e_0N^{(1)}_{(m)}({\Gamma}(M))
\end{eqnarray}
\noindent  where $e_0$ is a constant  (which provide the scale of
energy,
and which here and henceforth we set equal to one)
and
$N^{(1)}_{(m)}$ is the  number of edges in the graph
${\Gamma}(M)$,
{\it viz.}, the total number of pairs $(i,j)$
belonging to the intersection  pattern of the minimal $L(m)$-net
$\{p_1^{(0)}, \ldots , p_N^{(0)}\}$.
\noindent The corresponding Boltzmann weight (as a lattice gas) is
then
\begin{eqnarray}
\exp [- {\beta}E({\Gamma}(M))] = \exp [{\beta}
N^{(1)}_{(m)}({\Gamma}(M))]
\end{eqnarray}
The existence of the thermodynamical limit of the statistical
sum
(\ref{dueuno}),(which exists owing to the compactness of $\Ricco$, as
we shall see), along with the boundary condition
$\lim_{m \to \infty}b(m)=0$, allows us to
 identify   ${\Xi}(m,z)$ with
 the (grand canonical ensemble) partition function of a lattice
 gas evaluated at a temperature  proportional to the average
degree of the graph representing the $1$-skeleton of the manifolds in
$\Ricco$, {\it viz.},
 \begin{eqnarray}
 KT \sim  \langle  {N_{(m)}^{(0)}}^{-1}N^{(1)}_{(m)} \rangle
\label{temp}
\end{eqnarray}
\noindent  where $K$ denotes Boltzmann's constant, and at a chemical
potential $i_c \equiv -cKT$ proportional to the
 average volume, (as measured by
$\langle N_{(m)}^{(0)}\rangle {Vol B^e(1)m^{-n}}$), of the manifolds
in $\Ricco$, averages being understood with respect to the
statistical sum (\ref{dueuno}).\par
\subsection{Gromov-Hausdorff compactness and the thermodynamical
limit}
Roughly speaking,
according to the correspondence just established the
configurations of the  lattice gas considered  are labelled by random
graphs representing the possible $1$-skeletons  ${\Gamma}_{(m)}$
of manifolds with bounded geometry, and,
for a sufficiently large $m$, we get a  grand-ensemble of
such configurations whose
 thermodynamical parameters keep track of the average volume
 and curvature  of the manifolds in $\Ricco$.
We can exploit such a correspondence in discussing the thermodynamical
limit $m \to \infty$ for (\ref{dueuno}).
 From a  technical  side,
one  can prove the existence of such limit  by exploiting the
compactness properties of $\Ricco$. These properties easily
allow to show that $\lim_{m \to \infty} {\Xi}(m,z)$ exists even
if in general this limit is not unique. To be more specific,
let us remark that if (\ref{dueuno}) is truly to be interpreted
as the grand-partition function of a lattice gas statistical
mechanics on the denumerable graph, ${\Omega}_{\infty}$, resulting
from
$\lim_{m \to \infty}{\Omega}_{(m)}$, and if this statistics
is to be connected to geometry, then we must have a measure, of
geometrical origin,
on the possible configuration space of the system : the set
of all possible subgraphs of ${\Omega}_{\infty}$, set which we rewrite
as
$\{0,1\}^{{\Omega}_{\infty}}$, (this is the set of functions
from the set of vertices of ${\Omega}_{\infty}$ to $\{0,1\}$,
expressing occupation or not of the vertices).\par
In order to discuss the existence of such a measure,
we start by noticing that since $\Ricco$ is compact,
we can easily turn it into a measure space,
(proofs of  results below  not explicitly
given are simple adaptations of proof that may be
found, for example,
in the book of K.R.Parthasarathy\cite{Yama}).\par
\begin{Gaiad}
The smallest ${\sigma}$-algebra  of  subsets of $\Ricco$ which
contains
all $d_G$-open subsets of $\Ricco$ is denoted by ${\cal  B}$ and is
called
the Borel ${\sigma}$-algebra of $\Ricco$.
\label{miaundici}
\end{Gaiad}
\begin{Gaiad}
By a measure   ${\mu}$ on $\Ricco$ we shall understand a measure
(or the
completion of a measure)  defined on ${\cal B}$. If the measure is
normalized, (${\mu}[\Ricco ]=1$), and complete we call it a
probability
measure
over $\Ricco$. The set of all measures on $[\Ricco , {\cal B}]$ will
be denoted
by ${\cal M}(\Ricco)$.
\label{miadodici}
\end{Gaiad}
In connection with the latter definition we recall the if
${\cal C}(\Ricco)$ denotes
the Banach space of all  the $d_G$-continuous real (complex) valued
functions on
$\Ricco$, (with  the {\it sup} norm), then ${\cal M}(\Ricco)$ inherits
a topology,
the weak (or vague) topology, defined by taking as a basis of open
neighborhoods
for $\mu \in {\cal M}(\Ricco)$ the sets
\begin{eqnarray}
\left \{ \nu \in {\cal M} \colon \big \vert \int f_jd{\mu} - \int f_j
d{\nu} \big \vert  < {\epsilon}_j  \right \}
\end{eqnarray}
where $j= 1, \ldots, k$ , ${\epsilon}_j >0$ and $f_j \in
{\cal C}(\Ricco)$. \par
In particular, we say that a sequence $\{P_n \colon n=1,2,\ldots\}$
in ${\cal M}(\Ricco)$  {\it converges weakly} to
$P\in {\cal M}(\Ricco)$, and write $P=\lim_{n\to\infty}^wP_n$ if
$\int f dP_n \to \int f dP$ for every $f\in {\cal C}(\Ricco)$.\par
\vskip 0.5 cm
An  important consequence of the compactness of $\Ricco$ is
that any measure
$\mu \in  {\cal M}$ is regular and that ${\cal M}$ itself is a compact
space. \par
\begin{Gaiap}
${\cal M}(\Ricco)$ is a compact metrizable space in the weak
topology.
\label{miatredici}
\end{Gaiap}
 Finally for what concerns the characterization of the
support of a measure
 in $(\Ricco, {\cal B})$ we get \par
\begin{Gaiad}
Let ${\mu} \in {\cal M}$. The set of all manifolds (homology
manifolds)
$M \in \Ricco$ with the property that ${\mu}(U) >0$ for any open set
$U$
containing $M$ is called the support of ${\mu}$. A measure ${\mu}$
whose
support reduces to a single manifold is called a point (or atomic)
measure.
Conversely a measure ${\mu}$ on $\Ricco$ is said to be non-atomic
(or diffuse) if ${\mu}(\{M\}) = 0$ for all manifolds $M\in \Ricco$.
\label{miaquattordici}
\end{Gaiad}
Notice that in the weak topology diffuse measures are
prevalent over point measures.\par
\vskip 1 cm
According to Gromov's precompactness
theorem\cite{Grom}, (see theorem \ref{miaotto}),
given  the cut-off length scale $L(m)=  1/m$,  we can  always
choose a
finite  number (depending on $L(m)$) of riemannian  manifolds,
$\{M^*_i\}$,$i=1,  \ldots,k(m)< \infty$, such  that any manifold
in $\Ricco$ is, on length scales larger  than the  cutoff $L$,
metrically similar to one of the model  manifolds $M^*_i$.\par
We blend this basic remark with measure theory on $\Ricco$,
by considering, for any given $m$,  the convex combination of point
measures defined
on $\Ricco$ by
\begin{eqnarray}
{\mu}_{(m)}({\cal B}) \equiv [{\Xi}(m,z)]^{-
1}\sum_{M^*_i}z^{N_{(m)}^{(0)}(M^*_i)}\exp
[\beta N^{(1)}_{(m)}(M^*_i)]{\delta}(M^*_i)({\cal B}) \label{cucu}
\end{eqnarray}
\noindent where ${\delta}(M^*_i)$ denotes the atomic  measure based
on the generic model manifold $M^*_i$, {\it viz.}, the
measure defined by ${\delta}(M^*_i)({\cal B})=1$ if the manifold
$M^*_i \in {\cal B}$, and $0$ otherwise, (${\cal B}$ being a Borel
subset of $\Ricco$). Obviously,
$N_{(m)}^{(0)}(M^*_i)$ and $N^{(1)}_{(m)}(M^*_i)$ respectively denote
the filling function and the size of the one-skeleton
${\Gamma}_{(m)}(M^*_i)$ associated with the minimal $L(m)$-covering
of $M^*_i$.\par
With this choice of a measure on $\Ricco$, which is the most natural
one
in the framework we are considering, the following results hold
true\par
\begin{Gaiap}
For any given $m$, let $f\colon \Ricco \to {\Omega}_{(m)}$,
$M \mapsto {\Gamma}_{(m)}(M) \subset {\Omega}_{(m)}$, the map that
with each $M$ associates its $L(m)$-geodesic balls one-skeleton,
(up to combinatorial isomorphisms). Then:\par
(i) The induced measure on $\{0,1\}^{{\Omega}_{(m)}}$ defined by
\begin{eqnarray}
P_{(m)}({\cal A}) \equiv {\mu}_{(m)}(f^{-1}({\cal A}))
\end{eqnarray}
(where ${\cal A}$ is the generic Borel subset of
$\{0,1\}^{{\Omega}_{(m)}}$), gives the probability law on
$\{0,1\}^{{\Omega}_{(m)}}$ determined by
\begin{eqnarray}
 [{\Xi}(m,z)]^{-1}\sum_{M^*_i}z^{N_{(m)}^{(0)}(M^*_i)}\exp
[\beta N^{(1)}_{(m)}(M^*_i)]{\delta}(M^*_i)(f^{-1}({\cal A}))
\label{lattice}
\end{eqnarray}
which is the natural lattice gas grand-ensemble probability law
on the graph ${\Omega}_{(m)}$ with the conditioning that the
subgraphs ${\Gamma}\subset {\Omega}_{(m)}$ which are
$L(m)$-geodesic balls one-skeletons occur with probability
one.\par
(ii)The marginal functions $P_{(m)}$ so defined tend to a
(possibly not-unique) limit distribution as $m \to \infty$.
\label{miaquindici}
\end{Gaiap}
{\it PROOF}.
The first part of {\it (i)} follows immediately from the
definition of induced measure associated with a measurable
mapping between two measure spaces.\par
Consider now the (marginals of the)
 probability laws yielding for
a grand-ensemble statistical mechanical description of
a lattice gas on the  sequence of graphs ${\Omega}_{(m)}$,
$0<m<\infty$. These laws are given by
the probability measures on $\{0,1\}^{{\Omega}_{(m)}}$ provided by
\begin{eqnarray}
P_{\Omega}({\cal A}) \equiv
 [{\Xi}(m,z)^*]^{-1}\sum_{\Gamma \in {\cal
A}}z^{N_{(m)}^{(0)}(\Gamma)}\exp
[\beta N^{(1)}_{(m)}(\Gamma)]
\label{gas}
\end{eqnarray}
where ${\cal A}$ denote the generic borel subset in the probability
space $\{0,1\}^{{\Omega}_{(m)}}$, {\it e.g.}, any given collection
of subgraphs $\Gamma$ of ${\Omega}_{(m)}$.
(Recall that the set $\{0,1\}$ is topologized by the
discrete topology, and  $\{0,1\}^{{\Omega}_{(m)}}$
by the product topology; the Borel $\sigma$-field of
$\{0,1\}^{{\Omega}_{(m)}}$ is generated by the open sets of the
product topology).
 The order $N_{(m)}^{(0)}(\Gamma)$
and the size $N^{(1)}_{(m)}(\Gamma)$ have, in Eq.(\ref{gas}),
 the usual graph-theoretical
meaning (and they are not {\it a priori} connected with any geodesic
balls covering; the use of the same symbols adopted for graphs
associated
to one-skeletons is mantained, in Eq.(\ref{gas}) and in
other  similar expressions below, for notational convenience only).
Finally,
the
grand-partition function ${\Xi}(m,z)^*$ is given by an expression
formally analogous to (\ref{dueuno})
\begin{eqnarray}
 {\Xi}(m,z)^* \equiv \sum_{\forall\Gamma \subset
\Omega}z^{N_{(m)}^{(0)}(\Gamma)}\exp
[\beta N^{(1)}_{(m)}(\Gamma)]
\end{eqnarray}
where the summation is now extended to {\it all} subgraphs of
${\Omega}_{(m)}$, (whilst in (\ref{dueuno}) the sum is a priori
restricted to the subgraphs of ${\Omega}_{(m)}$ which are one-
skeletons
of $L(m)$-geodesic balls coverings of manifolds in $\Ricco$).\par
Let ${\cal B} \equiv f[\Ricco]$ denote the $P_{\Omega}$-measurable
subset of ${\Omega}_{(m)}$ consisting of graphs, $\Gamma$, which
are realized as one-skeletons of $L(m)$-geodesic balls coverings
of $M \in \Ricco$. Since $P_{\Omega}({\cal B})>0$, the
conditional probability of ${\cal A}$ given ${\cal B}$,
$P_{\Omega}({\cal A} \vert {\cal B})=
P_{\Omega}({\cal A}\cap {\cal B})/P_{\Omega}({\cal B})$ is
provided by
\begin{eqnarray}
 P_{\Omega}({\cal A}\vert {\cal B})
=\frac{\sum_{\Gamma \in {\cal A}\cap {\cal
B}}z^{N_{(m)}^{(0)}(\Gamma)}\exp
[\beta N^{(1)}_{(m)}(\Gamma)]}
{\sum_{\Gamma \in{\cal B}}z^{N_{(m)}^{(0)}(\Gamma)}\exp
[\beta N^{(1)}_{(m)}(\Gamma)]}
\end{eqnarray}
{}From the definition of the subset ${\cal B}$ it follows that we can
equivalently rewrite the sum
$\sum_{\Gamma \in {\cal A}\cap {\cal B}}\ldots$ as the sum
$\sum_{M_i^* \in f^{-1}({\cal A}\cap {\cal B})}\ldots$, while the
sum $\sum_{\Gamma \in {\cal B}}\ldots$ reduces to ${\Xi}(m,z)$. Thus
\begin{eqnarray}
P_{\Omega}({\cal A}\vert {\cal B})=
 [{\Xi}(m,z)]^{-1}\sum_{M^*_i}z^{N_{(m)}^{(0)}(M^*_i)}\exp
[\beta N^{(1)}_{(m)}(M^*_i)]
{\delta}(M^*_i)(f^{-1}({\cal A}\cap {\cal B}))
\end{eqnarray}
which proves the second part of {\it (i)}.\par
For what concerns statement {\it (ii)}, which addresses the exixtence
of the limit distribution associated with (\ref{lattice}) as
$m \to \infty$, we can proceed as follows.\par
Let $\{m_i\}$ be a
numerical sequence with
$m_i \to \infty$, and let $\{{\mu}_{(m_i)}\}$ the corresponding
sequence
of measures defined by Eq.(\ref{cucu}). Since $\Ricco$ is a compact
metric space, its associated space of measures
${\cal M}(\Ricco)$
 besides being not-empty, it
is also (weakly) compact, (see proposition \ref{miadodici}).In
particular, convex combinations of atomic
measures such as (\ref{cucu}) are dense in ${\cal M}(\Ricco)$. It
follows that  from the sequence of
measures $\{{\mu}_{(m_i)}\}$ we can  extract, (by chosing
appropriate subsequences if necessary), a converging subsequence as
$m_i \to \infty$. In particular, we may consider the set of limits
\begin{eqnarray}
{\cal F}_{\mu} \equiv \{\mu \in {\cal M}(\Ricco) \colon \mu =
\lim_{\Gamma \uparrow \Omega_{\infty}}^w {\mu}_{(m)}\}
\label{limited}
\end{eqnarray}
where $\{\Gamma \}$ is any increasing sequence of graphs,
yielding for ${\Omega}_{\infty}$, as $m_i\to\infty$.\par
As noticed before, since ${\cal M}(\Ricco)$ is weakly compact,
${\cal F}_{\mu}$ is non-empty, and its closed convex hull may
consist of more than one measure. This latter eventuality
will indicate the possible onset of phase transitions in the
statistical system of manifolds (coverings) described in
$\Ricco$ by (\ref{lattice}).$\clubsuit$\par
\vskip 1.5 cm
The result discussed above is a general existence result, and as such
it does not accomplish very much since it can not provide us with
any detailed information on the structure of the thermodynamical
limit of (\ref{dueuno}).
In particular the  measure-theoretic proof that the closed convex hull
of
${\cal F}_{\mu}$  actually contains more than one measure
is not so obvious, since we are not working with graphs on a
regular integer lattice ${\bf Z}^d$; (critical behavior for lattice
gases on regular ${\bf Z}^d$ lattices is a considerably well-developed
part of rigorous statistical mechanics. A reason for
such success is Peierls' argument, in
the refined version developed by Pirogov and Sinai, which provides the
key tool in addressing the phase analysis of the model, and which is
not easily extended to a generic graph).\par
When the lattice gas evolves on a general (denumerable) graph, phase
transition phenomena are still the rule, but critical parameters
are strongly graph dependent, and general results are less
accessible (at least to the authors). One may still draw interesting
consequences by exploiting {\it random graphs theory}, which is
a well developed field of modern graph theory. Typical problems
which are addressed and solved in such field are, (not surprinsingly),
strongly reminiscent, of the study of phase transitions, ({\it e.g.},
the existence of {\it threeshold functions} for the connectivity
of a graph, or for finding particular types of subgraphs).
But the mathematical techniques adopted are custom-tailored to graph
theory, and not easily adapted to lattice gas statistical
mechanics.\par
\subsection{Some algebraic properties of the partition function}
Either from general
properties of random graphs or from considerations based on the
physics of regular lattice gases, phase transitions for a
lattice gas on a general
graph (thought of as the union of a monotonically increasing
sequence of finite graphs $\{ {\Gamma}_i\}$), are expected to occur
when the graphs ${\Gamma}_i\subset {\Gamma}_{i+1}\subset \ldots$
 have a large percentage of vertices on their boundaries, and
when the graphs have a sufficiently large number of edges ({\it i.e.},
the
{\it interaction} between neighboring vertices is not too weak).
The rationale behind such remark being that in this case the
configurations of occupied vertices depend very much upon boundary
conditions. More explicitly, we consider the generic
$ {\Gamma}_i$ as included
within ${\Gamma}_{i+1}\subset {\Gamma}_{i+2}\subset \ldots$, and
fix {\it a priori} the occupation values of the vertices
$p_j \not\in {\Gamma}_i$. We modify accordingly the energy of
all configurations possible on ${\Gamma}_i$ and carry out the
thermodynamic limit on this conditioned system. If the limit
shows sensible dependence upon such boundary conditions then
we are in presence of a phase transition regime.\par
Thus our strategy in addressing the study of possible phase
transitions for the system described by (\ref{dueuno}) is to
redefine the law (\ref{lattice}) by conditioning the configurations
in ${\Omega}_{(m)}\backslash \Gamma$, and discuss the behavior
of the resulting probability distributions when $m \to \infty$.\par
As a first step in this direction we  rewrite the size
$N^{(1)}_{(m)}(\Gamma)$ of the generic
graph ${\Gamma}_{(m)}\subset {\Omega}_{(m)}$ as in the following\par
\begin{Gaial}
For any $\Gamma \subset {\Omega}_{(m)}$ we get
\begin{eqnarray}
N^{(1)}_{(m)}(\Gamma) = {1 \over 2}\sum_{p \in \Gamma} \left [
\sum_{q \in {\Omega}, q \not= p}A_{\Omega}(p,q) - \sum_{q \in
{\Omega}\backslash \Gamma}A_{\Omega\backslash \Gamma}(p,q) \right ]
\label{size}
\end{eqnarray}
where  $A_{(\cdot)}(p,q)$ is the adjacency matrix
of the graph indicated, ({\it i.e.},the matrix whose entries are
$1$ if the vertices $p$ and $q$ of the graph in question are
connected by an
edge, zero otherwise). And where, with a slight abuse of
notation, we have denoted by $\Omega \backslash \Gamma$
 the graph in $\Omega$ obtained by removing all the edges, (but not
the vertices), belonging to $\Gamma$.
\label{miasedici}
\end{Gaial}
 {\it PROOF}. The proof of (\ref{size}) is trivial, since the
first term in
the square brackets is simply
the number of all edges, in ${\Omega}_{(m)}$ issuing from $p$,
 ({\it i.e.}, the degree of $\Omega$),
while the second term provides the number of all edges
 issuing from $p$ and  not belonging to the graph $\Gamma$,
({\it i.e.}, the degree of $\Omega \backslash \Gamma$ at $p$).
The overall summation  over all points $p$ in $\Gamma$ is then halfed
owing to
the simmetry underlying the argument, (more explicitly, one
exploits the handshaking lemma according to which the sum of the
degrees of a graph is twice the size of
the graph itself). $\clubsuit$\par
\vskip 1.5 cm
Since the graph ${\Omega}_{(m)}$ is regular, the expression
(\ref{size}) can be put into the form
\begin{eqnarray}
N^{(1)}_{(m)}(\Gamma) = {1 \over
2}d({\Omega}_{(m)})N_{(m)}^{(0)}(\Gamma)
- {1 \over 2}\sum_{p \in \Gamma} \left [\sum_{q \in
{\Omega}\backslash \Gamma}A_{\Omega\backslash \Gamma}(p,q) \right ]
\end{eqnarray}
We can  now take advantage of this expression for the size of the
graph ${\Gamma}_{(m)}$ in order to rewrite
${\Xi}(m,z)$ as
the polynomial
\begin{eqnarray}
{\Xi}(m,z) = \sum_{\Gamma}{\hat z}^{N_{(m)}^{(0)}(\Gamma)}\prod_{p \in
\Gamma}\prod_{q \in \Omega \backslash \Gamma}
\exp [- {1 \over 2}\beta A_{\Omega \backslash \Gamma}(p,q)]
\label{Lee-Yang}
\end{eqnarray}
 where we have introduced a {\it normalized fugacity} ${\hat z}$
according to
\begin{eqnarray}
{\hat z} \equiv z \exp [{1 \over 2} \beta d(\Omega)]
\label{fugacity}
\end{eqnarray}
(Notice that in general ${\hat z}$ depends from the given $m$
through the expression of the degree $d({\Omega}_{(m)})$, however,
for $m$ sufficiently large this dependence will eventually disappear).
We wish to stress again that the sum (\ref{Lee-Yang}) is {\it
restricted}
only to those graphs ${\Gamma}_{(m)}$ which are $L(m)$-geodesic
balls one-skeletons for manifolds in $\Ricco$, and it is
{\it not extended}
to all possible subgraphs of ${\Omega}_{(m)}$. The unrestricted
sum ${{\Xi}(m,{\hat z})}^*$, already introduced during the proof
of proposition \ref{miaquattordici}, is  given by
\begin{eqnarray}
{{\Xi}(m,{\hat z})}^* = \sum_{\forall\Gamma \subset \Omega}
{\hat z}^{N_{(m)}^{(0)}(\Gamma)}\prod_{p \in
\Gamma}\prod_{q \in \Omega \backslash \Gamma}
\exp [- {1 \over 2}\beta A_{\Omega \backslash \Gamma}(p,q)]
\label{Lee-Yang2}
\end{eqnarray}
and we can formally write
\begin{eqnarray}
{{\Xi}(m,{\hat z})}^* = {\Xi}(m,{\hat z})+ \sum_{\Omega \backslash
f(\Ricco)}
{\hat z}^{N_{(m)}^{(0)}(\Gamma)}\prod_{p \in
\Gamma}\prod_{q \in \Omega \backslash \Gamma}
\exp [- {1 \over 2}\beta A_{\Omega \backslash \Gamma}(p,q)]
\label{star}
\end{eqnarray}
where the sum over ${\Omega \backslash f(\Ricco)}$ indicates summation
over
all graphs in ${\Omega}_{(m)}$ which are not $L(m)$-one-skeletons for
manifolds in $\Ricco$.\par
Notice that in the construction of ${\Omega}_{(m)}$ out of the
one-skeleton graphs ${\Gamma}_{(m)}(M)$ with the Erdos-Kelly algoritm,
(see (\ref{acca})), no edges are introduce between the graphs
${\Gamma}_{(m)}(M)$, new edges may occurr only between the $b_{(m)}$
added vertices and between these latter vertices and the graphs
${\Gamma}_{(m)}(M)$. From such remarks it follows that the graphs in
${\Omega}_{(m)}\backslash f(\Ricco)$ are defined by
the $b_{(m)}$ added vertices and by the possible edges between them
and the graphs ${\Gamma}_{(m)}$.\par
The relation (\ref{star}) between the unrestricted, ${\Xi}^*$, and the
restricted
statistical sum ${\Xi}$, is useful for suggesting the
boundary conditions which uncover the non-trivial phase structure of
$\lim_{m \to \infty}{\Xi}(m,{\hat z})$.\par
In order to proceed in this direction, let us notice that the
unrestricted statistical sum (\ref{Lee-Yang2}),
${{\Xi}(m,{\hat z})}^*$, takes on the classical polynomial
 Lee-Yang structure\cite{Ruel}.
It follows, according to the Lee-Yang circle theorem \cite{Ruel},
that the zeroes of ${{\Xi}(m,{\hat z})}^*$, thought of as a function
of the complex variable ${\hat z}$, all lie, for each given
value of $m$, on the circle
 $\{{\hat z} \colon \vert {\hat z} \vert =1 \}$ in the plane of the
complexified fugacity
${\hat z}$. In other words\par
\begin{Gaiap}
In the plane of complexified fugacity ${\hat z}$, all the zeros
of the polynomial ${{\Xi}(m,{\hat z})}^*$ are of module one.
\label{miadiciassette}
\end{Gaiap}
{\it PROOF}. Let $S \equiv \{i_1,\ldots, i_s\}$, and
$\bar{S}\equiv \{j_1, \ldots, j_{n-s}\}$ respectively denote the
generic subset of $\{1,\ldots,n\}$ and its complement. Also,
let $(F_{ij})_{i \not= j}$ be a family of real numbers such that
$-1 \leq F_{ij} \leq 1$, $F_{ij}=F_{ji}$ for $i,j=1,\ldots, n$.
Define the polynomial (of degree $n$) in $z$ by
\begin{eqnarray}
{\cal P}^{n}(z) = \sum_{S}{z}^{N(S)}\prod_{i \in {S}}
\prod_{j \in {\bar{S}}}F_{ij}
\end{eqnarray}
where $N(S)$ is the number of elements in $S$. Then, Lee-Yang circle
theorem asserts that the zeros of ${\cal P}^n(z)$ all lie on the
circle  $\{ z \colon \vert  z \vert =1 \}$. \par
In order to apply this result to ${\Xi}(m,\hat{z})^*$ it is
sufficient to label the vertices of the graph ${\Omega}_{(m)}$ and
to identify $S$ and $\bar{S}$ with the induced labelling of the
vertices of the generic subgraph $\Gamma \subset {\Omega}_{(m)}$
and of its complement ${\Omega}\backslash{\Gamma}$, respectively.
Finally take $F_{ij}\equiv \exp [- {1 \over 2}\beta A_{\Omega
\backslash
\Gamma}(p_i,q_j)]$. $\clubsuit$ \par
\vskip 1.5 cm
For each given
$m$,  ${{\Xi}(m,{\hat z})}^*$ is a polynomial in the variable
${\hat z}$ of degree
 $\lambda \equiv \vert \Omega (m)\vert$. Since
its zeroes all lie on the unit circle, and
${{\Xi}(m,{\hat z})}^* \leq {{\Xi}(m,1)}^*$,
for  $0< {\hat z} \leq 1$, it
follows that if we choose the determination of
$[{{\Xi}(m,{\hat z})}^*]^{1/\vert \Omega (m) \vert}$ which is real for
${\hat z}>0$, then $[{{\Xi}(m,{\hat z})}^*]^{1/\vert \Omega (m)
\vert}$
defines, for $\vert {\hat z}\vert <1$ and as $m$ varies, a uniformly
bounded family of analytic functions. We can now follow a
 standard argument, \cite{Ruel} to the effect that if $\{m_i\}$ is
a numerical sequence with $m_i \to \infty$ then
$\{[{{\Xi}(m_i,{\hat z})}^*]^{1/\vert \Omega (m_i) \vert}\}$ converges
for ${\hat z} >0$, and the convergence is uniform on any
disk $\{ {\hat z} \colon \vert {\hat z} \vert \leq {\alpha}, \alpha <1
\}$.
If we introduce the thermodynamic potential,
 (the pressure of the lattice gas),
\begin{eqnarray}
 {\Psi}({\hat z}, \beta) &=\nonumber\\
&= {\beta}^{-1} \lim_{m_i \to \infty}
 \left [ {1 \over {|{\Omega}(m_i)|}} \ln  \sum_{\Gamma \subset \Omega}
{\hat z}^{N_{(m)}(\Gamma)}
\prod_{p \in \Gamma} \prod_{q \in \Omega \backslash \Gamma} \exp [- {1
\over 2}{\beta}A_{\Omega \backslash \Gamma}(p,q)]  \right  ]
\end{eqnarray}
then, the above remarks show that such ${\Psi}({\hat z}, \beta)$ can
be analytically extended form the interval $0<{\hat z} <1$ to the
function
$p({\hat z},\beta)\equiv {\beta}^{-1}\ln\{ \lim_{m \to \infty}
[{{\Xi}(m,{\hat z})}^*]^{1/\vert \Omega (m) \vert}\}$ which is
analytic
if $\vert {\hat z} \vert <1$. \par
Similarly, by exploiting the fact that under the inversion ${\hat z}
\to 1/{\hat z}$, the
partition function ${{\Xi}(m, {\hat z})}^*$ goes over into
${\hat z}^{-\vert \Omega \vert}{{\Xi}(m,{\hat z})}^*$, we can
analytically extend ${\Psi}({\hat z}, \beta)$  from the interval
$1<{\hat z}<+ \infty$ to the function
$p({\hat z}^{-1},\beta)+\ln {\hat z}$.\par
Such well-known analysis implies that for the statistical system
described
by ${{\Xi}(m, {\hat z})}^*$, a phase transition can occur, in the
continuum limit $m \to \infty$, at most for ${\hat z}=1$.
Under the
lattice gas correspondence, the onset of such phase transitions
corresponds to the familiar behavior of a lattice gas, with
negative pair interaction,  for
which first-order phase transitions are present for sufficiently
small temperatures unless the interaction is essentially
one-dimensional.\par
\subsection{The phase structure in the thermodynamical limit}
Lee-Yang type results for regular ${\bf Z}^d$-lattice
gas
relates the presence or absence of a phase transition to the analycity
properties of the free energy associated with the grand-partition
function ${\Xi}^*$. Such result depends in a rather delicate fashion
upon
the way the thermodynamic limit is carried out, {\it i.e.},
the increasing sequence of finite volume configurations must tend
{\it regularly}\cite{Ruel} towards ${\bf Z}^d$. Typically one
requires that the limit is regular in the sense of van Hove, (roughly
speaking, the boundary of each finite volume region does not grow
as to display a very irregular geometric structure). \par
In order to extend such results to our case, we could exploit
the particular geometric origin of the one-skeletons graphs
${\Gamma}_{(m)}(M)$. Roughly speaking, the idea is to show that
for $m$ sufficiently large, we can always embed the generic
${\Gamma}_{(m)}(M)$ in a region of a regular lattice such
as ${\bf Z}^d$, provided that we add (a finite number of edges)
between some of the sites of the regular lattice. It must be
stressed that edges are added and not deleted, and that the
number of edges to be added does not depend on $m$ as $m \to
\infty$.\par
In this way one can establish a correspondence between the
lattice gas described by  ${{\Xi}(m,{\hat z})}^*$  and a regular
lattice gas whose attractive interaction between sites has
been here and there enhanced. The existence of phase transitions
for this system can be discussed by standard techniques.
Obviously, all this would be rather formal, and not particularly
illuminating from a geometrical point of wiev. Thus
we proceed differently.\par
\vskip 0.5 cm
If we fix $m$ then the probability,
$P_{\Omega}({\Gamma}_{(m)}(M))$,  of occurrence of a
one-skeleton graph (of a geodesic balls covering) of a manifold
of bounded geometry among all possible subgraphs of ${\Omega}_{(m)}$
is different from zero. This follows from the very definition
of the measure space $({\Omega}_{(m)}, P_{\Omega})$, and as we
let the parameter $\beta$ to vary we can also get configurations
whereby a particular one-skeleton graph dominates over another.
Both such properties are not so obvious when we carry out
the $m \to \infty$ limit, {\it i.e.}, when
${\Omega}_{(m)} \uparrow {\Omega}_{\infty}$.
Even if we take this limit with the exterior boundary condition
$b_{(m)}=0$ imposed upon the sequence of ${\Omega}_{(m)}$, we
have no {\it a priori} reasons of assuming that the limit
distribution $P_{\Omega}$ gives non-zero probability to geodesic
balls one-skeleton graphs. Moreover, even if this is the case, the
$P_{\Omega}$-dominance of a particular class of one-skeleton
graphs over another, {\it viz.}, the question of the existence
of phase transitions needs to be explicitly answered and related to
the geometry  of the manifolds in $\Ricco$. An affermative
answer to such questions is provided by the proposition below.
In particular, the nature of its proof suggests a
strategy for proving the existence of a non-trivial phase
structure for ${\Xi}({\hat{z}})$ and for understanding its
geometrical interpretation on $\Ricco$. \par
\vskip 0.5 cm
Before stating the result referred to above, few remarks are
in order.
Let us start by pointing out that, according to the Lee-Yang
characterization of the complex zeros of ${\Xi}^*$,
if there is a phase transition in the
system decribed by ${\Xi}(m,\hat {z})^*$, as $m \to \infty$,
then this can occur at most for $\hat {z}=1$. We can use this
condition for characterizing the value of the
chemical potential $i_c \equiv -(c/{\beta})$ for which a critical
behavior is possible. From the definition of the normalized
fugacity ${\hat{z}}$, (see (\ref{fugacity})), we get
$i_c = - (1/2)d({\Omega})$. Geometrically speaking, this condition
fixes the {\it average volume} of the manifolds sampled out in
$\Ricco$ according to the probability law $P_{\Omega}$.\par
A second remark concerns the boundary condition to be imposed
on the sequence of graphs ${\Omega}_{(m)}$ in carrying over
the limit ${\Omega}_{(m)} \uparrow {\Omega}_{\infty}$.
According to the relation (\ref{star}) between ${\Xi}^*$ and
${\Xi}$, the most natural exterior boundary condition which
may significantly affect the nature of the limit distribution
$P_{\Omega}$ as $m \to \infty$ is to keep empty the
 $b_{(m)}$ added vertices needed to regularize the set , $H_{(m)}$,
of geodesic balls one-skeleton graphs, (see Lemma \ref{miadieci}). In
other words, in what follows we take the limit $m \to \infty$,
through subgraphs of $H_{(m)}$, {\it viz.},
$H_{(m)} \uparrow {\Omega}_{\infty}$. For any finite $m$ this
choice tends to favor graphs which are geodesic balls one-skeleton
graphs rather than generic graphs. However, as remarked above,
this enhancement does not trivially extend to ${\Omega}_{\infty}$.
We shall see that this type of boundary condition is
strictly connected with the topology of the manifolds sampled
out by $P_{\Omega}$.\par
Phase transitions should manifest
themselves
by a sensible dependence upon the boundary conditions introduced
above. The resulting limit distribution $P_{\Omega}$ then must
describe some local distorsion of the system around some reference
configuration, or, in more geometrical terms, fluctuations around a
given
reference homotopy type ${\pi}^*$. \par
With these preliminary remarks along the way, we can prove the
following\par
\begin{Gaiap}
For any given $m$, (sufficiently large), let
$\{{\Gamma}^*_{(m)}\}$ denote the  (finite) collection
of one-skeleton graphs associated with minimal
$L(m)$-geodesic balls covering of manifolds in $\Ricco$
of a same homotopy type ${\pi}^*$.
As $m\to \infty$, let us denote respectively by
$H_{(m)}=\{ {\Gamma}^i_{(m)} \}$ and
${\Omega}_{(m)}$ the associated sequence of
$L(m)$-geodesic balls one-skeleton graphs and the Erdos-Kelly
minimal regular graph generated by such $\{{\Gamma}^i \}$.\par
Then, as $H_{(m)} \uparrow {\Omega}_{\infty}$ and for the given
value  $-(1/2)d({\Omega})$ of the chemical potential $i_c$, there
is a ${\beta}_{cr}$ such that
\begin{eqnarray}
\lim^w_{H_{(m)}\uparrow {\Omega}_{\infty}}
P_{\Omega}\{{\Gamma}^*_{(m)}\} >0
\end{eqnarray}
for all $\beta > {\beta}_{cr}$.\par
In other words, at sufficiently low temperatures, a (denumerable)
infinite graph ${\Gamma}^*$ representing the homotopy type
${\pi}^*$ occurs with non-zero probability
among all possible subgraphs of
${\Omega}_{(m)} \uparrow {\Omega}_{\infty}$.
\label{miadiciotto}
\end{Gaiap}

{\it PROOF. First part: an activity estimate}.\par
 For a given $m$, the Erdos-Kelly graph ${\Omega}_{(m)}$,
(with labelled vertices),
contains by construction the set of graphs
$\{{\Gamma}^*_{(m)}\}$ and graphs, (not
necessarily
decribing one-skeletons of geodesic balls coverings), which do not
contain $\{{\Gamma}^*_{(m)}\}$. Probabilities associated with these
latter set of graphs
can be estimated as follows.\par
For $\lambda$ a positive integer, let ${\Omega}_{(m)}(\lambda)$ be
the set of all graphs in ${\Omega}_{(m)}$, of order $\lambda$,
which do not contain the set of graphs $\{{\Gamma}^*_{(m)}\}$.
We have
\begin{eqnarray}
P_{\Omega}\{\Gamma \subset {\Omega}_{(m)} \colon \Gamma \not\supset
\{{\Gamma}^*_{(m)}\}\}\leq \sum_{\lambda}
\sum_{{\Gamma}_{\lambda}\in {\Omega}(\lambda)}
P_{\Omega}\{\Gamma \subset {\Omega}_{(m)} \colon \Gamma \supset
{\Gamma}_{\lambda}\}
\label{stima}
\end{eqnarray}
Next, we derive an (activity) estimate expressing the fact that
if ${\Gamma}_{\lambda}$ is a fixed graph,
(in ${\Omega}_{(m)}(\lambda)$), of large order, then the probability
$P_{\Omega}\{\Gamma \subset {\Omega}_{(m)} \colon \Gamma \supset
{\Gamma}_{\lambda}\}$ should be controlled by the activity
$z^{\lambda}$
of the graph ${\Gamma}_{\lambda}$. In more geometrical
terms let ${\Gamma}_{\lambda}$ be a geodesic balls one-skeleton graph
of a given manifold $M$. Since a large order for
${\Gamma}_{\lambda}$ means a large number of  disjoint geodesic balls
packing up in $M$, {\it i.e.}, a large volume for $M$, such
an estimate expresses the not surprising fact that the
$P_{\Omega}$-probability of a  manifold of large volume is small.\par
Thus, given a fixed (labelled) graph ${\Gamma}_{\lambda} \in
{\Omega}_{(m)}(\lambda)$,
let ${\Lambda}_{(m)}(\lambda)$ denote the set of graphs
${\Gamma} \subset {\Omega}_{(m)}$ containing ${\Gamma}_{\lambda}$.
We have
\begin{eqnarray}
 P_{\Omega}\{\Gamma \subset {\Omega}_{(m)} \colon \Gamma \supset
{\Gamma}_{\lambda}\}
=\frac{\sum_{\Gamma \in {\Lambda}}z^{N_{(m)}^{(0)}(\Gamma)}\exp
[\beta N^{(1)}_{(m)}(\Gamma)]}
{\sum_{\Gamma \in{\Omega}}z^{N_{(m)}^{(0)}(\Gamma)}\exp
[\beta N^{(1)}_{(m)}(\Gamma)]}
\label{attivita}
\end{eqnarray}
To get the desidered estimate, we map the generic graph $\Gamma$
in ${\Lambda}_{(m)}(\lambda)$ into a new graph $\tilde{\Gamma}$
in ${\Omega}_{(m)}$ obtained by deleting  ${\Gamma}_{\lambda}$
out of ${\Gamma}$. We can write
$N_{(m)}^{(0)}(\Gamma)
=N_{(m)}^{(0)}(\tilde{\Gamma})+N_{(m)}^{(0)}({\Gamma}_{\lambda})$, and
$N^{(1)}_{(m)}(\Gamma) =N^{(1)}_{(m)}(\tilde{\Gamma})+
N^{(1)}_{(m)}({\Gamma}_{\lambda})+
N^{(1)}_{(m)}(\tilde{\Gamma}\to {\Gamma}_{\lambda})$, where
$N^{(1)}_{(m)}(\tilde{\Gamma}\to {\Gamma}_{\lambda})$ is the number
of edges connecting
$\tilde{\Gamma}$ to ${\Gamma}_{\lambda}$.
In order to estimate the number
of such edges let us introduce the (average) {\it superficial degree}
of the graph ${\Gamma}_{\lambda}$
\begin{eqnarray}
d_S({\Gamma}_{\lambda})\equiv \frac{2N^{(1)}_{(m)}({\Omega}\to
{\Gamma}_{\lambda})}{N^S_{(m)}({\Gamma}_{\lambda})}
\end{eqnarray}
where $N^{(1)}_{(m)}({\Omega}\to {\Gamma}_{\lambda})$ is the number of
edges
connecting ${\Omega}_{(m)}$ to ${\Gamma}_{\lambda}$, and where
$N^S_{(m)}({\Gamma}_{\lambda})$ is the number of {\it boundary}
vertices of ${\Gamma}_{\lambda}$ in ${\Omega}_{(m)}$,
({\it viz.}, the number of vertices in ${\Gamma}_{\lambda}$
which are connected with vertices in ${\Omega}_{(m)}$).\par
Since ${\Gamma}\subset {\Omega}_{(m)}$, we get
\begin{eqnarray}
N^{(1)}_{(m)}(\tilde{\Gamma}\to {\Gamma}_{\lambda}) \leq
\frac{1}{2}d_S({\Gamma}_{\lambda})
\frac{N^S({\Gamma}_{\lambda})}{\lambda} \lambda
\end{eqnarray}
Thus, if we introduce the {\it effective average degree} of
the graph ${\Gamma}_{\lambda}$ according to
\begin{eqnarray}
d_{eff}({\Gamma}_{\lambda})\equiv d({\Gamma}_{\lambda}) +
d_S({\Gamma}_{\lambda})
\frac{N^S({\Gamma}_{\lambda})}{\lambda}
\end{eqnarray}
(where
$d({\Gamma}_{\lambda})$ is the average degree of the
graph ${\Gamma}_{\lambda}$),
we can bound the generic
term of the sum appearing in the numerator of (\ref{attivita}) by
\begin{eqnarray}
z^{\lambda}\exp
[\frac{1}{2}{\beta}d_{eff}({\Gamma}_{\lambda})\lambda]
z^{N(\tilde{\Gamma})}\exp [\beta N^{(1)}_{(m)}(\tilde{\Gamma})]
\end{eqnarray}
Whence
\begin{eqnarray}
 P_{\Omega}\{\Gamma \subset {\Omega}_{(m)} \colon \Gamma \supset
{\Gamma}_{\lambda}\}
 \leq z^{\lambda}\exp
[\frac{1}{2}{\beta}d_{eff}({\Gamma}_{\lambda})\lambda]
\frac{\sum_{\tilde{\Gamma} \in
{\Omega}}z^{N_{(m)}^{(0)}(\tilde{\Gamma})}\exp
[\beta N^{(1)}_{(m)}(\tilde{\Gamma})]}
{\sum_{\Gamma \in{\Omega}}z^{N_{(m)}^{(0)}(\Gamma)}\exp
[\beta N^{(1)}_{(m)}(\Gamma)]}
\end{eqnarray}
Since each graph $\tilde{\Gamma}$ is uniquely associated with a given
$\Gamma$, the ratio appearing in the above expression is not-greater
than one. Thus we get
\begin{eqnarray}
 P_{\Omega}\{\Gamma \subset {\Omega}_{(m)} \colon \Gamma \supset
{\Gamma}_{\lambda}\} \leq z^{\lambda}\exp [
\frac{1}{2}{\beta}d_{eff}({\Gamma}_{\lambda})\lambda]
\end{eqnarray}
which, on introducing the normalized fugacity ${\hat z}$, can
be rewritten as
\begin{eqnarray}
 P_{\Omega}\{\Gamma \subset {\Omega}_{(m)} \colon \Gamma \supset
{\Gamma}_{\lambda}\} \leq {\hat z}^{\lambda}\exp [-
\frac{1}{2}{\beta}(d(\Omega)-d_{eff}({\Gamma}_{\lambda}))\lambda]
\label{estimate}
\end{eqnarray}
This provides the required activity estimate. \par
\vskip 0.5 cm
{\it Second part: an entropy estimate}.\par
\subsection{Entropy estimates for geodesic ball coverings of three-
manifolds of bounded geometry}

In order to complete the proof of  the proposition we need
an {\it entropy} estimate which would provide an exponential bound to
the number
of combinatorially inequivalent graphs
contained in the set ${\Omega}_{(m)}(\lambda)$.
To this end, we shall estimate the number of isomorphism
classes of graphs in the set $H_{(m)}$, corresponding to
geodesic balls one-skeleton graphs of manifolds of a
given homotopy type ${\pi}^*$.\par
\vskip 0.5 cm
Let us consider $L(m)$-geodesic
balls coverings whose filling function $N_{(m)}^{(0)}$ takes on the
running (integer) value $\lambda$. Correspondingly
let us introduce the function
$B_{\lambda}(V,{\Gamma}_{(m)},{\pi}(M))$ which, at scale $L(m)$,
counts the number of
combinatorially inequivalent
$1$-skeletons ${\Gamma}_{(m)}$ with $\lambda$ vertices which can
 be generated by minimal geodesic balls coverings on a manifold $M$,
of given volume $V$, in the homotopy class $\{{\pi}(M)\}$. \par
If $m$ is sufficiently large, the function
$B_{\lambda}(V,{\Gamma}_{(m)},
{\pi}(M))$  depends, besides on the volume, only on the fundamental
group ${\pi}_1(M)$ of $M$. This remark follows by noticing that there
is a
relation between the presentation, associated
with the geodesic balls covering, of the fundamental group
${\pi}_1(M)$ and the
fundamental group
${\pi}_1[(\Gamma)(M)]$ of the graph ${\Gamma}_{(m)}(M)$. As for any
finite connected graph,
${\pi}_1[(\Gamma)(M)]$ is a free group on
$1+ N^{(1)}_{(m)}(\Gamma)-N^{(0)}_{(m)}(\Gamma)$ generators.
The inclusion map
${\Gamma}_{(m)}(M) \to {\Gamma}_{(m)}^{(2)}(M)$ induces an
epimorphism  ${\pi}_1[{\Gamma}(M)] \to  {\pi}_1({\Gamma}^{(2)})$
whose kernel is generated by the elements of ${\pi}_1[{\Gamma}(M)]$
which are killed off by pasting the faces $p^{(2)}_{ijk}$ of
the geodesic ball skeleton, while the further pasting
of higher-dimensional symplices has no effect on the presentation,
and, as stated,
 the function $B_{\lambda}(V,{\Gamma}_{(m)},{\pi}(M))$ can depend only
on the fundamental groups ${\pi}_1(M)$ realized by
 the manifolds in $\Ricco$.\par
\vskip 0.5 cm
{}From the algebraic properties of ${\Xi}(m,{\hat z})^*$, (see
proposition
$7$ and the associated remarks) it easily follows that
$B_{\lambda}(V,{\Gamma}_{(m)},{\pi}(M))$ can have, at worst,
an exponential growth with a possible subleading asymptotics,
(this latter being compatible with the possible development
of singularities in ${\Xi}(m,{\hat z})^*$, as ${\hat z}\to 1$,
consequence of the zeros of ${\Xi}(m,{\hat z})^*$ accumulating on the
unit circle).
This remark would suffice our purposes, however it is largely
unsatisfactory since it does not make explicit the ${\pi}_1$-
dependence of $B_{\lambda}(V,{\Gamma}_{(m)},{\pi}(M))$.\par
In order to make explicit such dependence, let us start by noticing
that, by construction, the function
$B_{\lambda}(V,{\Gamma}_{(m)},{\pi}(M))$ is continuous under
Gromov-Hausdorff convergence, in the sense that if
$\{M_{(i)}\}$ is a sequence of manifolds in
$\Ricco$  $d_G$-converging to a (homology) manifold $M$ then,
for a given $\lambda$ sufficiently large, we have
\begin{eqnarray}
\lim_{M_{(i)}\to M}
B_{\lambda}(V(M_{(i)}),{\Gamma}_{(m)}(M_{(i)}),{\pi}(M_{(i)}))=
B_{\lambda}(V(M),{\Gamma}_{(m)}(M),{\pi}(M))
\end{eqnarray}
By relaxing the volume constraint in $\Ricco$ so to allow
for sequence of manifolds $\{M_{(i)}\}$ with three-dimensional volume
going to zero, we may have $\{M_{(i)}\}$ collapsing
to a lower dimensional manifold. The classical example in this
direction is afforded by the Berger sphere: let $g_{can}$ the
standard metric on $S^3$, and consider the Hopf fibration
$\pi\colon{S^3}\to{S^2}$. Define
$g_{\epsilon}(v,v)={\epsilon}\cdot{g_{can}}(v,v)$ if
${\pi}_*v=0$, and
$g_{\epsilon}(v,v)={g_{can}}(v,v)$ if the vector $v\in{TS^3}$ is
perpendicular to the fibre of $\pi$. It is easily checked that
$(S^3,g_{\epsilon})\in {\cal R}(n=3,D=1,V=0)$ for any
$\epsilon\leq 1$, and that
\begin{eqnarray}
\lim_{\epsilon\to 0}d_G[(S^3,g_{\epsilon}),(S^2,{\hat g})]=0
\end{eqnarray}
where ${\hat g}$ is the round metric on the two-sphere with
curvature $4$.\par
In such a case, and more in general when three-dimensional manifolds
collapse to surfaces, (see the paper by K.Fukaya quoted
in\cite{Grom} for a very clear account of such topic), the counting
function
$B_{\lambda}(V_{\epsilon},{\Gamma}_{(m)},{\pi}(M_{\epsilon}))$
approaches, as $\epsilon\to 0$, the corresponding function on the
surface $\Sigma$ resulting from the collapse, namely
\begin{eqnarray}
B_{\lambda}({\Gamma}_{(m)},{\pi}(\Sigma)) {\buildrel {\lambda \to
\infty}
\over \longrightarrow}
({\Lambda})^{\lambda}
{\lambda}^{(\frac{{\chi}(\Sigma)}{2})({\gamma}-
2)-1}
\cdot{\rho}(1+O(\frac{1}{\lambda}))
\label{Superficie}
\end{eqnarray}
where ${\Lambda}$, ${\gamma}$, and ${\rho}$ are suitable
constants,(as stressed in the introductory remarks, this
asymptotics can be obtained in a number of inequivalent
ways\cite{Pari}).  \par
\vskip 0.5 cm
Notice that the Euler characteristic, in the above expression,
characterizes from a topological point of view, the subleading
asymptotics of the graph counting function. Its role here is that
of providing the {\it homotopy cardinality} of the complex determining
the surface $\Sigma$, (this point of view on the Euler characteristic,
together with a similar remark on the
cardinality meaning of the Reidemeister torsion,
has been suggested to us by the paper of D.Fried on dynamical systems
quoted in\cite{RayS}.   We find it quite illuminating for
determining the asymptotics of the counting function for three-
dimensional manifolds). As a matter of fact, the Euler characteristic
of a finite complex is the only homotopy invariant that satisfies
${\chi}(A\cup B)={\chi}(A)+{\chi}(B)-
{\chi}(A\cap B)$ for any two sub-complexes $A$, $B$ of $\Sigma$,
with the normalization ${\chi}(point)=1$.
This remark suggests that one may prove (\ref{Superficie}) by
induction on the costruction of $\Sigma$ by joining handles
to a sphere. Similarly, we can try to determine the asymptotics
of $B_{\lambda}(V(M),{\Gamma}_{(m)}(M),{\pi}(M))$ for three-
dimensional manifolds by induction on the subcomplexes giving rise to
$M$. To be more precise, we must look for an expression for
$B_{\lambda}(V(M),{\Gamma}_{(m)}(M),{\pi}(M))$ which satisfies the
following requirements:\par
\noindent {\it (i)} its leading asymptotics is exponential in
$\lambda$,  this requirement follows from Proposition 7;\par
\noindent {\it (ii)} upon collapsing of the manifold $M$ to a
surface, it must be consistent with the surface asymptotics as
expressed by (\ref{Superficie});\par
\noindent {\it (iii)} it must take into account that homology
manifolds are allowed for in $\Ricco$;\par
\noindent {\it (iv)} if $A$ and $B$ are any two subcomplexes,
(with $\lambda$ vertices),
of the geodesic ball skeleton yielding for $M$ then we must have
\begin{eqnarray}
B_{\lambda}(M|A\cup B)B_{\lambda}(M|A\cap B)=B_{\lambda}(M|A)
B_{\lambda}(M|B)
\end{eqnarray}
where $B_{\lambda}(M|\ldots)$ stands for
$B_{\lambda}(V(M),{\Gamma}_{(m)}(M),{\pi}(M))$ evaluated for
$M$ restricted to the subcomplex $\ldots$ considered. The rationale
underlying such requirement lies in noticing that
the counting function $B_{\lambda}(V(M),{\Gamma}_{(m)}(M),{\pi}(M))$
can be interpreted as a {\it measure} on the set of graphs
considered;\par
\noindent {\it (v)}
A detailed analysis of the collapsing
mechanism\cite{Grom} underlying the transition
from $\{M_{(i)}\}$ to a lower dimensional manifold shows that
the manifold resulting from the collapse in general is a
quotient of the original three-dimensional manifold by  free
circle actions acting on the $\{M_{(i)}\}$ by isometries. This
is particularly clear in the quoted example of the Berger sphere,
where $S^2=S^3/S^1$, (more general situations illustrating this
point can be found in the quoted paper by Fukaya\cite{Grom}). Thus,
the three-dimensional asymptotics for
$B_{\lambda}(V(M),{\Gamma}_{(m)}(M),{\pi}(M))$ must take into account
not only the homotopic cardinality of the
geodesic ball  two-skeleton of the manifold $M$, but also the
homotopic cardinality of the circle fibers in  $M$.\par
\vskip 0.5 cm
The above requirements suggest the following strategy for
characterizing\par \noindent
$B_{\lambda}(V(M),{\Gamma}_{(m)}(M),{\pi}(M))$. Let us start by
recalling that according
to a theorem by Fukaya\cite{Grom},(in particular see Th.11.1), if
$\{M_{(i)}\}$ is a sequence of riemannian manifolds in
${\cal R}(n=3, D, V=0)$ converging to a length space $(X,d)$, then
there exist a $C^{\infty}$ manifold $M$ endowed with a
metric $g_M$ of Lipschitz class $C^{1,\alpha}$,
on which there is a
smooth and isometric action of the orthogonal group
$O(n=3)$ such that \par
\noindent {\it (a)} $(X,d)$ is isometric to $(M,g_M)/O(n=3)$.\par
\noindent {\it (b)} For each $p\in M$, the isotropy group
$I_p\equiv \{\eta \in O(n)|{\eta}p=p\}$ is an extension of a torus
$T^k$ by a finite group. \par
\vskip 0.5 cm
This result on the structure of the collapsed boundary points of
$\Ricco$, (with $V=0$), shows that orthogonal representations
of the fundamental group of manifolds in $\Ricco$ ought to play a
basic role in determining an expression of
$B_{\lambda}(V(M),{\Gamma}_{(m)}(M),{\pi}(M))$ which is to be
consistent with the known surface asymptotics provided by
(\ref{Superficie}). As a matter of fact, given any such
representation,
${\theta}\colon {\pi}_1(M)\to O(n=3)$, there is a topological
invariant which exactly counts circle fibers and which satisfies the
required cardinality law.
This invariant is the Reidemeister-Franz representation torsion
of the manifold $M$ associated with the given representation:
${\Delta}^{\theta}(M)$. If $A$ and $B$ are subcomplexes of the
geodesic ball nerve yielding for $M$, then
\begin{eqnarray}
{\Delta}^{\theta}(M|A\cup B){\Delta}^{\theta}(M|a\cap B)=
{\Delta}^{\theta}(M|A){\Delta}^{\theta}(M|B)
\end{eqnarray}
with ${\Delta}^{\theta}$ normalized to one over $S^1$,
(the definition of the representation torsion is not explicitly
needed here, a detailed discussion of its geometrical meaning as far
as the notion of {\it simple homotopy} is concerned
can be found in the quoted references\cite{{Cohe},{DeRham},{RayS}}, and
in the second part of this paper where
its role in $3$-
dimensional simplicial quantum gravity is discussed in great details).\par
\vskip 0.5 cm
With these preliminary remarks along the way, let ${\theta}\colon
{\pi}_1(M)\to O(n=3)$ be
an orthogonal representation of ${\pi}_1(M)$ and let
${\Delta}^{\theta}_{(a)}(M)$, with $a=1,\ldots$ denote the
corresponding (finite) set of representation torsions, (notice that
manifolds in $\Ricco$ realize a finite number of simple homotopy
types, this remark implies that, for a given fundamental group, the
number of inequivalent representation torsion actually realized is
finite).\par
Given an orthogonal representation, we let
$B_{\lambda}({\Gamma}_{(m)}(M),{\pi}(M), {\Delta}^{\theta}(M))$ denote
the function
which counts the number of combinatorially inequivalent
one-skeletons ${\Gamma}^{(1)}_{(m)}$ with ${\lambda}$ vertices
which can be generated by minimal geodesic balls coverings on a
manifold $M$, of given volume $V$, and given $R-torsion$ in the given
representation ${\theta}\colon {\pi}_1(M)\to O(n=3)$.\par
We shall
prove the asymptotic estimate for the counting function\par\noindent
$B_{\lambda}(V(M),{\Gamma}_{(m)}(M),{\pi}(M), {\Delta}^{\theta}(M))$
by an
inductive argument exploiting  geodesic ball coverings and
the cardinality laws for the Euler characteristic and
the representation torsion.\par
\vskip 0.5 cm
Let $1/m=2\epsilon$, and correspondingly let
$\{B_{\epsilon}(p_i)\}_{i=1,\ldots,\lambda}$ a minimal
$\epsilon$-net in $M\in \Ricco$, with
$B_{\epsilon}(p_i)\cap B_{\epsilon}(p_j)=\emptyset$, for every
$i$,$j$. \par
 Let us denote by
$M_{(1)}\equiv \cup_i^{\lambda}B_{\epsilon}(p_i)$, and more in
general, for any integer $k\geq 1$ we set
\begin{eqnarray}
M_{(k)}\equiv \cup_i^{\lambda}B_{k\epsilon}(p_i)
\end{eqnarray}
Note that for $k=2$ the set $\{B_{2\epsilon}(p_i)\}$ covers
$M$, while for $k$ such that $k\epsilon\geq diam(M)$
each ball $B_{k\epsilon}(p_i)$ covers $M$. Thus
for $k$ large enough
${\chi}(B_{k\epsilon}(p_i))={\chi}(M)$.\par
\vskip 0.5 cm
Since the balls $\{B_{\epsilon}(p_i)\}$ are disjoint
${\pi}_1(M_{(1)})=\oplus_i{\pi}_1(B_{\epsilon}(p_i))$,
with the generic ${\pi}_1(B_{\epsilon}(p_i))$ reducing to the unit
element only if the corresponding ball is contractible.\par
Each of the $\lambda$ groups ${\pi}_1(B_{\epsilon}(p_i))$
can be thought of as providing a colouring of the corresponding
ball, and therefore
there are ${\lambda}!$ ways of distributing the {\it labels}
${\pi}_1(B_{\epsilon}(p_i))$,
over the unlabelled  balls
$\{B_{\epsilon}(p_i)\}$ in $M$, (the {\it coordinate} labelling of the
balls arising from $\{p_i\}$ must be factored out),
consequently we set
$B_{\lambda}({\Gamma}_{(m)}(M_{(1)}),{\pi}(M_{(1)}),
{\Delta}^{\theta}(M_{(1)}))=
m^n{\rho}_1{\lambda}!$,
where ${\rho}_1$ is a constant determined by the riemannian
volume of the balls $\{B_{\epsilon}(p_i)\}$, and where
${\Gamma}_{(m)}(M_{(1)})$ is the graph with
$\lambda$ vertices and no edges generated by
$\{B_{\epsilon}(p_i)\}$.\par
On applying Stirling's formula, we get
\begin{eqnarray}
B_{\lambda}({\Gamma}_{(m)}(M_{(0)}),{\pi}(M),
{\Delta}^{\theta}(M_{(1)})=m^n{\rho}_1
\sqrt{2\pi}\exp[-\lambda +\frac{a}{12\lambda}]{\lambda}^{\lambda
+1/2}
\end{eqnarray}
where $a$ depends on $\lambda$ but it is
such that $0<a<1$.\par
 For
large $\lambda$ we can rewrite
this expression asymptotically in a more
geometrical way, (we drop the inessential factor
$\exp(a/12{\lambda})$.\par
To begin with, let us remark that if the balls
$\{B_{\epsilon}(p_i)\}$ are contractible then
${\chi}(B_{\epsilon}(p_i))=1$ and
$\lambda={\chi}(M_{(1)})$. In general, for manifolds
$M\in\Ricco$, arbitrarily small geodesic balls are not
contractible and consequently
${\chi}(B_{\epsilon}(p_i))\not=1$. Thus, it is natural to
introduce a parameter ${\gamma}_1$ according to
\begin{eqnarray}
{\chi}(M_{(1)})=
\sum_i^{\lambda}{\chi}(B_{\epsilon}(p_i))=2{\lambda}/({\gamma}_1-2)
\end{eqnarray}
(The particular choice of the ratio $2/({\gamma}_1-2)$ is
for later convenience. Roughly speaking, this ratio measures
to what extent the local $\epsilon$-balls fail to be
contractible).\par
Since the ${\pi}_1(B_{\epsilon}(p_i))$ are non-trivial, it follows
also that, for a given orthogonal
representation $\theta$ of
${\pi}_1(M_{(1)})$, the corresponding R-torsions are, in general,
non-trivial,
{\it i.e.},  ${\Delta}^{\theta}(M_{(1)})\not=1$.
Thus, for any given value of ${\Delta}^{\theta}(M_{(1)})$, of
such torsions, it is convenient to introduce a parameter,
${\Lambda}_1$, measuring the deviation from triviality of the
torsion considered, and defined according to
\begin{eqnarray}
{\Lambda}_1\equiv [e{\Delta}^{\theta}(M_{(1)})]^{-1}
\end{eqnarray}
(the presence of $e$ and the exponent $-1$ are here to provide a
symmetric match with Stirling's formula).\par
In terms of the parameters ${\gamma}_1$, ${\Lambda}_1$ so introduced
we can rewrite the counting function for $M_{(1)}$ as
a function of the Euler characteristic and of the R-torsion
according to
\begin{eqnarray}
B_{\lambda}({\Gamma}_{(m)}(M_{(1)},{\pi}(M_{(1)}),{\Delta}^{\theta}(M_
{(1)}))& {\buildrel
{\lambda \to
\infty}
\over \longrightarrow}\nonumber\\
m^n({\Lambda}_1{\Delta}^{\theta}(M_{(1)})^{\lambda}&
({\lambda}=m^nV)^{(\frac{{\chi}(M_{(1)})}{2})({\gamma}_1-
2)+1/2}
\cdot{\rho}_{1}(1+O(\frac{1}{\lambda}))
\label{ostep}
\end{eqnarray}
\vskip 0.5 cm
We now extend (\ref{ostep}) to the generic $M_{(k)}$
by induction on $k$, namely
we assume that (\ref{ostep}), (suitably adapted), holds true
for the geodesic ball covering $M_{(k-1)}$ and then show that it
holds true also for the covering $M_{(k)}$.\par
\noindent We put
\begin{eqnarray}
B_{\lambda}({\Gamma}_{(m)}(M_{(k-1)}),{\pi}(M_{(k-1)}))&{\buildrel
{\lambda
\to \infty}
\over \longrightarrow}\nonumber\\
m^n({\Lambda}_{k-1}{\Delta}^{\theta}(M_{(k-1)}))^{\lambda}
&({\lambda}=m^nV)^{(\frac{{\chi}(M_{(k-1)})}{2})({\gamma}_{k-1}-
2)+1/2}
\cdot{\rho}_{k-1}(1+O(\frac{1}{\lambda}))
\label{kstep}
\end{eqnarray}
where now ${\Lambda}_{k-1}$, ${\gamma}_{k-1}$, and ${\rho}_{k-1}$ are
suitable constants.\par
The inductive step easily follows
by rewriting $M_{(k)}=M_{(k-1)}\cup(M_{(k)}\backslash M_{(k-1)})$. A
trivial application of the cardinality laws for the
Euler characteristic and the torsion immediately yields
\begin{eqnarray}
B_{\lambda}({\Gamma}_{(m)}(M_{(k)}),{\pi}(M_{(k)}))&{\buildrel
{\lambda
\to \infty}
\over \longrightarrow}\nonumber\\
m^n({\Lambda}_{k}{\Delta}^{\theta}(M_{(k)}))^{\lambda}
&({\lambda}=m^nV)^{(\frac{{\chi}(M_{(k)})}{2})({\gamma}_{k}-
2)}
\cdot{\rho}_{k}(1+O(\frac{1}{\lambda}))
\label{step}
\end{eqnarray}
where
\begin{eqnarray}
{\gamma}_k={\gamma}_{k-1}(1-
\frac{\chi(M_{(k)}\backslash M_{(k-1)})}{\chi(M_{(k)})})+
2\frac{\chi(M_{(k)}\backslash M_{(k-1)})}{\chi(M_{(k)})}
\end{eqnarray}
and
\begin{eqnarray}
{\Lambda}_k={\Lambda}_{k-1}
({\Delta}^{\theta}(M_{(k)}\backslash M_{(k-1)}))^{-1}
\end{eqnarray}
while ${\rho}_k={\rho}_{k-1}$. \par
In order to show that this induction mechanism yields the required
asymptotics, we prove that (\ref{step}) stabilizes
for $k$ large  enough, ({\it i.e.}, (\ref{step}) does not
depend on $M_{(k)}$ any longer but just from the underlying manifold
$M$). To this end, it is sufficient to chose
any $k> k_c$ with
$k_c$ such that $(k_c-1)\epsilon > diam(M)$. Under such
hypothesis, it is sufficient to apply again the cardinality laws
for the Euler characteristic and for the representation
torsion so as to get
\begin{eqnarray}
{\chi}(M_{(k)})&={\chi}(\cup_iB_{k\epsilon}(p_i))=
{\chi}(B_{k\epsilon}(p_1))={\chi}(M)={\chi}(M_{(k-1)})\nonumber\\
{\Delta}^{\theta}(M_{(k)})&={\Delta}^{\theta}(\cup_iB_{k\epsilon}(p_i)
)={\Delta}^{\theta}(B_{k\epsilon}(p_1))={\Delta}^{\theta}(M)={\Delta}^
{\theta}(M_{(k-1)})
\end{eqnarray}
Moreover
\begin{eqnarray}
{\gamma}_k = {\gamma}_{k-1}= {\gamma}_{k_c-1}\equiv {\gamma}_c
\end{eqnarray}
and similarly
\begin{eqnarray}
{\Lambda}_k={\Lambda}_{k-1}={\Lambda}_{k_c-1}\equiv {\Lambda}_c
\end{eqnarray}
These relations, once introduced in (\ref{step}), immediately
yield the asymptotics for the one-skeleton graphs counting function
on a manifold $M\in \Ricco$, namely
\begin{eqnarray}
B_{\lambda}({\Gamma}_{(m)}(M),{\pi}(M),{\Delta}^{\theta}(M))&
{\buildrel {\lambda \to
\infty}
\over \longrightarrow}\nonumber\\
m^n({\Lambda}_{c}{\Delta}^{\theta}(M))^{\lambda}&
({\lambda}=m^nV)^{(\frac{{\chi}(M)}{2})({\gamma}_{c}-
2)-1/2}
\cdot{\rho}_{c}(1+O(\frac{1}{\lambda}))
\label{Itzy}
\end{eqnarray}
where the added factor $-1$, (yielding for $-1/2$),  in
the power determining the subleading asymptotics,
$(\frac{{\chi}(M)}{2})({\gamma}_{c}-2)+1/2$, comes about by noticing
that since
\begin{eqnarray}
 {\Gamma}_{(m)}(M_{(k)})=\cup_i^{\lambda}({\Gamma}_{(m)}
(B_k{\epsilon}(p_i)))=\cup_i^{\lambda}({\Gamma}_{(m)}(M))
\end{eqnarray}
we get
\begin{eqnarray}
B_{\lambda}({\Gamma}_{(m)}(M_{(k)}),{\pi}(M),
{\Delta}^{\theta}(M_{(k)}))=
{\lambda}B_{\lambda}({\Gamma}_{(m)}(M),{\pi}(M),{\Delta}^{\theta}(M))
\end{eqnarray}
\vskip 0.5 cm
It is easily checked that, up to the natural arbitrariety connected
with the parameters ${\Lambda}_c$, ${\gamma}_c$, and
${\rho}_c$, (\ref{Itzy}) is consistent
with all the requirements {\it (i)-(v)}. \par
\vskip 0.5 cm
In order to obtain the full counting function
$B_{\lambda}({\Gamma}_{(m)}(M),{\pi}(M))$ from
(\ref{Itzy}) one should first sum over all representation torsion
associated with a given representation ${\theta}$ of the
given fundamental group in $O(n=3)$, (this sum is finite owing to
the simple homotopy finiteness theorem which holds true for spaces of
bounded geometry), then we should
{\it average} over all possible inequivalent representations of
${\pi}_1(M)$ in the orthogonal group, this a rather delicate technical
point that we do not address here.  It
 will be  (partially) settled down only in the final part of the
paper, where we show  how such representations of the fundamental
group are deeply rooted in the structure of three-dimensional
simplicial gravity. \par
\vskip 0.5 cm
With the above remarks in mind, we are now in position for deriving
the required entropy estimate
for the number of isomorphism classes of graphs in
${\Omega}_{(m)}(\lambda)$. \par
\vskip 0.5 cm
According to the homotopy finiteness theorem recalled above, only
a finite number of homotopy types of manifolds are realized
in $\Ricco$. Thus, by the definition of ${\Omega}_{(m)}(\lambda)$,
and of the counting function
$B_{\lambda}(V,{\Gamma}_{(m)},{\pi}(M))$, the number,
$\vert {\Omega}_{(m)}(\lambda)\vert$, of isomorphism
classes of graphs of order $\lambda$, contained in the set
${\Omega}_{(m)}(\lambda)\cap {H}_{(m)}$ is bounded  above, for
a given $V$, by,
(recall that we are conditioning the allowable configurations
in ${\Omega}_{(m)}$ by having the $b_{(m)}$ {\it boundary}
vertices empty)
\begin{eqnarray}
\vert {\Omega}_{(m)}(\lambda) \vert \leq \sum_{\pi}
B_{\lambda}({\Gamma}_{(m)},{\pi}(M))
\end{eqnarray}
where $\sum_{\pi}$ denotes the finite sum over all
homotopy types in $\Ricco$.
Thus, in the $m \to \infty$ limit where $\lambda \to \infty$,
$\vert {\Omega}_{(m)}(\lambda) \vert$ can, according to
(\ref{Itzy}), be bounded above by
\begin{eqnarray}
\vert {\Omega}_{(m)}(\lambda) \vert \leq
\sum_{\pi}
({\Lambda}_c{\Delta}^{\theta}(M))^{\lambda}
{\lambda}^{(\frac{{\chi}(M)}{2})({\gamma}_c-
2)-1}
\cdot{\rho}_{c}(1+O(\frac{1}{\lambda}))
\label{ultima}
\end{eqnarray}
which provides the required entropy estimate.\par
\vskip 1 cm
If we put together (\ref{stima}), (\ref{estimate}), and
(\ref{ultima}) we get for the probability of not-sampling
$\{{\Gamma}^*_{(m)}\}$ out of ${\Omega}_{(m)}$ the estimate
\begin{eqnarray}
& P_{\Omega}\{\Gamma \subset {\Omega}_{(m)} \colon \Gamma \not\supset
\{{\Gamma}^*_{(m)}\}\}\leq \nonumber \\
& \leq \sum_{\pi}\sum_{\lambda}
z^{\lambda}\exp{[\frac{1}{2}{\beta}d_{eff}({\Gamma}_{\lambda})
{\lambda}]}
{\rho}_{c}
({\Lambda}_c{\Delta}^{\theta}(M))^{\lambda}
{\lambda}^{(\frac{{\chi}(M)}{2})({\gamma}_c-
2)-1}
\label{watussi}
\end{eqnarray}
It follows that  if for each given $m_0$ and each $\epsilon >0$,
there exists some $m > m_0$ such that
\begin{eqnarray}
{\beta}[(\frac{c}{\beta}) -\frac{1}{2}d_{eff}({\Gamma}_{\lambda})]
\geq \log ({\Lambda}_{c}{\Delta}^{\theta}(M)) + \epsilon
\label{regimecrit}
\end{eqnarray}
for $\lambda$ sufficiently large,then the sum appearing at the right
side of (\ref{watussi}) is finite
and converges to zero as $\beta \to \infty$, for
$(c/\beta)=(1/2)d({\Omega})$ and
$d_{eff}({\Gamma}_{\lambda}) < d({\Omega})$.\par
Notice that this latter condition implies that
$\lim_{\lambda \to \infty}d_S({\Gamma}_{\lambda})
\frac{N^S({\Gamma}_{\lambda})}{\lambda}$ should be sufficiently small,
namely that ${\Gamma}_{\lambda}$ is not a {\it thin} graph,
like a necklace of occupied sites scattered in ${\Omega}_{(m)}$.
Critical behavior requires graphs ${\Gamma}_{\lambda}$
whose {\it volume} growth term $N_{(m)}^{(0)}({\Gamma}_{\lambda})$
dominates
over the {\it surface} growth term
$N^S_{(m)}({\Gamma}_{\lambda})$.\par
In such a case, given
$0 < {\delta} < (1/2)$,
the above convergence properties for (\ref{watussi}) imply that
there exists an inverse temperature $\bar{\beta}$, which is
independent from $m$, (since $d({\Gamma}_{\lambda})$ is bounded
above by $d(\Omega)$ in terms of the parameters
$n$, $r$, $D$, and $V$), such that
\begin{eqnarray}
P_{\Omega}\{\Gamma \subset {\Omega}_{(m)} \colon \Gamma \not\supset
\{{\Gamma}^*_{(m)}\}\} \leq \frac{1}{2} - \delta < \frac{1}{2}
\label{mbe}
\end{eqnarray}
for all $\beta \geq \bar{\beta}$.\par
Now, if we denote by ${\bf E}(\{{\Gamma}^*_{(m)}\})$ the mathematical
expectation  for the $P_{\Omega}$ occurrence of the set of one-
skeleton
graphs $\{{\Gamma}^*_{(m)}\}$ in ${\Omega}_{(m)}$, then we can write
\begin{eqnarray}
& {\bf E}(\{{\Gamma}^*_{(m)}\}) = \nonumber \\
& =P_{\Omega}\{\Gamma \subset {\Omega}_{(m)} \colon \Gamma \supset
\{{\Gamma}^*_{(m)}\}\}-
P_{\Omega}\{\Gamma \subset {\Omega}_{(m)} \colon \Gamma \not\supset
\{{\Gamma}^*_{(m)}\}\}= \nonumber \\
& = 1-
2P_{\Omega}\{\Gamma \subset {\Omega}_{(m)} \colon \Gamma \not\supset
\{{\Gamma}^*_{(m)}\}\} > 2{\delta}
\label{dimostrazione}
\end{eqnarray}
where, in the last two lines, we have used the normalization of
the probability $P_{\Omega}$ and  the uniform bound (\ref{mbe}).
Since the bound (\ref{dimostrazione}) is independent from $m$,
we can apply the dominated convergence theorem and
the proof of the proposition is completed. $\clubsuit$.\par
\vskip 1 cm
The above proof  directly shows the existence
of distinct limit distributions in the
limit set ${\cal{F}}_{\mu}$, (see (\ref{limited}) and
the associated remarks), and provides the geometrical
meaning of the various phases described by
$\lim_{m \to \infty}{\Xi}(m,\hat{z})$. \par
\begin{Gaiap}
Let ${\pi}^*_{(i)}$, (with
$i=1,\ldots, |{\pi}^*|<\infty$),
denote the finite collection of
distinct homotopy types realized in $\Ricco$.
Then, as $H_{(m)} \uparrow {\Omega}_{\infty}$, there
exists a critical inverse temperature ${\beta}_{cr}$ such that
for ${\beta} > {\beta}_{cr}$ we have $|{\pi}^*|$ limit
distributions in the limit set ${\cal{F}}_{\mu}$, namely
\begin{eqnarray}
{\mu}^{(i)} = \lim_{H_{(m)} \uparrow {\Omega}_{\infty}}
E(\{{\Gamma}^*_{(m)}\}^{(i)})
\end{eqnarray}
where  $\{{\Gamma}^*_{(m)}\}^{(i)}$ denotes the set of $L(m)$-one-
skeleton
graphs of manifolds with homotopy type ${\pi}^*_{(i)}$,
and where $E$ denotes expectation with respect to the
measure $P_{\Omega}$.
\label{miadiciannove}
\end{Gaiap}
{\it PROOF}. This is an immediate consequence of the
proof of the previous proposition. $\clubsuit$ \par
\vskip 0.5 cm
The phase structure of the
statistical system described by the partition
function ${\Xi}(m,\hat{z})^*$ results from the classical
mechanism expressing the competition between energy
and entropy. This provides a first (but incomplete) explanation of
how topology enters  in describing the various
phases: the entropy estimate (\ref{ultima}) is parametrized
by the Euler-Poincar\'e characteristic, to the effect that the number
of configurations
accessible to a geodesic balls one-skeleton graph depends on the
parameter ${\gamma}_c$. As long as ${\gamma}_c-2>0$, the largest
entropic contribution to the partition function comes from
three-manifolds , (${\chi}(M)$=0), while, if
${\gamma}_c-2<0$, pseudo-manifolds, (${\chi}(M)<0$), dominate.\par

\section{Simple homotopy types and the nature of
three-dimensional lattice quantum gravity}
The analysis of the previous paragraphs shows that the thermodynamical
limit $m \to \infty$ of the grand-partition function
${\Xi}({\hat z},m)$, associated with geodesic balls two-skeletons,
provides a natural framework for discussing
fluctuations of the homotopy type of the manifolds in $\Ricco$.
As we shall see in the remaining part of this paper,
the need of controlling, in a similar statistical
way, either the topology or the smoothness of the three-manifolds
sampled
by ${\Xi}$, naturally yields to (euclidean) simplicial
three-gravity. We wish to stress that now, we do not start
by directly introducing the generalization to dimension three
of ${\Xi}({\hat z},m)$, but rather we derive the partition function
of simplicial three-gravity from few elementary first principles
connected with a natural statistical reconstruction of the nerve
of a geodesic balls covering out of the presentations of the
fundamental group associated with the geodesic balls two-skeleton.\par
Instrumental to such a derivation is, again, the use of the
appropriate finiteness theorems which hold true for $\Ricco$. In this
connection let us remark that\cite{Pete} for $n=3$, spaces of
bounded geometries such as $\Ricco$ besides containing at most
finitely
many homotopy types, contain also finitely many
{\it simple homotopy} types, (actually, this result holds true
in any dimension).
We recall that simple homotopy is a particular homotopy equivalence
between
two spaces, obtained through a finite sequence of (elementary)
 expansions and collapses of cells, (given any cellular decomposition
of the spaces in question; see below for the technical
aspects associated with these definitions). \par
That simple homotopy ought to play a significant role in simplicial
gravity is also indicated by the structure of the entropy estimate
(\ref{Itzy}),
where the representation torsions are connected to the leading
asymptotics of the gedesic ball one-skeleton graphs. Such torsion, as
will be explained more in details below, are invariants of simple
homotopy rather than of hommotopic equivalences. \par
Loosely speaking, we can
exploit  simple homotopy in order to reconstruct the nerve of a
geodesic ball covering out of its two-skeleton, (within a given
homotopy type). Notice that since the boundary points
in $\Ricco$ obtained by Gromov-Hausdorff completion are not
topological
manifolds, (they are $n$-dimensional Homology manifolds, {\it i.e.},
$H_*(M,M-p)\simeq H_*(R^n,R^n-0)$ for all $p \in M$), nothing else
can be said in full generality on how manifolds topologies
are distributed in $\Ricco$. Thus the first
step in order to discuss to what extent we can statistically
control manifold
topologies in $\Ricco$, is to understand how simple homotopy interacts
with minimal geodesic balls coverings.
\vskip 0.5 cm
\subsection{Formal deformations of geodesic ball skeletons and
Whitehead torsions}

With the foregoing preliminary remarks along the way,
let us denote by $\{\pi (M)\}$ the homotopy class
associated with the generic manifold $M \in \Ricco$. Notice that if
$m$ is sufficiently
large  then the corresponding nerve is in the
same homotopy class of the manifold, and one may use such a
nerve in order to label the associated homotopy type. \par
The interaction between simple homotopy theory and the properties of
nerves of coverings of manifolds is naturally framed in the theory of
CW-complexes.\par
We recall that a CW-complex $K$ is
a Hausdorff space along with a family, $\{e_{\alpha}\}$, of open
topological cells of various dimensions such that, if we set
$K_j = \cup\{e_{\alpha}\vert dim({e}_{\alpha}) \leq j\}$, then the
following conditions are satisfied : \par
{\it (i)} $K= \cup_{\alpha}e_{\alpha}$ and
$e_{\alpha}\cap e_{\beta}=\emptyset$ whenever $\alpha \not= \beta$,
\par
{\it (ii)} for each cell $e_{\alpha}$ there is a map
${\psi}_{\alpha} \colon Q^n \to K$, where $Q^n$ is a topological
ball, (homeomorphic to $[0,1]^n$), of dimension $n= dim (e_{\alpha})$,
such that ${\psi}_{\alpha}$ restricted to the interior of $Q^n$ is a
homeomorphism onto $e_{\alpha}$, while the image, under
${\psi}_{\alpha}$
of the boundary $\partial Q^n$ of $Q^n$ is contained in $K_{n-1}$.
The map ${\psi}_{\alpha}$ is called the {\it characteristic map} for
the
cell $e_{\alpha}$, while ${\psi}\vert \partial Q^n$ is called the
attaching map for $e_{\alpha}$. The CW-complex $K$ is $n$-dimensional
if there are no cells of dimension greater than $n$.
 Finally, a CW-pair $(K,L)$ is a CW-complex $K$ together
with a subset $L \subset K$ which is also a CW-complex.\par
\vskip 0.5 cm
Let us consider the pair
 $({\cal N},{\Gamma}_2)$, where ${\cal N}$ and
${\Gamma}_2$ respectively denote the nerve and the associated
two-skeleton of a sufficiently fine geodesic balls covering of
the generic manifold $M$ in $\Ricco$. According to the
above remarks, we  can consider $({\cal N},{\Gamma}_2)$ as a
CW-pair generated by adjoining cells to the graph
${\Gamma}_{(m)}$. If the parameter $m$ is sufficiently large we  can
further assume that the nerve ${\cal N}$ has the same dimension of the
(homology) manifold which it approximates, and here we
restrict our attention  to dimension  equal to three,
$n=3$.\par
Under such assumptions, the CW-pair $({\cal N},{\Gamma}_2)$
is {\it homotopically trivial}, ${\cal N} \searrow {\Gamma}_2$,
namely ${\Gamma}_2$ is a strong deformation retract of ${\cal N}$.
This can be easily proven by noticing that the pair
$({\cal N},{\Gamma}_2)$ is such that
${\pi}_j({\cal N},{\Gamma}_2)=0$, for each
$j \leq dim ({\cal N}-{\Gamma}_2)$, where
${\pi}_j({\cal N},{\Gamma}_2)$ denotes the relative homotopy
groups of $({\cal N},{\Gamma}_2)$.\par
\vskip 0.5 cm
Our first step is to rewrite the
generic pair $({\cal N},{\Gamma}_2)$ in a simplified form, which,
roughly speaking, directly tells us how ${\cal N}$ is
homotopically built out
by adjoining cells to the generic vertex of ${\Gamma}_2$.\par
The building block for such construction is a geometrical operation
called a {\it formal deformation}, which is defined by a finite
sequence of either elementary collapses or elementary expansions.
One says that
$K$ collapses to $J$ if $K=J\cup e^n \cup e^{n-1}$ and if the cells
$e^n$ and $e^{n-1}$ are not in $J$ ,
and where the characteristic maps of $e^n$ and $e^{n-1}$,
respectively ${\phi}\colon {Q}^n \to K$ and
${\phi}|{Q}^{n-1}\to K$, are such that
${\phi}(\partial{Q}^n-{Q}^{n-1})\subset J$.\par
Conversely, an elementary expansion is the inverse operation which
corresponds to  attaching  a cell to $J$ along a face of the cell
by a suitable map, so as to generate a new complex $K$.\par
 If there is  formal deformation between two complexes
$K$ and $J$, we write $K \nearrow \searrow J$, and $K$ and $J$ are
said to have the same {\it simple-homotopy type}. If $K$ and
$J$ have a common subcomplex ${\hat K}$, whose cells are not
removed during the process of formal deformation, then we
write $K \nearrow\searrow J$ {\it rel} ${\hat K}$.\par
\vskip 0.5 cm
There is a basic result in simple homotopy theory,
(see {\it e.g.}, the book of Cohen\cite{Cohe}, chap.II), which along
with the above notational remarks, can be easily established for
the CW-pair $({\cal N},{\Gamma}_2)$. Let $h= dim({\cal N}-
{\Gamma}_2)$,
and let $q \geq h-1$ be an integer.Let $e^0$ be a $0$-cell of
${\Gamma}_2$, (a vertex). Then there is a formal deformation from
${\cal N}$ to a new CW-complex  ${\cal M}$, which does not alter
the underlying ${\Gamma}_2$, {\it i.e.},
${\cal N}\nearrow\searrow {\cal M}$, {\it rel} ${\Gamma}_2$, such that
\begin{eqnarray}
{\cal M}={\Gamma}_2
\cup\bigcup_{j=1}^ae^q_j\cup\bigcup_{i=1}^ae^{q+1}_i
\label{Schifo}
\end{eqnarray}
where the $e^q_j$ and $e^{q+1}_i$ have characteristic maps
${\psi}^{(q)}_j\colon Q^q \to {\cal M}$ and
${\psi}^{(q+1)}_i\colon Q^{q+1} \to {\cal M}$, respectively. And
where ${\psi}^{(q)}_j(\partial Q^q)= e^0$, namely the $q$-cells are
trivially
attached at the vertex $e^0$. If in the relation (\ref{Schifo}), we
choose the integer $q$ in such a way that  $q\geq 2$, then we say that
the pair $({\cal M},{\Gamma}_2)$ is in {\it simplified form}.\par
\vskip 0.5 cm

We recall that the {\it relative homotopy groups} ${\pi}_q(X,A,x)$,
with $q\geq 2$, of a triple $(X,A,x)$, ($X$ a topological space,
$A$ a closed subspace of $X$ containing the point $x$), are defined
by the collection of arcwise-connected components of the
function space $F_q(X,A,x)$, generated by all mappings
$f\colon Q^q \to X$ such that $f(\partial {Q}^q)$ lies in $A$,
and $f[Q^1\times \partial (Q^{q-1})\cup (0\times Q^{q-1})]=x$,
(the  set inside the square brackets
$J^{q-1} \equiv [Q^1\times \partial (Q^{q-1})\cup (0\times Q^{q-1})]$
is  the boundary of the hypercube $Q^q$ minus the open top face).
It is known that the relative  homotopy groups
${\pi}_q(X,A,x)$ are abelian groups as soon as $q\geq 3$, they
can be provided also with a ${\bf Z}{\pi}_1$-module structure.
More in details\cite{Cohe}, if $\alpha$ and $\phi$
respectively represents the homotopy classes $[\alpha]$ and
$[\phi]$ in ${\pi}_1(A,x)$ and ${\pi}_q(X,A,x)$, then
${\pi}_1$ acts on the relative homotopy groups by
$[\alpha]\cdot [\phi]=[\bar {\phi}]$, where $[\bar {\phi}]$ is
generated by dragging ${\phi}[J^{q-1}]$ along the
loop ${\alpha}^{-1}$. This action yields for the
${\bf Z}{\pi}_1$-module structure of
${\pi}_q(X,A,x)$ if we define the multiplication by
\begin{eqnarray}
(\sum_i {\lambda}_i [{\alpha}_i])[\phi]=\sum_i{\lambda}_i
([{\alpha}_i]\cdot [\phi])
\end{eqnarray}
In other words, each class $[\alpha] \in {\pi}_1$ characterizes
an automorphism of the group ${\pi}_q(X,A,x)$ by mapping the
homotopy class $[\phi]$ into $[\bar {\phi}]$.\par
\vskip 0.5 cm
The relative homotopy groups of interest to us are
obviously those generated by the homotopy classes of
the characteristic maps ${\psi}^{q+1}_i$ and ${\psi}^q_i$,
{\it viz.},
${\pi}_{q+1}({\cal M},{\Gamma}_2
\cup\bigcup_{j=1}^ae^q_j)$ and
${\pi}_q({\Gamma}_2
\cup\bigcup_{j=1}^ae^q_j,{\Gamma}_2)$, respectively. If we
consider them as  ${\bf Z}{\pi}_1$-modules in the sense recalled
above, then it can be shown that\cite{Cohe}
${\pi}_{q+1}({\cal M},{\Gamma}_2
\cup\bigcup_{j=1}^ae^q_j)$ and
${\pi}_q({\Gamma}_2
\cup\bigcup_{j=1}^ae^q_j,{\Gamma}_2)$ are
{\it free}-${\bf Z}{\pi}_1$-modules with bases
$\{[{\psi}^{q+1}_i]\}_{i=1,\ldots,a}$ and
$\{[{\psi}^{q}_i]\}_{i=1,\ldots,a}$, respectively.\par
\vskip 0.5 cm
The Whitehead torsion characterizing the simple
homotopy type of the pair $({\cal M}, {\Gamma}_2)$,
is read out  through a boundary
operator. Recall that, for $q\geq 2$, the boundary function
$\partial \colon F_q(X,A,x)\to F_{q-1}(X,A,x)$ defined
by $(\partial {f})(t_1, t_2, \ldots, t_q)=
f(1, t_2, \ldots, t_q)$ induces a homomorphism, (which we
denote again by $\partial$), of ${\pi}_q(X,A,x)$ into
${\pi}_{q-1}(A,x)$. In our case, we actually get an isomorphism,
(since ${\cal{M}}$ retracts on ${\Gamma}_2$),
\begin{eqnarray}
0\to {\pi}_{q+1}({\cal M},{\Gamma}_2
\cup\bigcup_{j=1}^ae^q_j) {\to}^{\partial}
{\pi}_q({\Gamma}_2
\cup\bigcup_{j=1}^ae^q_j,{\Gamma}_2)\to 0
\end{eqnarray}
 given by
\begin{eqnarray}
\partial [{\psi}^{q+1}_i] = \sum_j w_{ij}[{\psi}^{q}_j]
\label{vincolo}
\end{eqnarray}
where $w_{ij}$ is the $(a\times {a})$ non-singular
${\bf Z}{\pi}_1$-incidence matrix defining associated with
the pair $({\cal M}, {\Gamma}_2)$.\par
\vskip 0.5 cm
The matrix $w_{ik}$ is naturally acted upon by a set of operations
which consists in: {\it (i)} Multiply on the left the $i$-th
row of the matrix by  (plus or minus) an element of the
fundamental group ${\pi}_1({\Gamma}_2,e_0)$; {\it (ii)}
Add a left group-ring multiple of one row to another; {\it (iii)}
Expand the matrix by adding a corner identity matrix. \par
Such operations give rise to a new incidence matrix which generates
a simplified pair in the same simple-homotopy class of
$({\cal M}, {\Gamma}_2)$.\par
The equivalence classes, ${\tau}(w_{ik})$, under the
operations {\it (i), (ii), (iii)},
generated by the non-singular incidence  matrices $w_{ik}$,
 are the Whitehead torsions associated
with the matrices $w_{ik}$. The collection of such ${\tau}(w_{ik})$
gives rise to a group, the
Whitehead group of ${\pi}_1({\Gamma}_2)$, $Wh({\pi}_1({\Gamma}_2))$.
\vskip 0.5 cm
In general, the Whitehead group is an infinite-dimensional
(abelian) group, however, similarly to what happens to the
number of distinct homotopy types, the number of
inequivalent simple homotopy types realized by manifolds
in $\Ricco$ is {\it finite}. Thus,
independently from $m$, there are only a finite number
of inequivalent Whitehead torsions ${\tau}(w_{ik})$ realized as
we consider the totality of finer and finer geodesic balls
coverings of manifolds in $\Ricco$.\par

\subsection{Gaussian insertion of three-cells and simplicial gravity}

With the foregoing remarks along the way, our strategy
will be aimed to providing a statistical procedure for generating,
a simplified pair $({\cal M}, {\Gamma}_2)$
out of the presentation of the fundamental group ${\pi}_1(M)$
associated with
the generic $L(m)$-geodesic balls two skeleton ${\Gamma}_2$.
To this end,
let ${\tau}_{(i)}({\pi}_1, w)$, with $i=1,\ldots, k <\infty$, denote
the
distinct Whitehead torsions realized by the
inequivalent, (in the simple homotopy sense), pairs
$({\cal M}, {\Gamma}_2)$ introduced above.
As recalled, any such torsion is described, (modulo
the trasformations described by {\it (i), (ii), (iii)}),by
a non-singular $a_{(m)}\times a_{(m)}$ matrix,
$w_{jl}({\pi}_1({\Gamma}_2, e_0))$, whose entries are elements
of the group ring ${\bf Z}({\pi}_1({\Gamma}_2, e_0))$.\par
\vskip 0.5 cm
Consider an orthogonal representation,
${\theta}$,
of  ${\pi}_1({\Gamma}_2, e_0)$, say by orthogonal $p\times p$
matrices,
turning the $p$-dimensional Euclidean space ${\bf R}^p$ into a right
${\bf R}({\pi}_1({\Gamma}_2, e_0))$-module. Correspondingly,
we denote by ${\theta}_*[w_{jl}({\pi}_1({\Gamma}_2, e_0))]$,
(if no confusion arises we shall write ${\theta}_*(w_{il})$),
the image through $\theta$ of the Whitehead torsion $w_{jl}$,
and with
${\Delta}^{\theta}({\cal M}, {\Gamma}_2)=
\det ({\theta}_*(w_{jl})$, the associated  Reidemeister-Franz
representation torsion in the represention  ${\theta}$.\par
Notice that
${\theta}_*(w)$ is a $a_{(m)}\times a_{(m)}$ matrix with entries in
${Mat}_p({\bf R})$, namely a matrix of order $pa_{(m)}$
with real entries, (${Mat}_p({\bf R})$ denotes the ring
of all $p\times p$ matrices with real entries).\par

\vskip 0.5 cm
According to the above remarks, the relative homotopy groups
describing
the {\it attachement} of the two-dimensional and three-dimensional
cells to ${\Gamma}_2$, which gives rise to
$({\cal M}, {\Gamma}_2)$, respectively are
${\pi}_2({\Gamma}_2
\cup\bigcup_{j=1}^ae^2_j, {\Gamma}_2)$ and
${\pi}_{3}({\cal M}, {\Gamma}_2
\cup\bigcup_{j=1}^ae^2_j)$,
(notice that  ${\pi}_2({\Gamma}_2
\cup\bigcup_{j=1}^ae^2_j, {\Gamma}_2)$ is abelian  since the
two-cells $e^2_j$ are trivially attached to ${\Gamma}_2$).\par
 Through the orthogonal representation
${\theta}$ of the fundamental group ${\pi}_1({\Gamma}_2, e_0)$,
these modules, (thought of as ${\bf R}({\pi}_1({\Gamma}_2))$
modules), give rise to the  real vector spaces
\begin{eqnarray}
{\pi}^{\theta}_{3}({\cal M}, {\Gamma}_2
\cup\bigcup_{j=1}^ae^2_j)& ={\bf R}^p{\otimes }_{R{\pi}_1}
{\pi}_{3}({\cal M}, {\Gamma}_2
\cup\bigcup_{j=1}^ae^2_j)\nonumber\\
{\pi}^{\theta}_2({\Gamma}_2
\cup\bigcup_{j=1}^ae^2_j, {\Gamma}_2) &=
{\bf R}^p{\otimes }_{R{\pi}_1}
{\pi}_2({\Gamma}_2
\cup\bigcup_{j=1}^ae^2_j, {\Gamma}_2)
\end{eqnarray}
which can be endowed with a preferred base, respectively
${\bf x}_{\alpha}\otimes [{\psi}^3_k]$ and ${\bf x}_{\alpha}\otimes
[{\psi}^2_k]$,
generated by
an orthogonal base of ${\bf R}^p$, (here denoted by ${\bf
x}_{\alpha}$), and
the distinguished base of the relative homotopy groups ${\pi}_3$
and ${\pi}_2$ consisting of the characteristic maps of the cells
generating ${\cal M}$ out of ${\Gamma}_2$. \par
\vskip 0.5 cm
Such choice of a preferred base can be used to
characterize an inner product, in the spaces ${\pi}^{\theta}_2$, in
which
$[{\theta}_*(w_{li})^{tr}{\theta}_*(w_{kl})]^{-1}$ , (where
${\theta}_*(w_{li})^{tr}$ is the transpose of ${\theta}_*(w_{li})$),
is a symmetric positive bilinear form.\par
The construction of this inner product has a natural
physical interpretation that will be significant in what follows.\par
To begin with, let us remark that
the generic element of ${\pi}_2^{\theta}({\Gamma}_2
\cup\bigcup_{j=1}^ae^2_j, {\Gamma}_2)$,
say $\xi =\sum_{{\alpha},i} {\xi}^{{\alpha}(i)}({\bf x}_{\alpha}
\otimes [{\psi}^2_i])$, can be equivalently thought of as a
configuration
of $a_{(m)}$ ordinary vectors, ${\xi}^{(i)}={\xi}^{{\alpha}(i)}$ in
${\bf R}^p$, each such vectors
being associated with a corresponding two-cell $e_i^2$.
The non-trivial attachement of the three-cells $e^3_k$ to ${\Gamma}_2$
is described by the isomorphism ${\theta}_*(\partial )^{-1}\colon
{\pi}^{\theta}_2 \to {\pi}^{\theta}_3$ which with each vector
${\xi}\in {\pi}^{\theta}_2$ associates the vector
${\theta}_*(\partial)^{-1}{\xi} \in {\pi}^{\theta}_3$ given by
\begin{eqnarray}
{\theta}_*(\partial)^{-1}{\xi}=\sum_{{\alpha},k,l}
{\xi}^{{\alpha}(k)}{\bf x}_{\alpha}
[{\theta}_*(w_{kl})]^{-1}\otimes [{\psi}^3_l]
\end{eqnarray}
Again, this vector can be thought of as a collection of $a_{(m)}$
vectors in ${\bf R}^p$, each such vector, say the one
associated with the three-cell $e_l^3$,
$[{\theta}_*(\partial)^{-1}{\xi}]_{(l)}$, is
obtained from the vectors ${\xi}^{(k)}$ through the action of
the $a_{(m)}$  real
valued $p\times p$ matrices $[{\theta}_*(w_{kl})]^{-1}$,
(the cell index $l$ being fixed, while $k=1,2,\ldots, a_{(m)}$).
Namely
\begin{eqnarray}
[{\theta}_*(\partial)^{-1}{\xi}]_{(l)}\equiv \sum_k
{\xi}^{(k)}
[{\theta}_*(w_{kl})]^{-1}
\end{eqnarray}
With these preliminary remarks,
let us consider two generic vectors
$\xi =\sum_i {\xi}^{{\alpha}(i)}({\bf x}_{\alpha}
\otimes [{\psi}^2_i])$ and
$\eta =\sum_j {\eta}^{{\nu}(j)}({\bf x}_{\nu}
\otimes [{\psi}^2_j])$ in
${\pi}_2({\Gamma}_2
\cup\bigcup_{j=1}^ae^2_j, {\Gamma}_2)$, (${\alpha},{\nu}=1,\ldots, p$
and $i,j =1, \ldots, a_{(m)}$; here and in what follows,
summation over repeated ${\bf R}^p$-indexes is assumed, while
summation over cell indices is always explicitly stated). \par
According to the previous remarks, we can define an
inner product, $<{\xi}|{\eta}>$,
between ${\xi}$ and ${\eta}$ by setting
\begin{eqnarray}
<{\xi}|{\eta}>_w \equiv \sum_l [{\theta}_*(\partial)^{-
1}{\xi}]_{(l)}\cdot
[{\theta}_*(\partial)^{-1}{\eta}]_{(l)}
\end{eqnarray}
where  $\cdot $ denotes the ordinary euclidean
inner
product in ${\bf R}^p$, between the euclidean vectors
$[{\theta}_*(\partial)^{-1}{\xi}]_{(l)}$ and
$[{\theta}_*(\partial)^{-1}{\eta}]_{(l)}$.
More explicitly, we get
\begin{eqnarray}
<{\xi}|{\eta}>_w &\equiv
\sum_{i,l,k}
\{[{\theta}_*(w_{li})^{tr}{\theta}_*(w_{kl})]^{-1}\}_{{\alpha}{\nu}}
{\xi}^{{\alpha}(i)}{\eta}^{{\nu}(k)}
\end{eqnarray}
where, (for each given value of the cell-indices $i$, $l$, $k$),
$\{[{\theta}_*(w_{li})^{tr}{\theta}_*(w_{kl})]^{-1}\}_{\alpha \nu}$
is the $({\mu}, {\alpha})$ entry in the $p \times p$ matrix
obtained by matrix product between the $p\times p$ matrices
$[{\theta}_*(w_{li})^{tr}]^{-1}$ and $[{\theta}_*(w_{kl})]^{-1}$.\par
\vskip 0.5 cm
The geometrical construction just described shows connections
with the physics of magnetic systems (or, if you prefer, with
the formalism of lattice field theory). As a matter of fact,
the simplified pair $({\cal M}, {\Gamma}_2)$ is described
by a collection of ${\bf R}^p$-vectors ${\xi}^{(i)}$ distributed
on a set, ${\Lambda}$, of $a_{(m)}$ sites, with the vector
${\xi}^{(i)}$ at  the site $(i)$ and the vector
${\xi}^{(k)}$ at the site $(k)$,
interacting through  a {\it site-dependent} interaction matrix
$A^{w}_{(i)(k)}$ whose components are given
by $(A^{w}_{(i)(k)})_{{\mu}{\nu}}=-
\{[{\theta}_*(w_{li})^{tr}{\theta}_*(w_{kl})]^{-
1}\}_{{\mu}{\nu}}$.\par
With such a choice of
the interaction term, it follows that the
inner product we have introduced is
the total interaction
energy, $H_{w}(\Lambda)$, associated with the corrisponding
configuration
of the ${\bf R}^p$-vectors ${\xi}^{(i)}$
on the collection of sites ${\Lambda}_{(m)}$, namely
\begin{eqnarray}
H_{w}(\Lambda)[\xi] \equiv -\frac{1}{2}\sum_{i,j \in
\Lambda}{\xi}^{{\mu}(i)}(A_{(i)(j)})_{{\mu}{\nu}}
{\xi}^{{\nu}(j)} =\frac{1}{2} <{\xi}|{\xi}>
\label{energia}
\end{eqnarray}

This remark suggests that
the  properties of the collection of cells $\{e^3_i\}$
can treated statistically by assigning to each configuration of
cells associated with the generic vector
${\bf {\xi}}$ of
${\pi}^{\theta}_{2}({\Gamma}_2
\cup\bigcup_{j=1}^ae^2_j, {\Gamma}_2)$ an equilibrium
Boltzmann-Gibbs distribution at (inverse) temperature $J$,
given by
\begin{eqnarray}
dP_w^{(m)}[{\xi}]\equiv (Z^{(m)}_w)^{-1}
 \exp (-JH_w({\Lambda})[\xi])\prod_{i\in \Lambda}
d{\mu}_{i}({\xi})
\end{eqnarray}
where $d{\mu}_i$ is the Lebesgue measure on ${\bf R}^p$, and
where  the normalizing factor, (the partition function), $Z^{(m)}_w$
is given by
\begin{eqnarray}
Z^{(m)}_w \equiv \int_{\Lambda}
\exp[- JH_{w}(\Lambda)]\prod_{i\in \Lambda}
d{\mu}_{i}({\xi})
\end{eqnarray}
More explicitly
\begin{eqnarray}
Z^{(m)}_w \equiv
\int_{{\pi}^{\theta}_{2}}
\exp[-
\frac{1}{2}J<{\xi}|{\xi}>]
D[{\mu}(\xi)]
\end{eqnarray}
where
$D[{\mu}(\xi)]
=\prod_{i=1}^{a_{(m)}}\prod_{\alpha =1}^p d{\xi}^{(i){\alpha}}$
is
the positive,
$O(p)$-invariant, measure on  ${\pi}^{\theta}_{2}({\Gamma}_2
\cup\bigcup_{j=1}^ae^2_j, {\Gamma}_2)$
associated with the reference inner product
$<{\xi}|{\eta}>_{I}=\sum_i{\xi}^{(i)}\cdot {\eta}^{(i)}$,
corresponding to the trivial pasting (trivial torsion),
(thus, $P_w[\xi]$ is the multinormal distribution with
variance $[JA^w_{(i)(k)}]^{-1}$).\par
\vskip 0.5 cm
In order to see if, in the thermodynamical
limit, the pair
$({\cal M}, {\Gamma}_2)$, as described by  $Z^{(m)}_w$,
occurs with a non-vanishing probability, it will be necessary to
give a precise mathematical definition of the functional measure
\begin{eqnarray}
\lim_{m\to \infty} dP_w^{(m)}[{\xi}]
\end{eqnarray}
and in particular, to discuss  the behavior of
$Z^{(m)}_w$ in the $m\to \infty$ limit.\par
In this connection, two remarks are in order. First of
all, as $m$ increases,
the $L(m)$-geodesic ball two-skeleton ${\Gamma}_2$ changes,
and in particular, the presentation of the fundamental
group ${\pi}_1({\Gamma}_2,e_0)$, within a given homotopy
class, is altered to the effect that the Whitehead incidence
matrices $w_{ik}$ vary with $m$ even if they may represent
a same Whitehead torsion ${\tau}(w_{ik})$. \par
Moreover, also the number $a_{(m)}$ of two-cells and three-cells,
(hence the number of ${\bf R}^p$-vectors ${\xi}^{(i)}$), grows
with $m$, and eventually $a_{(m)}\to \infty$. These two
facts are related to each other, since the generic simplified pair
$({\cal M}, {\Gamma}_2)$ is assumed to come from the formal
deformation of the pair $({\cal N}, {\Gamma}_2)$ generated from
a $L(m)$-geodesic ball nerve and its two skeleton.
This remark, to be made precise below, suggests
that in order to take into account, as $m\to\infty$, all
possible (equivalent) presentations of the fundamental group
sampled by $dP^{(m)}_w[\xi]$, we have to pass from a
{\it canonical} to a {\it grand-canonical} point of view. Formally
this means that, in order to consider the  limit $m\to\infty$
of $dP^{(m)}_w[\xi]$, we have to introduce a fugacity $z$
for the simplified pair $({\cal M}, {\Gamma}_2)$, and
define a grand-canonical partition function by weighting each
canonical function $Z_w^{(m)}$ with a factor
$z^{g^{\pi}_{(m)}}$, where $g^{\pi}_{(m)}$ denotes
the number of generators which occur in the presentation
of ${\pi}_1({\Gamma}_2,e_0)$ associated with the $L(m)$-geodesic
ball covering considered.\par
\vskip 0.5 cm
Each ${\pi}_1({\Gamma}_2)$ is finitely presented, and without loss in
generality, ${\pi}_1({\Gamma}_2)$ can be thought of as given by a
certain number of generators $x_1,\ldots, x_g$ and relations among
such generators $R_i(x_1,\ldots, x_g)=1$, ($i=1,\ldots,R^{\pi}$).
The number $g_{(m)}^{\pi}$ and $R_{(m)}^{\pi}$
of such generators and relations is connected to the particular one-
skeleton
graph ${\Gamma}_{(m)}$ which is contained in ${\Gamma}_2$.
In particular, the number of generators is provided by
$g^{\pi}_{(m)}=1+N^{(1)}_{(m)}(\Gamma)-N^{(0)}_{(m)}(\Gamma)$.
Given such generators, we can associate with the graph
${\Gamma}_{(m)}$, the wedge product of circles
\begin{eqnarray}
\vee ({\Gamma}_{(m)})=e^0\cup(e^1_1\cup \ldots \cup e^1_g)
\end{eqnarray}
where $e^0$ is a given vertex of ${\Gamma}_{(m)}$. We can use this
presentation of (the free fundamental group of) ${\Gamma}_{(m)}$ in
order
to generate a two-dimensional complex  which is
homotopically equivalent to ${\Gamma}_2$, and
which we shall  denote by ${\Gamma}^*_2$. Explicitly, if
$x_j$ is the element of the free group
${\pi}_1(\vee ({\Gamma}_{(m)}), e^0)$ which is represented by a
characteristic map for $e^1_j$, and if
${\phi}_i\colon {\partial}Q^2 \to \vee ({\Gamma}_{(m)})$ represents
the relation  $R_i(x_1,\ldots, x_g)$, then we set
\begin{eqnarray}
{\Gamma}^*_2 \equiv \vee ({\Gamma}_{(m)}){\cup}_{{\phi}_1}Q^2
{\cup}_{{\phi}_2}\ldots{\cup}_{{\phi}_k}Q^2
\end{eqnarray}
\vskip 0.5 cm
With these preliminary remarks,
let us assume that we are given with a non-singular
${\bf Z}({\pi}_1({\Gamma}^*_2))$ matrix, $w^*_{ij}$ to be thought of
as a
possible Whitehead torsion labelling the simple homotopy type in
$\Ricco$ defined by the $L(m)$-geodesic ball covering
$({\cal N}, {\Gamma}_2)$.
Let us assume that
$w^*_{ij}$ is an $a\times a$ matrix with
$a_{(m)}\leq N^{(3)}_{(m)}({\cal N})$, where
$N^{(3)}_{(m)}({\cal N})$ is the number
of three-dimensional faces $p^{(3)}_{ijkl}$
in the geodesic balls nerve ${\cal N}$.\par
We can always replace
the original, $a_{(m)}\times a_{(m)}$ torsion matrix $w^*_{lj}$
with an equivalent $N^{(3)}_{(m)}\times N^{(3)}_{(m)}$ torsion
matrix $w_{ik}$ which has a block form obtained by expanding
$w_{lj}$ by adding an identity
matrix $I$ on the corner diagonal.\par
According to the definition
of  simplified pair, this expansion
does not alter the Whitehead torsion, (see the operations
{\it (i), (ii), (iii)}, characterising the torsion of
an incidence matrix $w_{ik}$), and it is also possible\cite{Cohe}
to show that there is a
$CW$-pair in simplified form, that, by abuse of
notation, we shall continue to denote
by $({\cal M}, {\Gamma}_2 )$, whose
torsion is provided by this new incidence matrix $w_{ik}$.
This pair can be constructed
as follows, (see\cite{Cohe}, proposition 8.7). Since
$w_{ik}$ is an $N^{(3)}_{(m)}\times N^{(3)}_{(m)}$ matrix we let
$K_2\equiv {\Gamma}^*_2\cup e^2_1\cup\ldots \cup e^2_{N^{(3)}}$.
The two-cells, $e^2_j$, so introduced have characteristic
maps ${\psi}^2_j$ such that ${\psi}^2_j(\partial Q^2)=e^0$.
We denote by $[{\psi}^2_j]$ the
homotopy class, in the relative homotopy group
${\pi}_2(K_2, {\Gamma}^*_2)$, represented by the map
\begin{eqnarray}
{\psi}^2_j \colon (Q^2, Q^1, [Q^1\times\partial Q^1\cup (0\times
Q^1)])
\to (K_2, {\Gamma}^*_2, e^0)
\end{eqnarray}
Notice that the same map, thought of with range in $(K_2,e^0)$, can
be used so as to represent the homotopy class $\Vert {\psi}^2_j \Vert$
in ${\pi}_2(K_2, e^0)$. Thus, if
${\bf i} \colon (K_2, e^0; e^0)\hookrightarrow (K_2, {\Gamma}^*_2,
e^0)$
is the injection, we get
${\bf i}_*\Vert {\psi}^2_j \Vert = [{\psi}^2_j]$. \par
If we denote by $f_j\colon (Q^2,\partial Q^2)\to (K_2, e_0)$ the
map representing $\sum_q w_{jq}\Vert {\psi}^2_q \Vert$, let
us attach $3$-cells to $K_2$ so as to get
\begin{eqnarray}
{\cal M} \equiv K_2\cup e^3_1\cup\ldots \cup e^3_{N^{(3)}}
\end{eqnarray}
where the $3$-cells $e^3_j$ have characteristic maps
\begin{eqnarray}
{\psi}^3_j\colon (Q^3, Q^2, [Q^1\times\partial Q^2\cup (0\times Q^2)])
\to ({\cal M}, K_2, e^0)
\end{eqnarray}
such that  ${\psi}^3_j\vert Q^2= {\bf i}\circ f_j$. But
by definition $\partial [{\phi}^3_j]= [{\phi}^3_j\vert Q^2]$, thus we
get
\begin{eqnarray}
\partial [{\psi}^3_j]= [{\bf i}\circ f_j]=
{\bf i}_*(\sum_q w_{jq}\Vert {\psi}^2_q\Vert )= \sum_q
w_{jq}[{\psi}^2_q]
\end{eqnarray}
which shows that the CW-pair $({\cal M}, {\Gamma}_2)$ has boundary
operator provided by the matrix $w_{ij}$, and it is
in the same simple homotopy class of $({\cal N}, {\Gamma}_2)$.
(Notice that the fact that the
${\cal M}$ constructed with $w_{ik}$ actually retracts on
${\Gamma}_2$, is less trivial and the reader
is referred to\cite{Cohe} for a detailed argument, (second part of
proposition $8.7$, p. $34$)).
\vskip 0.5 cm
Once established that the incidence matrix of the simplified
pair associated with $({\cal N}, {\Gamma}_2 )$ can be assumed
to be a ${\bf Z}({\pi}_1)$-matrix of order
$N^{(3)}_{(m)}({\cal N})$, we can easily formalize the stated
relation between the canonical and the grand-canonical
point of view.\par
\vskip 0.5 cm
According to the properties of minimal geodesic ball coverings
recalled in the introductory remarks, we can always assume
that
\begin{eqnarray}
c \equiv \limsup_{m\to \infty}
\left (\frac{N^{(0)}_{(m)}}{N^{(1)}_{(m)}}\right )
\end{eqnarray}
is a function on $\Ricco$), bounded above in
terms of the parameters $n$, $r$, $D$, $V$.
This implies that the number
$g^{\pi}_{(m)}=1+N^{(1)}_{(m)}-N^{(0)}_{(m)}$ of generators of
${\pi}_1({\Gamma}_2,e_0)$ is such that
\begin{eqnarray}
g^{\pi}_{(m)}({\cal N})= {\cal O}(N^{(1)}_{(m)}({\cal N}))
\label{ventiquattro}
\end{eqnarray}
as $m \to \infty$.\par
Thus the natural grand-canonical
partition function providing the statistical description
of the fields ${\bf {\xi}}$ of
 ${\pi}^{\theta}_{3}({\cal M}, {\Gamma}_2
\cup\bigcup_{j=1}^ae^2_j)$, can be written as
\begin{eqnarray}
Z(m,J,{\pi}_1) \equiv
 \sum_{({\cal M},{\Gamma}_2)_{(m)}}\exp
{[{i_c}N^{(1)}_{(m)}({\Gamma}_2)]}
 \int_{{\pi}^{\theta}_{2}}
\exp[-\frac{1}{2}J<{\xi}|{\xi}>]
D[{\mu}(\xi)]
\label{generating}
\end{eqnarray}
\vskip 0.5 cm
where ${i_c}$ plays
the role of a  chemical potential  determining
the average number of generators associated with
the presentation
of ${\pi}_1({\Gamma}_2, e_0)_{(m)}$, and where
the sum $\sum_{({\cal M},{\Gamma}_2)}$ is the sum over
the  three-dimensional
simplified pairs, $({\cal M},{\Gamma}_2)_{(m)}$,
generated by $L(m)$-geodesic balls coverings
of manifolds in $\Ricco$ with isomorphic fundamental groups
${\pi}_1({\Gamma}_2)$. Correspondingly, the integral
$\int_{{\pi}^{\theta}_{2}}\ldots D[{\mu}(\xi)]$
is to be evaluated, over  the vector space
${\pi}^{\theta}_{2}({\Gamma}_2
\cup\bigcup_{j=1}^ae^2_j, {\Gamma}_2)$, by using the
$N^{(3)}_{(m)} \times N^{(3)}_{(m)}$ incidence matrices
$w_{ik}$  associated with each given
$({\cal M},{\Gamma}_2)$, ($w_{ik}$ and the corresponding
$({\cal M},{\Gamma}_2)$ being explicitly realized following
the construction delineated above).
\par
\vskip 0.5 cm
According to the definition of the scalar product $<{\xi}|{\xi}>$
between the fields
${\xi} \in {\pi}^{\theta}_{2}({\Gamma}_2
\cup\bigcup_{j=1}^ae^2_j, {\Gamma}_2)$, it immediately follows that
the canonical partition function $Z^{(m)}_w({\Gamma}_2)$, describing
the gaussian
statistical pasting of the three-cells $e^3_j$
into a given ${\Gamma}_2$, can be explicitly evaluated
\begin{Gaiap}
For a given $L(m)$-geodesic ball covering presentation
of ${\pi}_1({\Gamma}_2, e_0)$, and a corresponding
${\bf Z}({\pi}_1({\Gamma}_2, e_0))$ incidence matrix
$w_{ik}$,
the canonical partition function $Z_w({\Gamma}_2)$  is given by
\begin{eqnarray}
Z_w({\Gamma}_2)=
\left({\frac{J}{2{\pi}}}\right)^{-(p/2)N^{(3)}_{(m)}}
{\Delta}^{\theta}[({\cal N},
{\Gamma}_2)]
\label{Eulero}
\end{eqnarray}
where ${\Delta}^{\theta}[({\cal N},
{\Gamma}_2)]$ denotes the Reidemeister-Franz representation
torsion of the pair $({\cal N}, {\Gamma}_2)$ corresponding to the
simple homotopy type labelled by $w_{ik}$.
\label{treuno}
\end{Gaiap}
{\it PROOF}. By direct computation of the Gaussian integral
\begin{eqnarray}
\int_{{\pi}^{\theta}_{2}}
 & \exp[-\frac{1}{2}J
\sum_{i,l,k}
\{[{\theta}_*(w_{li})^{tr}{\theta}_*(w_{kl})]^{-1}\}_{{\alpha}{\nu}}
{\xi}^{{\alpha}(i)}{\xi}^{{\nu}(k)}]
\prod_{i=1}^{N^{(3)}_{(m)}}\prod_{\alpha =1}^p d{\xi}^{(i){\alpha}}
= \nonumber\\
&= \left({\frac{J}{2{\pi}}}\right)^{-(p/2)N^{(3)}_{(m)}}
\left\{\det
[{\theta}_*(w))({\theta}_*(w))^{tr}]^{-1}\right\}^{-1/2}
\label{gaussian}
\end{eqnarray}
where we have used the expression of $<{\xi}|{\xi}>$
in terms of the representation
of the incidence matrix $w_{ik}$ defining
the simple homotopy type of $({\cal N}, {\Gamma}_2)$. Notice that
${\theta}_*(w)$ is a $N^{(3)}_{(m)}\times N^{(3)}_{(m)}$
matrix with entries in
${Mat}_p({\bf R})$, namely a matrix of order $pN^{(3)}_{(m)}$
with real entries, (${Mat}_p({\bf R})$ denotes the ring
of all $p\times p$ matrices with real entries).
The determinant resulting from the gaussian integration is
trivially reduced to
the Reidemeister torsion, (in the
orthogonal representation ${\theta}$ of ${\pi}_1({\Gamma}_2,e_0)$),
of the simplified pair $({\cal M}, {\Gamma}_2)$ corresponding
to the incidence matrix $w_{ik}$.\par
Since the Reidemeister-Franz torsion
${\Delta}^{\theta}({\cal M},{\Gamma}_2)$ is a
{\it combinatorial invariant}, and  the
simplified pair $({\cal M}, {\Gamma}_2)$ is combinatorially
equivalent, (in the simple homotopy sense), to the pair
$({\cal N}, {\Gamma}_2)$,
we can write\cite{Cohe}
\begin{eqnarray}
{\Delta}^{\theta}[({\cal M},
{\Gamma}_2)]= {\Delta}^{\theta}[({\cal N},{\Gamma}_2)]
\label{combinatorial}
\end{eqnarray}
which together with (\ref{gaussian}) yields the relation
(\ref{Eulero}).$\clubsuit$
\vskip 0.5 cm
The net effect of such result is that we can rewrite
$Z(m,J,{\pi}_1)$ directly in terms of combinatorial
quantities referring to $L(m)$-geodesic balls coverings
\begin{Gaial}
The statistical sum $Z(m,J,{\pi}_1)$ describing a
Gaussian-pasting of three-cells onto ${\Gamma}_2$ can be
rewritten as
\begin{eqnarray}
Z(m,J,{\pi}_1)=
\sum_{({\cal N},{\Gamma}_2)}{\Delta}^{\theta}({\cal N},{\Gamma}_2)
\exp {[{i_c}N^{(1)}_{(m)}({\Gamma}_2)
-{\sigma} N^{(3)}_{(m)}({\cal N})]}
\label{gravit}
\end{eqnarray}
where we have set ${\sigma}\equiv \frac{p}{2}\ln
(\frac{J}{2{\pi}})$. $\clubsuit$
\label{treg}
\end{Gaial}
\vskip 0.5 cm
It follows from (\ref{gravity}) that, up to the
torsion invariants, $Z(m,J,{\pi}_1)$ has the structure
of the partition function of {\it three-dimensional simplicial
quantum gravity} as written down for the simplicial
approximation generated by minimal $L(m)$-geodesic balls
coverings. Namely,
\vskip 0.5 cm
\begin{eqnarray}
Z(m,J,{\pi}_1) = \sum_{({\Gamma}_2,{\cal N})}
{\rho}({\Gamma}_2,{\cal N})\exp -S({\Gamma}_2,{\cal N})
\end{eqnarray}
\vskip 0.5 cm
where the {\it action} $S({\Gamma}_2,{\cal N})$ is given by
\vskip 0.5 cm
\begin{eqnarray}
S({\Gamma}_2,{\cal N})\equiv {\sigma}N^{(3)}_{(m)}({\cal N})-
{i_c}N^{(1)}_{(m)}({\Gamma}_2)
\label{cosmological}
\end{eqnarray}
\vskip 0.5 cm
and the {\it weight} ${\rho}({\Gamma}_2,{\cal N})$ is given by
\vskip 0.5 cm
\begin{eqnarray}
{\rho}({\Gamma}_2,{\cal N}) \equiv
{\Delta}^{\theta}({\cal N},{\Gamma}_2)
\end{eqnarray}
Before commenting on this connection between the theory
described by $Z(m,J,{\pi}_1)$ and three-dimensional
simplicial quantum gravity, it is useful to stress that here
the theory suggests a natural weight,
${\rho}({\Gamma}_2,{\cal N})$, for the triangulations occurring
in the statistical sum. As is known, the attribution of such a
weight is a problem in all discretized approaches to quantum
gravity. Typically one assumes ${\rho}({\Gamma}_2,{\cal N})=1$,
basing such choice on universality arguments and on the
experience gained in dimension two. The weight we
get here is the Reidemeister-Franz torsion, which, obviously
depends on the orthogonal representation of the fundamental
group that we are considering. This dependence, which is rather
manifest in the approach presented here, has interesting
consequences on which we shall comment in the concluding
remarks.\par
\vskip 0.5 cm
If we put together all such remarks and take into account the
Gaussian nature of the partition
function $Z(m,J,{\pi}_1)$ the we get as a general
result the following
\begin{Gaiap}
Let ${\cal T}$ denote the polytope, (or if you prefer,
the dynamical triangulation), associated with
the nerve ${\cal N}_{(m)}$ of the $L(m)$-minimal geodesic
ball covering of the generic manifold $M \in \Ricco$.
Define the corresponding simplicial quantum gravity partition
function as
\begin{eqnarray}
{\cal Z}({\cal T})\equiv
\sum_{{\cal T}}{\rho}({\cal T})
\exp {[{\alpha}N^{(1)}_{(m)}({\cal T})
-\gamma N^{(3)}_{(m)}({\cal T})]}
\end{eqnarray}
where ${\rho}({\cal T})$ is a weight for the triangulation ${\cal T}$,
${\alpha}$ is a {\it bare} gravitational coupling constant, and
${\gamma}$ is a {\it bare} cosmological constant.\par
Then,
for a given orthogonal representation
${\theta}\colon {\pi}_1({\cal T})\to O(p)$, of the fundamental group
${\pi}_1({\cal T})$,
such ${\cal Z}({\cal T})$ is equivalent to the
partition function $Z(m,J,{\pi}_1)$  which describes
the gaussian
insertion of  three-dimensional cells onto ${\Gamma}_2({\cal T})$.
\label{gravity}
\end{Gaiap}
{\it Proof}. The proposition is a formal restatement
of the above remarks. The only point to stress is that
apparently, we could  have dispensed
ourselves from considering just manifolds of bounded geometry
belonging
to some $\Ricco$. Notice however that in such a case we do not
have at our disposal the control on homotopy types and simple homotopy
types that we have for manifolds in $\Ricco$. In particular, the
full Whitehead group $W({\pi}_1({\Gamma}_2))$ comes into play. In
general
$W({\pi}_1({\Gamma}_2))$ has an infinite number of inequivalent
elements (torsions), and the above equivalence between
${\cal Z}({\cal T})$ and $Z(m,J,{\pi}_1)$ would be
rather formal, involving, for each ${\Gamma}_2$, a sum over an
infinite number of inequivalent gaussian distributions.
We see again that $\Ricco$ provides a very effective
regularization framework for question concerning discrete
models of euclidean quantum gravity. $\clubsuit$
\section{The thermodynamical limit of three-dimensional simplicial
gravity}
As recalled in the introductory remarks, when we implement
the random triangulation approach to three-dimensional quantum
gravity, we fix {\it a priori} the length of the edges of
the triangulation, (say, by setting their length to one), and
fix our attention to the connectivity properties of the simplicial
approximation considered. Then we discuss the resulting
{\it dynamical triangulation} model at longer and longer distance
scales, trying to extract information from the {\it scaling}
limit. Typically this is done with energy-entropy arguments,
by examining the critical behavior of the partition function and
of the correlation functions.\par
When using spaces of bounded geometries and geodesic ball coverings
as a regularization scheme, we are actually taking the dual point of
view. Here the {\it dynamical
triangulation}
has edges the lengths of which are not equal to one but rather
to $2L(m)=2/m$. Thus, the number of cells of the generic
$L(m)$-geodesic balls nerve ${\cal N}_{(m)}$ diverges as
$m\to\infty$. In this sense, the limit $m\to \infty$ actually
corresponds to a {\it thermodynamical} limit rather than
to a scaling limit. \par
More in details, if we have an
algorithm  for generating random triangulation, (with fixed edge-
length),
then the condensation of extended objects requires the critical
behavior of the partition function in the scaling regime, and then
we have to check that the resulting condensate is a {\it manifold}
thought of as obtained from a sequence of random triangulations
by rescaling lengths by a factor $s\to \infty$.\par
The point of view afforded by the use of the spaces of bounded
geometries $\Ricco$ is different and somehow simpler,
in the sense that in such a case
we know a priori that the simplicial approximations, arising from
geodesic ball coverings, yield for (homology) manifolds as
$m\to \infty$, (convergence is in the Gromov-Hausdorff topology).
This is a direct consequence of the compactness of $\Ricco$.\par
The problem here is not so much to prove that the triangulation
involved in the regularization yields in the limit a
(generalized) manifold, but rather to prove that the partition
function and the associated probability measure are
well defined, as $m\to \infty$, for a suitable range of
values of the couplings.\par
\vskip 0.5 cm
With these preliminary remarks along the way we prove the
following result which allows us to disclose the nature of
the $m\to\infty$ limit of the partition function
$Z(m,J,{\pi}_1)$,
\begin{Gaiap}
Let $||{\Delta}^{\theta}||({\pi}_1)$ denote the finite number
of distinct Reidemeister-Franz representation torsions associated
with those manifolds in $\Ricco$ whose fundamental group is
isomorphic to  a given group ${\pi}_1$, and let us consider
the function over $\Ricco$ defined by
\begin{eqnarray}
b(M)\equiv \limsup_{m\to \infty}
\left (\frac{N^{(1)}_{(m)}(M)}{N^{(3)}_{(m)}(M)}\right )
\end{eqnarray}
Then, for any given value of the ratio
$0<b<\infty$ there exists a finite value of
the chemical potential ${i_c}$,
\begin{eqnarray}
{i_c}=-\frac{1}{2} \frac{p}{b}
\end{eqnarray}
and a finite number of critical values
\begin{eqnarray}
J^{(i)}_{crit} =2{\pi}
{\left ({\Delta}^{\theta}_{(i)}\right )}^2
\end{eqnarray}
of the coupling $J$,
(with $i=1,\ldots, ||{\Delta}^{\theta}||({\pi}_1)$),
corrisponding to which the statistical system described by
the partition function
$Z(m,J,{\pi}_1)|_{J=J_{crit}}$, (obtained by $Z(m,J,{\pi}_1)$ by
restricting the summation
over incidence matrices with a given value of the torsion
${\Delta}_{(i)}$, and by setting $J$ equal to the corresponding
value of $J_{crit}$), admits a well defined limit
as $m\to \infty$.\par
\noindent Let $GL({Mat}_p({\bf R}))$ be the infinite general
linear group of non-singular matrices over the ring
${Mat}_p({\bf R})$, of all $p \times p$ matrices with
real entries.
If ${\cal E}_{(m)}\subset GL({Mat}_p({\bf R}))$ is the
set of all possible incidence matrices
${\tilde w} \equiv {\theta}_*(w)$, realized over
$\Ricco$ as $m$ varies, then $\lim_{m\to\infty}Z(m,J,{\pi}_1)$
is associated with a unique
probability measure $P^{(i)}_y(d{\tilde w})$ on
$GL({Mat}_p({\bf R}))$, supported on the incidence matrices
whose Reidemeister torsion is
${\Delta}^{\theta}_{(i)}$, and we get
\begin{eqnarray}
\lim_{m\to\infty}Z(m,J,{\pi}_1)=
\lim_{m\to\infty}\int_{{\cal E}_{(m)}}P^{(i)}_y(d{\tilde w})=
{\Delta}^{\theta}_{(i)}
\end{eqnarray}
As $J$ varies from one critical value
to another, the system exhibits a phase transition
associated with the changes of the corresponding simple-homotopy
types, and the set of manifolds in
$\Ricco$ with isomorphic fundamental groups
appears as a mixture of pure phases described
by the probability measure on $GL({Mat}_p({\bf R}))$ given by a convex
combination of the above measures
\begin{eqnarray}
P(d{\tilde w}) \equiv \frac{1}{\alpha}\sum_{i=1}^{\alpha}
P^{(i)}(d{\tilde w})
\end{eqnarray}
with $\alpha=||{\Delta}^{\theta}||({\pi}_1)$.
\label{bello}
\end{Gaiap}
{\it Proof}.
Notice that the function $b$ introduced above provides,
roughly speaking, the (inverse of the) number of
tetrahedra shared, on the average, by
the generic link in the polytope generated by the geodesic
ball nerves ${\cal N}_{(m)}(M)$. As the notation suggests,
$b$ is not constant on $\Ricco$, and by using it as a parameter in
controlling
the critical regime of the theory, (basically as a
chemical potential), we select among all possible manifolds in
$\Ricco$ with Reidemeister torsion ${\Delta}_{(i)}$, the class of
manifolds for which the critical regime occurs.\par
\vskip 0.5 cm
The proof of proposition \ref{bello} is divided in two parts. First,
by
exploiting few elementary properties of
the generic general linear group, $GL(q,{\bf R})$, we introduce
a Gaussian measure on the space $GL({Mat}_p({\bf R}))$. Then, such
measure is connected with the $m\to \infty$ limit of
$Z(m,J,{\pi}_1)|_{J=J_{crit}}$.\par
\vskip 0.5 cm
\noindent{\it First part}:\par
\vskip 0.5 cm
\subsection{Representation torsions and Gaussian measures on the
general linear group $GL({Mat}_p({\bf R}))$}
Let $GL(a,{\bf Z}{\pi}_1({\Gamma}_2))$ denote  the
general linear group of non-singular, ({\it i.e.}, with
two-sided inverse), $a \times a$ matrices over the ring
${\bf Z}{\pi}_1({\Gamma}_2)$.
Under the natural injection of $GL(a,{\bf Z}{\pi}_1({\Gamma}_2))$ into
$GL(a+1,{\bf Z}{\pi}_1({\Gamma}_2))$, one defines the infinite
general linear group $GL({\bf Z}{\pi}_1({\Gamma}_2))$ as the
direct limit
$\lim_{a\to\infty}\cup GL(a,{\bf Z}{\pi}_1({\Gamma}_2))$.\par
Similarly, we denote by $GL({Mat}_p({\bf R}))$ the infinite general
linear group of non-singular matrices over the ring
${Mat}_p({\bf R})$, of all $p \times p$ matrices with
real entries. (Notice that for any $a$ we have
$GL(a,{Mat}_p({\bf R}))\simeq GL(pa, {\bf R})$).\par
\vskip 0.5 cm
Let $bE_{ij}$ denote the ${\bf Z}{\pi}_1({\Gamma}_2)$-matrix
 with an entry $b$ in the
$(i,j)$-th place and zero elsewhere. Let us denote by
$E({\bf Z}{\pi}_1({\Gamma}_2))\subset GL({\bf Z}{\pi}_1({\Gamma}_2))$
the (commutator) subgroup generated by the matrices
of the form $(I+bE_{ij})$, (elementary matrices).
Also, let us introduce the group $E_T$ generated by
$E({\bf Z}{\pi}_1({\Gamma}_2))$ and by the group of
the trivial units of ${\bf Z}{\pi}_1({\Gamma}_2)$.
By means of such subgroups we can algebraically characterize
the Whitehead group $W({\pi}_1({\Gamma}_2))$ as the quotient
\begin{eqnarray}
W({\pi}_1({\Gamma}_2))=\frac{GL(Z{\pi}_1({\Gamma}_2))}{E_T}
\end{eqnarray}
A similar characterization holds true for $W(Mat_p({\bf R}))$
which can be written as
\begin{eqnarray}
W(Mat_p({\bf R}))=\frac{GL(Mat_p({\bf R}))}{E(Mat_p({\bf R}))}
\end{eqnarray}
where $E(Mat_p({\bf R}))$ is the commutator subgroup generated by
all elementary matrices in $GL(Mat_p({\bf R}))$.\par
Notice that the Whitehead group $W({Mat}_p({\bf R}))$ can be
identified
with the half-line of positive numbers, (thought of as a
multiplicative group), and we can turn it into a measure space
by considering on it
the measure $d{\mu}(\Delta)=d{\Delta}/{\Delta}$, (with $d{\Delta}$ the
Lebesgue measure on ${\bf R}^{\times}$), invariant under
multiplication. Let us also recall that the Haar measure
${\cal H}_q$ on $GL(q,{\bf R})$ is given by
\begin{eqnarray}
d{\cal H}_q({\tilde w})=\frac{\prod^q_{i,j}d{\tilde w}_{ij}}
{|\det {\tilde w}|^q}
\end{eqnarray}
Thus, given a (suitably normalized) left-invariant Haar measure
${\cal S}_q$ on the
special linear group $SL(q,{\bf R})$ generated by all
elementary matrices in $GL(q,R)$, we can write (by
abuse of notation)
\begin{eqnarray}
\int_{W({Mat}_p({\bf R}))}\left (\int_{SL(q,{\bf R})}f({\tilde w}h)
d{\cal S}_q(h)
\right )
d({\mu}(\Delta))({\tilde w})=
\int_{GL(q,{\bf R})}f({\tilde w})d({\cal H}_q)({\tilde w})
\label{quotient}
\end{eqnarray}
for every continuous function $f$ with compact support on
$GL(q,{\bf R})$. \par
\vskip 0.5 cm
Let $E_{ka}$ denote the $q\times q$ matrix which has the element
in the $(k,a)$ place equal to $1$ and all other elements equal to
$0$. If $k\not= a$ and $x^{k}_{(a)} \in {\bf R}$, define the elentary
matrix in $GL(q, {\bf R})$ as
\begin{eqnarray}
B_{ka}(x^{k}_{(a)})\equiv I_q+x^{k}_{(a)}E_{ka}
\end{eqnarray}
(no summation over the indexes $k$ and $a$), where $I_q$ denotes
the identity matrix in $GL(q, {\bf R})$. In terms of such elementary
matrices, each matrix ${\tilde w}\in GL(q,{\bf R})$ can be
represented as
\begin{eqnarray}
{\tilde w}={\tilde s}D =(\prod_{ka}B_{ka})D
\end{eqnarray}
where ${\tilde s}\in SL(q,{\bf R})$ is, as indicated,
a finite product of elementary matrices $B_{ka}$,
and $D$ is a diagonal matrix
of the form $D_{ij}={\delta}_{ij}$, with $i,j=1,\ldots, q-1$, and
$D_{qq}=\det {\tilde w}={\Delta}^{\theta}({\tilde w})$,
namely $D\equiv I_q+({\Delta}^{\theta}-1)E_{qq}$.
(Notice that multiplication on the left such as
$B_{ka}(x)D$, adds $x$ times the $a$-th row to the $k$-th
row of $D$).\par
For each given value of the index $a=1,\ldots, q$, let
${\cal G}_{t(q)}(d{\bf x}_{(a)})$ denote the $q$-dimensional
Gaussian measure of variance $t >0$ defined by
\begin{eqnarray}
{\cal G}_{t(q)}(d{\bf x}_{(a)}) = \frac{1}{(2{\pi}t)^{q/2}}
\exp\left [-\frac{||{\bf x}_{(a)}||^2}{2t}\right ]d{\bf x}_{(a)}
\end{eqnarray}
where $||{\bf x}_{(a)}||^2 \equiv \sum_k^q(x_a^k)^2$, and
$d{\bf x}_{(a)}$ means the Lebesgue measure on ${\bf R}^q$.\par
\vskip 0.5 cm
Let $C_0(GL(q,{\bf R}))$ and $C_0(W({Mat}_p({\bf R})))$
respectively denote the spaces
of continuous functions with compact support on the
groups $GL(q, {\bf R})$ and $W({Mat}_p({\bf R}))$), then
by means of the parametrization ${\tilde w}={\tilde s}D$
of $GL(q,{\bf R})$  we can consider the linear mapping from
$C_0(GL(q,{\bf R}))$ onto $C_0(W({Mat}_p({\bf R})))$, defined by
gaussian averaging over $SL(q, {\bf R})$ according to
\begin{eqnarray}
f({\tilde w})\mapsto {\hat f}_{\cal G}
(t(q),{\Delta}({\tilde w}))\equiv
\int f({\tilde w}={\tilde s}({\bf x}_{(a)})D)
\prod_a^q{\cal G}_{t(q)}(d{\bf x}_{(a)})
\label{quarantanove}
\end{eqnarray}
More explicitly, we choose the variance $t(q)$ according
to
\begin{eqnarray}
2{\pi}t(q) = \frac{1}{q}{\Delta}^2
\label{varianza}
\end{eqnarray}
so as to get
\begin{eqnarray}
{\hat f}_{\cal G}({\Delta}({\tilde w}))=
\int_{SL(q,{\bf R})}f({\tilde w}=
{\tilde s}D)q^{q^2/2}{\Delta}^{-q^2}
\exp\left [-{\pi}q{\Delta}^{-2}||{\bf x}_{(a)}||^2\right ]
\prod_{k,a}^qd{x}^{k}_{(a)}
\label{normalized}
\end{eqnarray}
\vskip 0.5 cm
Notice that the particular choice (\ref{varianza}) of the variance is
motivated by the fact that as
$q$ increases, we need to enhance  the occurrence
(with respect to the given,
gaussian measure),
of the matrices $B_{ka}(x^k_{(a)})$
which eventually reduce to the identity matrix, (such matrices
generate,
up to determinants, the full general linear group). More in details,
as $q\to\infty$, the variance $t(q)\to 0$ and the associated
Gaussian measure degenerates to the Dirac measure supported on the
zero sequence $\{x^k_{(a)}\}=\{0,\ldots,\}$. This yields for
$B_{ka}$ which reduce, almost everywhere with respect to the gaussian
measure, to identity matrices, as required.\par
Notice also that the variables $x^k_{(a)}$ with $k=a$ do not appear in
the factorization ${\tilde w}={\tilde s}D$, thus such variables are
integrated out to $1$ in (\ref{normalized}).
This normalization is needed in order to extend the
positive linear form
\begin{eqnarray}
f({\tilde w})\mapsto \int_{W(Mat_p(R))}
{\hat f}_{\cal G}({\Delta}({\tilde w}))d({\mu}(\Delta))
\label{semplice}
\end{eqnarray}
to a measure over the infinite-dimensional space
$GL(Mat_p({\bf R}))$. Obviously, the Gaussian factor
breaks the invariance of the measure, (whereas if in the above
expression, ${\hat f}_{\cal G}({\Delta}({\tilde w}))$ is
replaced with a Haar average over $SL(q,{\bf R})$, then the resulting
positive linear form is easily shown to define a left Haar measure
on each $GL(q, {\bf R})$). This extension is accomplished
as follows.\par
\vskip 0.5 cm
Let $C(GL(Mat_p({\bf R})))$ denote the algebra of all continuous
real-valued functions on $GL(Mat_p({\bf R}))$. From the definition of
$GL(Mat_p({\bf R}))$ as the direct limit group of the system $\{GL(q,
{\bf R}), i_{q,q+1}\}$, (where
$i_{q,q+1}$ denotes the canonical injection of $GL(q,{\bf R})$ into
$GL(q+1,{\bf R})$), it follows that if $w$ is an element of
$GL(Mat_p({\bf R}))$ then there exists an integer $q$ and a
matrix $w_{(q)} \in GL(q,{\bf R})$ such that
$i_q(w_{(q)})=w$, (with $i_q$ being the injection of
$GL(q,{\bf R})$ into $GL(Mat_p({\bf R}))$). Thus if
$f\in C(GL(Mat_p({\bf R})))$ we can always write
$f(w)=f[i_q(w_{(q)})]$ for some $w_{(q)}\in GL(q,{\bf R})$.\par
These remarks show that, as claimed, the measure (\ref{semplice}) can
be naturally extended to a regular measure
$P(d{\tilde w})$ on $GL(Mat_p({\bf R}))$ by defining
\begin{eqnarray}
&\int_{GL(Mat_p({\bf R}))}f({\tilde w})dP({\tilde w})
\equiv \int_{W({Mat}_p({\bf R}))} \nonumber\\
&\left (
\int_{SL(q,{\bf R})}f[i_q({\tilde w}_{(q)})]q^{q^2/2}
{\Delta}^{-q^2}
\exp\left [-{\pi}q{\Delta}^{-2}\sum_a||{\bf x}_{(a)}||^2
\right ]
\prod_{k,a}^qd{x}^{k}_{(a)}\right )
d({\mu}(\Delta))
\label{nuovamu}
\end{eqnarray}
\vskip 0.5 cm
\noindent for every continuous function $f$ in
$GL(Mat_p({\bf R}))$, and for a $q$, (possibly depending on $f$),
sufficiently large. \par
\noindent In order to prove that such an extension is unique (and
well-defined
in the $q\to\infty$ limit), let us inject the matrix
$w_{(q)}\in GL(q,{\bf R})$ into $GL(q+k,{\bf R})$  for
$k>0$. Denoting by $i_{q,q+k}\colon GL(q,{\bf R})\to GL(q+k,{\bf R})$
the injection mapping, and noticing that
\begin{eqnarray}
i_{q+k}\circ i_{q,q+k}=i_q
\end{eqnarray}
we can write
\begin{eqnarray}
f[i_q(w_{(q)})]=f[i_{q+k}(w_{(q+k)})]
\end{eqnarray}
where we have set $w_{(q+k)}\equiv i_{q,q+k}(w_{(q)})$.\par
If we notice that the factorization
${\tilde w}_{(q)}={\tilde s}_qD =(\prod_{ja}^qB_{ja})D$
yields
\begin{eqnarray}
{\tilde w}_{(q+k)}={\tilde s}_{q+k}D =
(\prod_{ja}^{q+k}B_{ja})D
\end{eqnarray}
with $x^j_{(a)}=0$ for $j>q$, and $a>q$, then we can write
\begin{eqnarray}
&\int_{SL(q,{\bf R})}f[i_q({\tilde w}_{(q)})]q^{q^2/2}
{\Delta}^{-q^2}
\exp\left [-{\pi}q{\Delta}^{-2}||{\bf x}_{(a)}||^2\right ]
\prod_{j,a}^qd{x}^{j}_{(a)}=\nonumber\\
&\int_{SL(q+k,{\bf R})}f[i_{q+k}
({\tilde w}_{(q+k)})](q+k)^{(q+k)^2/2}
{\Delta}^{-(q+k)^2}
\exp\left [-{\pi}(q+k){\Delta}^{-2}||{\bf x}_{(a)}||^2\right ]
\prod_{j,a}^{q+k}d{x}^{j}_{(a)}\nonumber
\end{eqnarray}
where we have exploited the fact that the extra variables
$x^j_{(a)}$ for $j$, $a$ ranging over $q+1,\ldots, q+k$, not
being explicitly present in the function
$f[i_{q+k}({\tilde w}_{(q+k)}=
{\tilde s}_{q+k}D)]$, integrate
out to $1$ owing to the Gaussian nature of the integration.
According to standard results these remarks imply that
$P(d{\tilde w})$ is indeed the unique extension of
(\ref{semplice}) to $GL(Mat_p({\bf R}))$.\par
\vskip 0.5 cm
According to the measure  $P(d{\tilde w})$, for any $q$,
the $q^2$ variables $x_{(a)}^k$ have
the multinormal distribution with average $0$ and variance
matrix $t(q){\delta}_{ef}$, ($e$ and $f$ ranging over
$1,\ldots, q^2$). In particular, as $q\to\infty$, the variance
goes to $0$, and $P(d{\tilde w})$ degenerates to the Dirac measure
concentrated on the zero sequence in ${\bf R}^{\infty}$,
the infinite product of real lines.
 This fact depends on our particular
choice, (\ref{varianza}), of the variance $t(q)$.
Recall  that this choice  of the variance is motivated by the fact
that as
$q$ increases, we need to emphasize  the role of the matrices
which eventually reduces to the identity matrix, so as to
generate, {\it in measure}, the general linear group.\par
As a consequence of such remarks, it follows that
even if we are  considering the $q^2$ variables
$x_{(a)}^k$,  we should be actually interested  in the $q$ random
variables
\begin{eqnarray}
y_a \equiv \sum_{k=1}^q x^k_{(a)}
\end{eqnarray}
with $a=1,\ldots,q$. \par
The variables $y_a$  are  multinormally distributed with a zero mean,
but now with a variance matrix given by
\begin{eqnarray}
qt(q){\delta}_{ij}=(2{\pi})^{-1}{\Delta}^2
\end{eqnarray}
(with $i$,$j=1,\ldots,q$). We shall denote by
$P_y(d{\tilde w})$ the distribution  of the variables $y_a$
with respect to
$P(d{\tilde w})$.\par
 As follows from standard manipulations of gaussian integrals, this
distribution follows,
for $q$ generic, by replacing the original set $x^k_{(a)}$ with
the equivalent set of variables $(y_a, {\hat y}_i=x^i_{(a)})$ where
$i\not= a$, and integrating out to $1$
the $q^2-q$
extra variables ${\hat y}_i$. \par
Notice that, owing to the presence of the integration over
$W({Mat}_p({\bf R}))$, the measures $P({\tilde w})$ as well as the
associated distribution
$P_y({\tilde w})$ are not normalized to one.
\vskip 0.5 cm
\noindent{\it Second part}:\par
\vskip 0.5 cm
\subsection{The partition function in the thermodynamical limit}

Now we  establish a
connection between the partition function
$Z(m,J,{\pi}_1)$, and the joint distribution
$P_y(d{\tilde w})$ of the variables $y_a$  .\par
\vskip 0.5 cm
For each given $L(m)$-geodesic balls two-skeleton
${\Gamma}_2$, the
orthogonal representation ${\theta}$ of the fundamental group
${\pi}_1({\Gamma}_2)$ extend to a unique ring homomorphism
$\theta \colon {\bf Z}{\pi}_1({\Gamma}_2) \to Mat_p({\bf R})$,
which induces a group homomorphism between
$GL({\bf Z}{\pi}_1({\Gamma}_2))$ and
$GL({Mat}_p({\bf R}))$
and between the corresponding
Whitehead groups,
${\theta}_*\colon  W({\pi}_1({\Gamma}_2))
\to W(Mat_p({\bf R}))$, given by
\begin{eqnarray}
{\theta}_*[{\tau}(w_{ik})]={\tau}{\theta}_*(w_{ik})=
|\det ({\theta}(w_{ik}))|
\end{eqnarray}
Thus, since there are only a finite number of inequivalent simple
homotopy types realized in $\Ricco$, there is only a finite number
of inequivalent Reidemeister-Franz representation torsion
${\Delta}^{\theta}_{(i)}= |det({\theta}(w_{jk}))|$ realized by
manifolds
in $\Ricco$, ($i=1,\ldots, a <\infty$).\par
\vskip 0.5 cm
For a given value of the parameter $m$, let us denote by
${\cal I}_{(m)}$ the set of matrices  which are
the image, under the homomorphism
${\theta}_*$,
of the set of incidence matrices realized in
$\Ricco$. In particular, for any given value of $q$,
let ${\cal E}^{(i)}_{(m,q)}$ be the finite set of matrices
${\tilde w}^{(\alpha)}$ in ${\cal I}_{(m)}\cap GL(q,{\bf R})$
whose Reidemeister torsion is ${\Delta}_{(i)}$.  \par
\noindent Let
$U\subset GL(q,{\bf R})$ denote
a neighborhood of  ${\cal E}^{(i)}_{(m,q)}$, and let
$f\in C^{\infty}(GL({Mat}_p({\bf R})))$ be
a function such that the support
of $f\circ i_q$ is contained in $U$. \par
By localizing $f$ around each
${\tilde w}^{(\alpha)} \in {\cal E}^{(i)}_{(m,q)}$
so that $f\circ i_q \downarrow {\delta}_{\cal E}$, (the
Dirac measure supported on ${\cal E}^{(i)}_{(m,q)}$),
and integrating over $GL({Mat}_p({\bf R}))$ with
respect to $P_y(d{\tilde w})$  we get
\begin{eqnarray}
\int_{GL(Mat_p({\bf R}))}f({\tilde w})
{\phi}({\tilde w})dP_y({\tilde w})\to
\sum_{{\tilde w}^{(\alpha)} \in {\cal E}^{(i)}_{(m,q)}}
\exp[-{\pi}{\Delta}^{-2}\sum_{a=1}^q||y_a||^2
({\tilde w}^{(\alpha)})]
{\Delta}_{(i)}^{1-q}{\phi}({\tilde w}^{(\alpha)})\nonumber
\end{eqnarray}
for any function ${\phi}\in C^{\infty}(GL({Mat}_p({\bf R})))$
compactly supported in $U$.\par
More in general, let
${\cal E}^{(i)}_{(m)}\equiv \cup_q {\cal E}^{(i)}_{(m,q)}$,
as $q$ varies in the finite set of the possible values, over
$\Ricco$, of $pN^{(3)}_{(m)}$. Such ${\cal E}^{(i)}_{(m)}$ is
the finite set of matrices
${\tilde w}^{(\alpha)}$ in ${\cal I}_{(m)}$
whose Reidemeister torsion takes on the given
value ${\Delta}_{(i)}$. The foregoing analysis implies that
the measure of  ${\cal E}^{(i)}_{(m)}$ with respect
to $P_y(d{\tilde w})$ is given by
\begin{eqnarray}
{\mu}^{(i)}_{(m)}({\cal E}^{(i)}_{(m)})
\equiv &\int_{GL(Mat_p({\bf R}))}
({\tilde w})dP_y({\delta}_{\cal E}({\tilde w}))=\nonumber\\
&\sum_{{\tilde w}^{(\alpha)} \in {\cal E}^{(i)}_{(m)}}
\exp[-{\pi}{\Delta}^{-2}\sum_{a=1}^
{pN^{(3)}_{(m)}({\tilde w}^{(\alpha)})}||y_a||^2
({\tilde w}^{(\alpha)})]
{\Delta}_{(i)}^{1-pN^{(3)}_{(m)}({\tilde w}^{(\alpha)})}
\label{cinquantaquattro}
\end{eqnarray}
where $dP_y({\delta}_{\cal E}({\tilde w}))$ is the distribution
of the {\it counting} measure ${\delta}_{\cal E}({\tilde w})$
of the set ${\cal E}({\tilde w})$ with respect to $P_y$.\par
\vskip 0.5 cm
This result for the $P_y(d{\tilde w})$-measure of the set
${\cal E}^{(i)}_{(m)}$ is rather similar to the expression
of the partition function $Z(m,J,{\pi}_1)$, which
according to  proposition \ref{treuno} can
be written as
\begin{eqnarray}
\sum_{{\Delta}_i \in W(Mat)}
\sum_{{\tilde w}^{(\alpha)} \in {\cal E}^{(i)}_{(m,q)}}
\exp {[{i_c}N^{(1)}_{(m)}({\Gamma}_2)]}
\left (\frac{J}{2{\pi}}\right )^{-(p/2)N^{(3)}_{(m)}
({\tilde w}^{(\alpha)})}
{\Delta}^{\theta}({\tilde w}^{(\alpha)})
\label{otto}
\end{eqnarray}
In order to compare this expression with
(\ref{cinquantaquattro}),
let us remark that since the random variables $y_a$ are multinormally
distributed with  variance ${\Delta}^{2}/2{\pi}$, then
according to the strong law of large numbers, we have
\begin{eqnarray}
\lim_{q\to\infty}\frac{1}{q}\sum_{a=1}^q
\left (||y_a||^2 \right )=\frac{{\Delta}_{(i)}^2}{2{\pi}}
\end{eqnarray}
for $P_y$-almost all vectors $y_a$.
It follows that, for $q$, (and thus $m$), sufficiently large (notice
that
$q=pN^{(3)}_{(m)}$), we can write (\ref{cinquantaquattro}),
with a slight abuse of notation,
as
\begin{eqnarray}
{\mu}^{(i)}_{(m)}=
\sum_{{\tilde w}^{(\alpha)} \in {\cal E}^{(i)}_{(m)}}
\exp[-\frac{p}{2}N^{(3)}_{(m)}
({\tilde w}^{(\alpha)})]
{\Delta}_{(i)}^{1-pN^{(3)}_{(m)}({\tilde w}^{(\alpha)})}
\label{resemblance}
\end{eqnarray}
which, upon setting $N^{(3)}_{(m)}\simeq bN^{(1)}_{(m)}$,
bears a strong resemblance with (\ref{otto}). Explicitly,
by comparing the expression of the measure
${\mu}^{(i)}_{(m)}({\cal E}^{(i)}_{(m)})$, (see (\ref{resemblance})),
and
the right side of (\ref{otto}), (by restricting the sum
over the torsions to a given value, say ${\Delta}_{(i)}$),
it immediately follows that if
the parameter $J$  takes the value
\begin{eqnarray}
\frac{J_{crit}}{2{\pi}}=
{\left ({\Delta}^{\theta}_{(i)}\right )}^2
\label{valcrit}
\end{eqnarray}
and if
\begin{eqnarray}
{i_c}= {i_c}_{(m)}\equiv -\frac{p}{2}
\frac{N^{(3)}_{(m)}}{N^{(1)}_{(m)}}
\end{eqnarray}
then, for $m$ sufficiently large, and for any given value of the
ratio $N^{(3)}_{(m)}/N^{(1)}_{(m)}$, we can write
\begin{eqnarray}
Z(m,J,{\pi}_1)|_{J=J_{crit}}=
{\mu}^{(i)}_{(m)}({\cal E}^{(i)}_{(m)})
\end{eqnarray}
where $Z(m,J,{\pi}_1)|_{J=J_{crit}}$ denotes the partition function
$Z(m,J,{\pi}_1) $ evaluated by restricting the summation
over the torsions ${\Delta}_{(k)}$ to a given value ${\Delta}_{(i)}$,
and by setting $J$ equal to the corresponding
value of $J_{crit}$.\par
Notice that, as $m\to\infty$, we get
\begin{eqnarray}
{i_c}=-\frac{p}{2}
\frac{N^{(3)}_{(m)}}{N^{(1)}_{(m)}}
\to_{m\to \infty} - \frac{1}{2}\frac{p}{b}
\end{eqnarray}
(where $b\equiv \limsup_{m\to \infty}
(N^{(1)}_{(m)}/N^{(3)}_{(m)})$).\par
\vskip 0.5 cm
Let us denote by $P^{(i)}_y(d{\tilde w})$  the
measure obtained by restricting the measure $P_y$
to the subset of matrices in $GL(Mat_p({\bf R}))$ with torsion
${\Delta}^{\theta}_{(i)}$, namely
\begin{eqnarray}
P^{(i)}_y(d{\tilde w})\equiv
{\delta}({\Delta}_{(i)})\otimes P_y(d{\tilde w})
\end{eqnarray}
where ${\delta}({\Delta}_{(i)})$ is the Dirac delta, (with
respect to the invariant measure $d({\mu}(\Delta))$ on
$W({Mat}_p({\bf R}))$. Notice that this measure is normalized
to ${\Delta}_{(i)}$ rather than to $1$. (The factor
${\Delta}_{(i)}$ is originated by by that
${\delta}({\Delta}_{(i)})$ is a Dirac distribution with respect to the
multiplicative measure on ${\bf R}^+$).\par
\vskip 0.5 cm
According to such remarks  it follows that, corresponding to the
critical values
$J=J_{crit}$ and $i_c=-p/2b$, we have
\begin{eqnarray}
\lim_{m\to\infty}
Z(m,J,i_c, {\pi}_1)|_{J=J_{crit}}=
\lim_{m\to\infty}
\int_{GL(Mat_p({\bf R}))}P^{(i)}_y({\delta}_{\cal E}({\tilde w}))
\end{eqnarray}
\vskip 0.5 cm
We can explicitly evaluate the above limit and show that
$\lim_{m\to\infty}Z(m,J,i_c,{\pi}_1)
|_{J=J_{crit}}={\Delta}_{(i)}$ by noticing
that, as
$m\to\infty$, the mapping from the set of incidence matrices
${\cal E}_{(m)}$ and
the subset of matrices in $GL(Mat_p({\bf R}))$ with torsion
${\Delta}^{\theta}_{(i)}$ is, up to formal deformations,
surjective.\par
In order to prove this latter statement let
${\tilde w}$ be a matrix in
$GL(Mat_p({\bf R}))$ which is  not in
${\cal E}_{(m)}$, for some given value of the parameter $m$.
This means that  for an integer $q$, (depending on $m$), the matrix
${\tilde w}^{(q)}\in GL(q,{\bf R})$,  with
$i_q({\tilde w}^{(q)})={\tilde w}$, is not in ${\cal E}_{(m)}$.
However, through formal deformations we can always map such
${\tilde w}^{(q)}$ into a new ${\tilde w}^{(p)}$, in general
with $p\not= q$, which is realized as the incidence matrix,
(in the given orthogonal representation), of some geodesic ball
covering of manifolds in $\Ricco$. Thus, as $m\to\infty$, and up to
formal deformations, we can always assume that a given incidence
matrix ${\tilde w}$ is actually realized by some geodesic ball
covering for some value, say ${\hat m}$ of the cut-off parameter. We
stress that this is no longer true for a fixed
$m$, (since, as recalled above,  a formal deformation may require to
go  from a matrix of order $q$, corresponding to the given $m$, to a
matrix of order $p$, with $p\not= q$). \par
It is natural to define ${\cal E}$ as the direct limit of the
${\cal E}_{(m)}$ or more explicitly, ${\tilde w} \in {\cal E}$ if
there exists a value of the cut-off parameter $m$, a corresponding
integer $q=q(m)$, and
 a matrix ${\tilde w}^{(q)}\in {\cal E}_{(m)}$ such that $i_q({\tilde
w}^{(q)})={\tilde w}$. \par
\vskip 0.5 cm
According to the above remarks, ${\cal E}$ can be identified
with the subset of matrices in $GL(Mat_p({\bf R}))$ with torsion
${\Delta}^{\theta}_{(i)}$, set which supports the  measure
$P^{(i)}_y(d{\tilde w})$. Thus we get
\begin{eqnarray}
&\lim_{m\to\infty}
Z(m,J,i_c, {\pi}_1)|_{J=J_{crit}}=
\lim_{m\to\infty}
\int_{GL(Mat_p({\bf R}))}
P^{(i)}_y({\delta}_{{\cal E}_m}({\tilde w}))=\nonumber\\
&=\int_{GL(Mat_p({\bf R}))}P^{(i)}_y(d{\tilde w})=
{\Delta}_{(i)}
\end{eqnarray}
as stated.
\vskip 0.5 cm
This last results shows, as claimed, that
as $m \to \infty$, the statistical system described by
$Z(m,J,{\pi}_1)$ admits a well-defined thermodynamical
limit for ${i_c}=-p/2b$ and
$J=J_{crit}$, (where $J_{crit}$ is one of the
values (\ref{valcrit}), say
$2{\pi}{\left ({\Delta}^{\theta}_{(i)}\right )}^2$).\par
Owing to the combinatorial invariance of the representation torsion,
such limit depend only on the particular representation of the
fundamental group, and on the simple homotopy type
that we are considering.
As we shall see in a moment, such limit describes
(homology) manifolds $M$ in $\Ricco$ whose Reidemeister-Franz
torsion takes on the given value ${\Delta}^{\theta}_{(i)}$, and
which are generated, as $m \to \infty$, by rather regular arrays
of $L(m)$-geodesic balls.\par
\vskip 0.5 cm
Moreover, since, if $i\not= k$, we have trivially
\begin{eqnarray}
\lim_{m\to\infty}{\mu}^{(i)}_{(m)}
\left [\det {\tilde w} ={\Delta}_{(k)} \right ]=
\lim_{m\to\infty}P^{(i)}_y
({\cal E}^{(k)}_{(m)})= 0
\end{eqnarray}
this authorizes the conclusion that as $J$ varies from one
critical value to another,
the system exhibits a  phase transition
associated with the changes of the corresponding simple-homotopy
types, and, through the orthogonal representation ${\Theta}$,
the set of manifolds in $\Ricco$
with isomorphic fundamental group appears as a mixture of pure phases
described
by the probability measure on $GL(Mat_p({\bf R}))$
\begin{eqnarray}
P(d{\tilde w}) \equiv \frac{1}{\alpha}\sum_{i=1}^{\alpha}
P_{(i)}(d{\tilde w})
\end{eqnarray}
(with $\alpha=||{\Delta}^{\theta}||({\pi}_1)$).
$\clubsuit$
\vskip 1 cm
A somewhat surprising feature of the above results is the  way in
which the gaussian distribution $P^{(i)}_y$ over the general linear
group yields for the exact evaluation of the thermodynamical limit of
the partition function. Not quite so surprising is the fact that the
partition function is provided
by the corresponding value of the representation torsion.
This latter result, in particular, points toward a direct connection
with the Chern-Simons\cite{Witt} approach to quantize three-gravity,
where the
corresponding partition function is  provided by
the Ray-Singer torsion, the analytic couterpart of
${\Delta}_{(i)}$. There is equality\cite{RayS} between the two as long
as the Reidemeister-Franz torsion is, for a given representation of
the fundamental group, known to come from a smooth triangulation.\par
Either the nature of $P^{(i)}_y$ or the connection with Chern-Simons
approach follows from the role that
the orthogonal representation of the fundamental group has
in our results\cite{BBC}.\par
\vskip 0.5 cm
The use of a representation of the fundamental group is
forced upon us by the structure of the space of bounded geometries,
(recall Fukaya's theorem\cite{Grom} on the structure of
collapsed manifolds providing the $d_G$-boundary points of
${\cal R}(n,r,D,V=0)$). Actually, it is the
use of a representation that allows us to label the cells, (to
{\it colour} them), and thus to implement the rather standard
statistical mechanical approach yielding to the results
connecting $Z(m,J,{\pi}_1)$ to simplicial three-gravity. It is
clear that different representations, (say on more elementary groups
than the orthogonal group), may yield for different statistical
mechanical models, ({\it e.g.}, Ising-like models;
it must be noted that the $3D$-Ising model has been
recently related to $2+1$-dimensional quantum
gravity\cite{Martellini}), but, as recalled, the
orthogonal group suggests itself also because it captures, in an
essential
way, the non abelian character of the fundamental group. This is
an important point if one does not want to loose geometrical
information
during the process of (homotopical) reconstruction of the manifold,
(through formal deformations), which underlines the definition
of $Z(m,J,{\pi}_1)$. \par
\vskip 0.5 cm
For a given manifold $M\in \Ricco$, let $Hom({\pi}_1(M),O(p))$ denote
the set of all homomorphisms ${\pi}_1(M)\to O(p)$, {\it i.e.},
the space of all representations of ${\pi}_1(M)$ into the
orthogonal group $O(p)$. For a given $m$, (sufficiently large), the
fundamental group ${\pi}_1(M)$ is actually given,
as a finitely generated group, by the presentation associated
with the $L(m)$-geodesic ball two skeleton ${\Gamma}^{(2)}_{(m)}$, as
${\pi}_1(M)\simeq {\pi}_1({\Gamma}^{(2)}_{(m)})$. More explicitly, as
we have seen in a previous paragraph, we can assume that
${\pi}_1({\Gamma}^{(2)}_{(m)})$ has a presentation
\begin{eqnarray}
<{\gamma}_1,\ldots,{\gamma}_{l(m)}|R_1({\gamma}_1,\ldots,
{\gamma}_l)=\ldots =I>
\end{eqnarray}
where ${\gamma}_1,\ldots,{\gamma}_{l(m)}$ denote the
$m$-dependent generators, (with $l(m)=1-N^{(0)}_{(m)}+N^{(1)}_{(m)}$)
associated with the one-skeleton graph ${\Gamma}^{(1)}_{(m)}$, and
where $R_i=\ldots =I$ denote
the corresponding relations (associated with the pasting of the faces
$p^{(2)}_{ijk}$ into ${\Gamma}^{(1)}_{(m)}$). The map
${\gamma}_i\to {\phi}({\gamma}_i)$, with
${\phi}({\gamma}_i)\in O(p)$, and $i=1,\ldots,l(m)$, defines an
element of $Hom({\pi}_1(M),O(p))$, and as $m$ varies (and eventually
$m\to\infty$), the $m$-depending presentations
${\pi}_1({\Gamma}^{(2)}_{(m)})$ are associated with a corresponding
set of representations in $Hom({\pi}_1(M),O(p))$.\par
In our approach, leading to simplicial three-gravity, we are actually
dealing with the group ring ${\bf Z}{\pi}_1(M)$, and since every
linear representation of ${\pi}_1(M)$ can be extended to a
representation of its group ring, it is not too difficult to realize
that the measure $P^{(i)}_y(d{\tilde w})$ on
$GL({Mat}_p ({\bf R}))$ can be interpreted as a measure on the
representation space associated with the group ring
${\bf Z}{\pi}_1(M)$.\par
The space $Hom({\pi}_1(M),O(p))$, or rather, the associated orbit
space (under conjugation), $Hom({\pi}_1(M),O(p))/O(p)$, is the
deformation space of flat principal $O(p)$-bundles over $M$.
In such a sense, $P^{(i)}_y(d{\tilde w})$ induces a measure over such
orbit space. Now, the orbit space of flat bundles associated with the
space of flat connections, ($SO(2,1)$-connections for
$2+1$-gravity, $SO(3)$-connections for euclidean $3$-gravity), is
exactly the configuration space over which one functionally integrates
in defining Chern-Simons versions of three-gravity.
These remarks, which can be made more precise, make less misterious
either the role of $P^{(i)}_y(d{\tilde w})$ or the connection with the
standard field-theoretic approach to three-dimensional quantum
gravity.
\vskip 0.5 cm
Another non-trivial
aspect of the limit $\lim_{m\to \infty}Z(m,J,{\pi}_1)$
lies in noticing that such limit is actually well defined
in a convex region in the plane of all possible couplings
$(\sigma, i_c)$, and that  a first order phase transition
occurs precisely when such couplings approach the critical
values provided by proposition \ref{bello}. Such phase
transition does not describe the change from a simple homotopy
type into another, ( since it occurs within a given simple
homotopy type),  it rather signals the passage from an irregular
configuration,
(a three-dimensional homology manifold generated by sequences of
$L(m)$-minimal geodesic ball coverings with a low filling),
to a more regular
one describing a (homology) manifold generated by more
structured packings of $L(m)$-geodesic balls, (such nicer packings,
are {\it characterized} by the statistical domination of
configurations with a large filling function $N^{(0)}_{(m)}$).\par
\noindent In order to prove such statements we proceed as
follows.\par
\noindent Let us consider a given value, ${\Delta}$,
of the Reidemeister-Franz torsion, and rewrite
$Z(m,J,{\pi}_1)|_{\Delta}$ as
\begin{eqnarray}
Z(m,J,{\pi}_1)|_{\Delta}=
\sum_{({\cal N},{\Gamma}_2)}{\Delta}^{\theta}
({\cal N},{\Gamma}_2)
\exp
\left [-N^{(3)}_{(m)}({\cal N})\left (\sigma -
i_c \frac{N^{(1)}_{(m)}({\Gamma}_2)}
{N^{(3)}_{(m)}({\cal N})}\right )\right ]
\end{eqnarray}
For a given $m$, sufficiently large, and a running
integer $t$, let us denote  by
$Q_{(m)}({\Delta},b,t)$ the number of combinatorially inequivalent
$L(m)$-geodesic balls nerves ${\cal N}_{(m)}$ of given
torsion ${\Delta}$,  with
$N^{(3)}_{(m)}({\cal N})=t$ and with a given value of
the ratio  ${N^{(1)}_{(m)}({\Gamma}_2)}/
{N^{(3)}_{(m)}({\cal N})}= b$; ($Q_{(m)}({\Delta},b,t)$ is
basically the micro-canonical partition function).
Thus, for a given value of $b$, we can write
\begin{eqnarray}
Z(m,J,{\pi}_1)|_{\Delta}=
\sum_t{\Delta}^{\theta}
\left [Q_{(m)}({\Delta},b,t) \right ]
\exp[-t(\sigma - i_c b) ]
\label{power}
\end{eqnarray}
With these preliminary remarks, the following result obtains
\begin{Gaiap}
For any given value of $b$ and for a
given value ${\Delta}^{\theta}_{(i)}$ of the torsion ,
the partition function
$Z(m,J,{\pi}_1)|_{\Delta}$ is well defined, as $m \to \infty$, for all
$i_c$, and ${\sigma}$ in the convex region
${\cal D}(\sigma, i_c)$ defined, in the plane of possible couplings
$(\sigma, i_c)$, by the condition
\begin{eqnarray}
\sigma - b{i_c} \geq {\sigma}_{crit} + \frac{1}{2}p > 0
\label{crit2}
\end{eqnarray}
where
\begin{eqnarray}
{\sigma}_{crit}\equiv
\frac{p}{2}\ln \left (\frac{J_{crit}}{2{\pi}}\right )=
p \ln \left ({\Delta}^{\theta}_{(i)}\right )
\end{eqnarray}
is the critical value of the {\it bare} cosmological constant
${\sigma}$ corresponding to the given ${\Delta}^{\theta}_{(i)}$. \par
\noindent If
\begin{eqnarray}
g(b)\equiv \limsup_{t\to\infty}\frac{1}{t}
\ln {\left [Q_{(m)}({\Delta},b,t)\right ]}
\label{potential}
\end{eqnarray}
denotes the rate of exponential growth
in the number of geodesic ball nerves ${\cal N}_{(m)}$ as a
function of $b$, then $g(b)$
is a positive concave function of $b$.
Moreover, as $i_c \to - \frac{1}{2}p/b$, and
$\sigma \to {\sigma}_{crit}$, the function $g(b)$ reduces
to a constant.
Thus, the system described by the partition function
$Z(m,J,{\pi}_1)|_{\Delta}$ exhibits, as $m\to\infty$,
a first-order phase transition at
$\sigma = {\sigma}_{crit}$ and ${i_c}(b)=-\frac{1}{2}p/b$.
\label{transizione}
\end{Gaiap}
\vskip 0.5 cm
{\it Proof}. For a given value of the parameter $b$,
we can consider (\ref{power}) as a Dirichlet series
of type $t$, (or, equivalently, as a power series in the
variable $\exp [-\sigma +{i_c}b]$; there are a number of
advantages in considering (\ref{power})as a Dirichlet
series, advantages which are connected to the use of the
variables $\sigma$ and $i_c$). According to
proposition \ref{bello}, such series converges if
$\sigma -{i_c}b = {\sigma}_{crit}+ \frac{1}{2}p$. \par
\noindent  If ${\sigma}_{crit}+ \frac{1}{2}p>0$, or more explicitly,
if
\begin{eqnarray}
\ln ({\Delta}^{\theta}_{(i)})^2+1>0
\end{eqnarray}
\noindent then this convergence implies that
\begin{eqnarray}
\sum_t^k Q_{(m)}({\Delta},b,t)= {\cal O}\left (
\exp [{\sigma}_{crit}+ \frac{1}{2}p]k \right )
\label{grandeO}
\end{eqnarray}
as $t\to \infty$, and conversely it also implies that
$\lim_{m\to\infty}Z(m,J,{\pi}_1)|_{{\Delta},b}$ converges in the
half-plane  $\sigma -{i_c}b > {\sigma}_{crit}+ \frac{1}{2}p$.\par
Moreover, from (\ref{grandeO}) it follows that
\begin{eqnarray}
\sum_t^k Q_{(m)}({\Delta},b,t) < M
\exp [{\sigma}_{crit}+ \frac{1}{2}p]k
\end{eqnarray}
for some constant $M$, and $k=1,2,3,\ldots$. Thus
\begin{eqnarray}
\frac{\ln \sum_t^k Q_{(m)}({\Delta},b,t)}{k}< \frac{\ln M}{k}+
{\sigma}_{crit}+ \frac{1}{2}p
\end{eqnarray}
\noindent By taking the limit superior of the left-hand side we
get the abscissa of convergence of (\ref{power}), which is thus
bounded above by  ${\sigma}_{crit}+ \frac{1}{2}p$. Thus it follows
from the
foregoing arguments that the function
$g(b)$ defined by (\ref{potential}), exists and it is finite.
Its concavity trivially follows from Jensen inequality as applied
to the logaritmic function.\par
\vskip 0.5 cm
We can now show that, when ${i_c}=-\frac{1}{2}p/b$ and
$\sigma \to {\sigma}_{crit}$, an horizontal segment is present
in the graph of $g(b)$.\par
Again according to proposition \ref{bello}, as $m\to\infty$,
the partition function $Z(m,J,{\pi}_1)|_{\Delta}$ is well
defined for  $\sigma ={\sigma}_{crit}$ and all
$i_c$ varying on the curve, (in plane $(b,i_c)$), defined
by ${i_c}(b)= -\frac{1}{2}p/b$.
Corrisponding to such values of the couplings we get
\begin{eqnarray}
\sum_t &
\left [Q_{(m)}({\Delta},b_1,t)\right ]
\exp[-t({\sigma}_{crit} -{i_c}(b_1)b_1) ]=\nonumber \\
\sum_t &
\left [Q_{(m)}({\Delta},b_2,t)\right ]
\exp[-t({\sigma}_{crit} - {i_c}(b_2)b_2)]=\ldots\nonumber\\
\sum_t &
\left [Q_{(m)}({\Delta},b_i,t)\right ]
\exp[-t({\sigma}_{crit} -{i_c}(b_i)b_i) ]
\label{line}
\end{eqnarray}
for all pairs $(b_{(i)}, {i_c}(b_{(i)}))$ lying on the
curve ${i_c}(b)= -\frac{1}{2}p/b$. From
(\ref{line}), it follows that
$Q_{(m)}({\Delta},b_1,t)=Q_{(m)}({\Delta},b_2,t)=
\ldots=Q_{(m)}({\Delta},b_{(i)},t)$,
which, by reference to the definition, (\ref{potential}),
of the function $g(b)$, implies the presence of an horizontal
segment in the graph of $g(b)$. By standard arguments, (see
{\it e.g.}, the book by Ruelle\cite{Ruel}), the fact that $g(b)$
reduces to a constant on some interval contained in $[0,b_{max}]$,
implies the existence of a first-order phase transition at
$i_c = -\frac{1}{2}p/b$, and
$\sigma={\sigma}_{crit}$.\par
\vskip 0.5 cm
{}From a geometrical point of view, we recall that $b$ provides the
(inverse) density of tetrahedra for link in the polytope associated
with a
minimal geodesic ball covering.
For each given value of such density, the function $g(b)$ provides,
in the large $m$ limit, the rate of (exponential) growth in the number
of geodesic ball nerves ${\cal N}_{(m)}$, with such density of
tetrahedra,
actually realized for manifolds in $\Ricco$. The above result
on the existence of a first order phase transition, can be interpreted
by saying that, for $\sigma = {\sigma}_{crit}$, the rate of
exponential
growth $g(b)$ stays constant on the critical
line ${i_c}(b)= -\frac{1}{2}p/b$, while the density of
tetrahedra for link undergoes a finite change.\par
\vskip 0.5 cm
In order to discuss the geometrical meaning of this phase transition
more in details, let us remark that along the same lines of the
entropy estimates established in paragraph (3.5),
the microcanonical partition function $Q_{(m)}({\chi},{\Delta},b,t)$,
(for the value $b=3$ of the parameter $b$ which is actually realized,
for a given volume $V$, and for a given value of the Euler
characteristic), can be asymptotically estimated as $t\to\infty$
according to, (see \ref{euler}),
\begin{eqnarray}
Q_{(m)}({\chi},{\Delta},b,t)
{\buildrel {t \to \infty}
\over \longrightarrow}
m^n(A_c)^t
(t=m^nV)^{(\frac{{\chi}(M)}{2})({\zeta}_0-
2)-1}
\cdot{\eta}_{h}(1+O(\frac{1}{t}))
\label{asintQ}
\end{eqnarray}
where  $n=3$, and $A_c$, ${\zeta}_0$, and ${\eta}_{h}$ are suitable
constants possibly depending on the torsion ${\Delta}_{(i)}$.\par
Upon introducing this expression in $Z(m,J,{\pi}_1)|_{\Delta}$,
it follows that as $m\to\infty$, the behavior of the partition
function is dominated by the subleading power in (\ref{asintQ}).
This subleading power yields to a converging
$Z(m,J,{\pi}_1)|_{\Delta}$ as $m\to\infty$ only if
$(\frac{{\chi}(M)}{2})({\zeta}_0-2)<0$ which implies that
${\chi}(M)<0$ and ${\zeta}_0-2 >0$. While, if
${\chi}(M)=0$ then $Z(m,J,{\pi}_1)|_{\Delta}$ diverges.\par
Thus pseudo-manifolds dominate the thermodynamical limit of
$Z(m,J,{\pi}_1)|_{\Delta}$. $\clubsuit$
\vskip 0.5 cm
According to proposition \ref{transizione}
the transition which isolates the pure phase
$P_{(i)}$ among the mixture which is associated
with $\lim_{m\to\infty}Z(m,J,{\pi}_1)$ geometrically
corresponds to a pseudo-manifold.\par
One is tempted to see in such behavior of
the partition function the analytical counterpart of the
{\it vacuum}, (or rather of the {\it vacua}, each vacuum
being associated with one of the simple homotopy types realized),
whose existence is numerically suggested by more
direct computer-assisted approaches \cite{{Jana5},{Jana6}}. \vskip 0.5
cm
\section{Concluding remarks}

The  foregoing results show that even if we can define
coherently the thermodynamical limit of simplicial three-gravity
as described by $Z(m,J,{\pi}_1)$, we end up with configurations in
which pseudo-manifolds dominate. Obviously, we would be more
interested
in discussing the possibility for the existence of a continuum
limit of the theory, where the cell size, in properly defined
physical units, tends to zero, and where manifolds have a non-
vanishing probability of occurrence.\par
 It has been
argued\cite{{Jana5},{Jana6}}
that this latter limit presumably does not exists. The first order
nature of the transition from the {\it hot} phase to the
{\it cold} phase supports this point of view. In this sense,
further analytical support comes also from our results, again
owing to the first order nature of the transition accompanying
the realization of the {\it regular covering} phase out of
the phase describing homology manifolds associated with
finer and finer geodesic balls coverings. \par
As we have shown here, euclidean simplicial three-gravity
basically appears as a theory which {\it statistically
reconstructs} an extended three-dimensional
object out of the presentation of the fundamental group
${\pi}_1({\Gamma}_2,e_0)$ associated with a dynamical triangulation.
The basic role in the theory is thus played by the fundamental
group, and by its orthogonal representations.\par
We exploited both such roles by
resolving the thermodynamic limit of the
theory into extended object (homology manifolds) which corresponds
to the distinct ways, (up to the action of the fundamental group), one
may
get homology manifolds through collapses or expansions of cells.
The existence of a critical value for $\sigma$ is  associated
with the partition of the configuration space of the theory into
simple homotopy equivalence classes labelled by the representation
torsions ${\Delta}^{\theta}_{(i)}$. While the existence of a critical
value for the chemical potential $i_c$, given a value of the
parameter $b$, is related in a quite non-trivial way to the fact that
corresponding
to a generic $L(m)$-geodesic ball coverings there is a local {\it
optimal}
covering, {\it viz.}, the  {\it minimal}
$L(m)$-geodesic ball covering maximizing the associated filling
function $N^{(0)}_{(m)}(M)$.
  In particular,  since homogeneous and isotropic
three-manifolds have a strong chemical affinity for a
{\it geodesic balls gas}, (recall that the {\it best filling} is
realized on such manifolds), one may expect that (portions of)
space forms should be
the natural outcome of the phase transition mechanism discussed above.
Sometimes this may be the case, but in general this mechanism is
unrealistic. In order to explain this remark,
let us notice that
the  geodesic ball filling mechanism of riemannian manifolds bears a
strong analogy with the mechanism
governing space-filling procedures through
random close packing of small spheres\cite{Zallen}.
As is known, random close packing, while not as
efficient as crystalline close packing in filling a portion of
three-dimensional space,  provides a good compromise.
It corresponds to a metastable arrangement, (stable against small
displacements), associated with a local energy minimum in the
configuration space of the system. This stability
is so strong that there is no way
to deform continuously, through an increase of density, a random
close packing into a crystalline close packing.\par
\noindent These remarks suggest that minimal geodesic ball
coverings of riemannian manifolds in $\Ricco$ are the natural
counterpart of random close packing of ordinary space, and that
(local) crystalline close packing corresponds to the optimal geodesic
ball fillings of constant curvature space. In this sense, the
statistical behavior of geodesic ball coverings as described by
$Z(m,J,{\pi}_1)$ can be heuristically modelled after that
of random close packing. In particular, the difficulties in
having a continuum limit of the theory can be related to the
known lack of correlation between short-range dense packing
and long-range crystalline order which is typical of the
dimension three.\par
\noindent The  first order phase transition appearing in the
thermodynamic limit of the geodesic ball packing may then
have its natural counterpart in the sudden deformation
of random close packed spheres, (deformation into Wigner-Seitz
cells\cite{Zallen}; by definition, a Wigner-Seitz cell surronding a
given site, is the cell containing all points closer to that site
than to any other), as they extend to fill a space of fixed
volume. This transition is accompanied by a jump of the coordination
number between the cells, (number which, roughly speaking, corresponds
to our parameter $b$), but it is not associated with a deformation
of the random close packing filling into a crystalline close
packing, {\it i.e.}, long range order is not activated.\par
\vskip 0.5 cm
The above remarks should be compared with the situation in
dimension four. Leaving aside the heuristics of random close
packings, the point is again
in the interplay between homotopy theory and the differentiable
structures a manifold can carry.\par
\noindent For spaces of bounded geometries such as $\Ricco$, there is,
as often quoted, a finiteness theorem for the distinct homotopy types
that can be realized. In dimension three and four, this is
all the topological information we can exploit while keeping
ourselves in the Gromov-Hausdorff compactification of the
set of riemannian structures considered. This is the price we
have to pay so as to be able to exploit the compactness
of $\Ricco$ for measure-theoretical purposes ({\it i.e.}, for
having a thermodynamical limit). As remarked above, the effect
of this state of affairs is that in dimension three, the
control of $Z(m,J,{\pi}_1)$ is determined by the fundamental
group. However, in dimension four there is a subtle interplay
between homotopy theory and differentiable structures. To make
a significant example, one knows, by a theorem of Whitehead,
that the homotopy type of a compact simply-connected four-manifold
is determined by the isomorphism class of the intersection form
of the manifold\cite{Whitehead}. Moreover, through the results of
Freedman\cite{Freedman} and Donaldson\cite{Donaldson}, one can
establish a direct connection between the algebraic properties
of the intersection forms and the topological and differentiable
structures that can live on compact (simply-connected) four-manifolds.
Thus, one may reasonably try to carry over the previous
analysis to the four-dimensional case,
hoping to reach for a sensible continuum limit.
In the four-dimensional case, the labelling of simple homotopy
types through the Reidemeister-Franz representation torsion, does
not provide much information. Such torsions are trivial,
({\it i.e.}, ${\Delta}^{\theta}=1$), for any smooth four dimensional
manifolds. However, the homotopical approach, which drives
our formalism, can be put to work  quite
effectively also in dimension four. As recalled,
homotopical information is now provided,
(say in the simply connected case, for simplicity), by
the intersection form, and it is
possible to adopt the strategy described here by exploiting the
reconstruction of (simply connected) four manifolds out of a
wedge of two-spheres on which a four cell is pasted through a
suitable attaching map. \par

\vfill \eject

\section*{References}
\begin{description}
\bibitem[1] {Pfister}
J.Fr\"{o}hlich, C.E. Pfister, T.Spencer, {\it On the
statistical mechanics of surfaces}, Lecture Notes in Physics
{\bf 173}, 169-199 (Springer-Verlag, 1982)
\bibitem[2] {Garrido}
J.Fr\"{o}hlich,{\it The statistical mechanics of surfaces},
in {\it Applications of Field theory to statistical
mechanics}, ed. L.Garrido, 31-57 (Springer Lecture Notes
in Physics {\bf 216}, 1984)
\bibitem[3] {David}
F.David, Nucl.Phys.{\bf B257}, 543-576 (1985); Phys.Lett.
{\bf 159B},303-306 (1985)
\bibitem[4] {Dur}
J.Ambj\o rn, B.Durhuus, J.Fr\"{o}hlich, Nucl.Phys.{\bf B257},
433-449 (1985);\par
J.Ambj\o rn, B.Durhuus, J.Fr\"{o}hlich, P.Orland, Nucl.Phys.
{\bf B270}, 457-482 (1986)
\bibitem[5] {Kazakov1}
V.A.Kazakov, I.K.Kostov, A.A.Migdal, Phys.Lett. {\bf 157B},
295-300 (1985)
\bibitem[6] {Brez}
E.Br\'ezin and V.Kazakov,Phys.Lett. {\bf B236},144 (1990).
\bibitem[7] {Doug}
M.Douglas and S.Shenker, Nucl.Phys. {\bf B235}, 635 (1990).
\bibitem[8] {Gros}
D.J.Gross and A.A.Migdal, Phys.Rev.Lett. {\bf 64}, 127 (1990);
Nucl.Phys. {\bf B340}, 333 (1990).
\bibitem[9] {Ambj}
J.Ambj\o rn, {\it Random surfaces: a non-perturbative
regularization of strings} in {\it Probabilistic Methods in
Quantum Field Theory and Quantum Gravity} P.H.Damgaard {\it et al.}
eds. (Plenum Press, New York, 1990).
\bibitem[10] {Kazakov2}
V.A.Kazakov, Mod.Phys.Lett. {\bf A4},2125-2139 (1989)
\bibitem[11] {Kazakov3}
V.A.Kazakov, A.A.Migdal, Nucl.Phys. {\bf B311}, 171-190 (1989)
\bibitem[12] {Kostov1}
I.K.Kostov, A.Krzywicki, Phys.Lett.{\bf B187}, 149-152 (1987)
\bibitem[13] {Kostov2}
I.K.Kostov, M.L.Mehta, Phys.Lett. {\bf B189}, 118-124 (1987)
\bibitem[14] {Kostov3}
I.K.Kostov, Nucl.Phys.B(Proc.Suppl.) {\bf 10A}, 295-322 (1989)
\bibitem[15] {Pari}
D.Bessin, C.Itzykson and J.B.Zuber, Adv.Appl.Math. {\bf 1},109(1980);
E.Br\'ezin, C.Itzykson, G.Parisi, J.B.Zuber, Commun.Math.Phys.
{\bf 59}, 35 (1978); D.Bessis, Commun.Math.Phys. {\bf 69},
143 (1979);  W.J.Tutte, Canad.Journ.Math. {\bf 14}, 21 (1962).
\bibitem[16] {Fernandez}
R.Fernandez, J.Fr\"{o}hlich, A.Sokal, {\it Random walks, critical
phenomena, and triviality in quantum field theory},TMP
(Springer-Verlag, Berlin Heidelberg 1992)
\bibitem[17] {Regge}
T.Regge, Il Nuovo Cimento {\bf 19}, 558 (1961);
\bibitem[18] {Witten}
E.Witten, Surveys in Diff.Geom. {\bf 1} 243-310 (1991).
\bibitem[19] {Konse}
M.Kontsevitch, Commun.Math.Phys. {\bf 147} 1 (1992).
\bibitem[20] {Penner}
R.Penner, Bull.Amer.Math.Soc. {\bf 15} 73 (1986);
see also: J.Harer,D.Zagier, Invent.Math. {\bf 185} 457 (1986)
\bibitem[21] {Jana1}
J.Ambj\o rn, Nucl.Phys.{\bf B}(Proc.Suppl.){\bf 25A}
(1992)8;
\bibitem[22] {Jana2}
J.Ambj\o rn, B.Durhuus, T.J\'onsson, Modern Phys.Lett.{\bf A 6}
(1991)1133;
\bibitem[23] {Jana3}
 M.E.Agishtein and A.A.Migdal,
Mod.Phys.Lett. {\bf A 6} (1991)1863;
\bibitem[24] {Jana4}
J.Ambj\o rn and S.Varsted {\it A new approach
to three-dimensional quantum gravity} NBI-HE-91-40 (1991);
\bibitem[25] {Jana5}
D.V.Boulatov,A.Krzywicki, Mod.Phys.Lett.{\bf A6}, 3005 (1991);
\bibitem[26] {Jana6}
J.Ambj\o rn, D.V.Boulatov, A.Krzywicki , S.Varsted,
{\it The vacuum in three-dimensional simplicial quantum
gravity} NBI-HE-91-46, LPTHE Orsay 91/57, (1991);
\bibitem[27] {Jana7}
J.Ambj\o rn and S.Varsted {\it Three-dimensional simplicial
quantum gravity} NBI-HE-91-45 (1991);
\bibitem[28] {Jana8}
J.Ambj\o rn and J.Jurkiewicz {\it Four-dimensional
simplicial quantum gravity} Niels Bohr Inst.Preprint
NBI-HE-91-47 (1991)
and revised version NBI-HE-91-60;
\bibitem[29] {Jana9}
N. Godfrey and M.Gross,
Phys.Rev. {\bf D43} (1991) R1749;
\bibitem[30] {Jana10}
M.Gross and S.Varsted {\it elementary
moves and ergodicity in D-dimensional simplicial quantum gravity}
NBI-HE-91-33;
\bibitem[31] {Jana11}
M.Gross, Nucl.Phys.{\bf B20}(Proc.Suppl.)(1991)724.
\bibitem[32] {Witt}
A.Schwarz, Lett.Math.Phys.{\bf 2}(1978)247;\par
E.Witten, Commun.Math.Phys. {\bf 117}(1988)353.
\bibitem[33] {Frohlich}
J.Fr\"{o}hlich,
{\it Regge calculus and discretized gravitational functional
integrals} Preprint IHES (1981), reprinted in {\it Non-perturbative
quantum field theory-Mathematical aspects and applications},
Selected Papers of J.Fr\"{o}hlich, (World Sci. Singapore 1992);
\bibitem[34] {Dancis}
J.Dancis, Topology and its Applic. {\bf 18}(1984)17.
\bibitem[35] {Grov}
K.Grove and P.V.Petersen, Annals of Math. {\bf 128}, 195 (1988).
\bibitem[36] {Grom}
M.Gromov, {\it Structures m\'etriques pour les vari\'et\'es
Riemanniennes} (Conception Edition Diffusion Information
Communication Nathan, Paris, 1981);see also
S.Gallot, D.Hulin, J.Lafontaine {\it Riemannian Geometry},
(Springer Verlag, New York,1987). A particularly clear account
of the results connected with Gromov-Hausdorff convergence of
riemannian manifolds is provided by the paper of
K.Fukaya {\it Hausdorff convergence of riemannian manifolds and its
applications}, Advanced Studies in Pure Math. $18$-I,
{\it Recent topics in differential and analytic geometry}, 143-238,
(1990). A recent review on properties of geodesic ball coverings is
provided by the paper: J.Cheeger {\it Critical points of distance
functions and applications to Geometry} in {\it Geometric Topology:
Recent Developments}, P.de Bartolomeis, F.Tricerri eds. Lect.Notes in
Math. {\bf 1504}, 1-38, (1991).
\bibitem[37] {Carf}
M.Carfora and A.Marzuoli, Class.Quantum Grav. {\bf 9}(1992)595;
M.Carfora and A.Marzuoli, Phys.Rev.Lett. {\bf 62} (1989)1339.
\bibitem[38] {popa}
M.Carfora, A.Marzuoli {\it Finiteness theorems in Riemannian
geometry and lattice quantum gravity},
Contemporary Mathematics {\bf 132} (proceedings of AMS research
meeting {\it Mathematical Aspects of Classical Field Theory },
Am.Math.Soc. Eds. M.J.Gotay, J.E.Marsden, V.Moncrief (1991).
\bibitem[39] {IJMPA}
M.Carfora and A.Marzuoli, {\it Reidemeister torsion and
simplicial quantum gravity} to appear in Intern.Journ.Modern Phys.A
\bibitem[40] {Marz}
M.Carfora, A.Marzuoli, M.Martellini {\it Combinatorial
and topological phase structure of non-perturbative n-dimensional
quantum gravity}, Intern.Journ.Modern Phys.B, {\bf 6} 2109-2121
(1992).
\bibitem[41] {Grunbaum}
B.Gr\"{u}nbaum, {\it Convex Polytopes}, Interscience Publishers,
(New York, 1967)
\bibitem[42] {Cohe}
M.M.Cohen, {\it A course in simple homotopy theory}, GTM 10,
(Springer Verlag, New York, 1973); see also
J.W.Milnor, Bull.Amer.Math.Soc. {\bf 72} 358 (1966);
C.P.Rourke, B.J.Sanderson {\it Introduction to Piecewise-Linear
Topology}, (Springer Verlag, New York, 1982).
\bibitem[43] {DeRham}
G.deRham, S.Maumary, and M.A.Kervaire {\it Torsion et Type Simple
d'Homotopie}, Lect.Notes in Math.{\bf 48}(Springer Verlag 1967).
\bibitem[44] {RayS}
D.B.Ray and I.M.Singer, Advances in Math. {\bf 7}(1971)145;
D.B.Ray, Advances in Math. {\bf 4}(1970)109;
D.Fried, {\it Lefschetz formulas for flows}, Contemp.Math.
{\bf 58}, Part III, 19-69, (1987);
J.Cheeger, Ann.Math.{\bf 109}(1979)259.
\bibitem[45] {Pete}
K.Grove, P.V.Petersen and J.Y.Wu, Bull.Am.Math.Soc. {\bf 20}, 181
(1989); Invent.Math. {\bf 99}, 205 (1990) (and its Erratum).
\bibitem[46] {Pans}
P.Pans\'u {\it Effondrement des varietes riemanniennes}
Sem.Bourbaki, Asterisque {\bf 121-122}(1985)63 ; T.Yamaguchi,
Ann.of Math., {\bf 133}(1991)317
\bibitem[47] {Erdo}
P.Erdos, P.Kelly, Amer.Math.Monthly {\bf 70} 1074 (1963);
see also chap.10 of F.Harary (ed.) {\it A Seminar on Graph
Theory} (Holt, Rinehart and Winston, Inc.,New York, 1967).
\bibitem[48] {Yama}
K.R.Parthasarathy, {\it Probability Measures on Metric Spaces}
(Academic, New York, 1967); Y.Yamasaki, {\it Measures on infinite
dimensional spaces} (World Scientific, Singapore, 1985).
\bibitem[49] {BBC}
C.Bartocci, U.Bruzzo, M.Carfora, A.Marzuoli,
{\it Homotopy theory and simplicial quantum gravity},
in preparation
\bibitem[50] {Martellini}
G.Bonacina, M.Martellini, M.Rasetti, {\it $2+1$ dimensional
quantum gravity as a Gaussian fermionic system and the
$3D$-Ising model}, Preprint n.855, Univ.Roma {\it La
Sapienza}, (1992)- to appear in Knots and its ramifications;
\bibitem[51] {Ruel}
D.Ruelle,{\it Statistical Mechanics: rigorous results}
(W.A.Benjamin, New York, 1974).
\bibitem[52] {Zallen}
R.Zallen, {\it Stochastic Geometry: Aspects of Amorphous Solids},
in {\it Fluctuation Phenomena}, eds. E.W.Montroll and
J.L.Lebowitz, (North-Holland, Amsterdam, 1979).
\bibitem[53] {Whitehead}
J.H.C.Whitehead, Comment. Math. Helv., {\bf 22},48-92 (1949)
\bibitem[54] {Freedman}
M.H. Freedman, J.Differential Geom. {\bf 17}, 357-453 (1982),
see also M.Freedman and F.Quinn, {\it Topology of 4-manifolds},
Princeton Math.Ser., vol.39, (Princeton Univ.Press, Princeton,
NJ, 1990)
\bibitem[55] {Donaldson}
S.K.Donaldson, J.Differential Geom. {\bf 18}, 279-315 (1983),
see also S.K.Donaldson and P.B.Kronheimer, {\it The
geometry of four-manifolds}, Oxford Math. Monographs, (Oxford
Univ. Press, 1990)

\end{description}
\vfill\eject
FIGURE CAPTIONS\par
\vskip 1 cm
{\it Fig.1}. A small array of closely packed (geodesic) balls is
here shown in three-dimensional Euclidean space. Even in such a
case, such packings do not yield for a regular tessellation of
the ambient space. In general, a $L(m)$-minimal geodesic ball
covering has a rather irregular pattern, also because for
manifolds in $\Ricco$ arbitrarily small metric balls need not
be contractible. However, for a given
(sufficiently) large value of $m$, either the number of geodesic
balls or their intersection pattern, (obtained upon doubling
the radius of the balls), embody non-trivial information on the
geometry and topology of the underlying riemannian manifold.\par
\vskip 1 cm
{\it Fig.2}. A portion of a $1$-skeleton graph corresponding to
a $L(m)$-geodesic ball covering. The length of each link is
proportional to $1/m$.\par
\vskip 1 cm
{\it Fig.3}. A pictorial representation of the compactness of
Gromov's space. Notice that while the number of equivalence
classes ${\cal O}^{(m)}_{\Gamma}$ varies with $m$, the number of
distinct homotopy types (and of simple homotopy types) is
finite and independent from $m$. Such number can be estimated in
terms of the parameters $n$, $r$, $D$, $V$ characterizing the
particular space of bounded geometries, $\Ricco$, considered.\par
\vfill\eject
{\it Fig.4}. The Erd\"{o}s-Kelly minimal graph generated by distinct
geodesic ball $1$-skeletons
\vskip 1 cm
{\it Fig.5}. The lattice gas associated with a geodesic ball covering.
\vskip 1 cm
{\it Fig.6}. The exterior boundary condition assumed in carrying out
the thermodynamical limit is to keep empty the $b_{(m)}$
added sites associated with the Erd\"{o}s-Kelly minimal graph
generated by the geodesic ball $1$-skeletons.
\vfill\eject
{\it Fig.7}. $K$ collapses to $J$, (across the tetrahedron $e^n$
onto its face $e^{n-1}$). More in general, $K$ collapses to
$J$, and we write $K\downarrow J$, if there is a sequence of
elementary collapses, as the one described above, from $K$ to
$J$. An elementary expansion is the inverse operation, and a
sequence of elementary collapses and expansions define a formal
deformation from $K$ to $J$. The equivalence relation, in
homotopy, associated with such notion of formal deformation
yields for the simple homotopy type of a CW-complex.\par
\vskip 1 cm
{\it Fig.8}. The CW-pair $({\cal N}, {\Gamma}^{(2)})_{(m)}$,
(here shown in dimension three), is equivalent, under formal
deformation, to the two-skeleton ${\Gamma}^{(2)}_{(m)}$ at a
vertex of which, $e^{(0)}$, one attaches a suitable number of
two-cells $Q^2$ and three-cells $Q^3$. Some metrical information
concerning the original pair
$({\cal N}, {\Gamma}^{(2)})_{(m)}$ is embodied in the number of
three-cells attached. \par
\vskip 1 cm
{\it Fig.9}. Trivially added cells, (here represented by the
surfaces of the dotted tetrahedrons), associated with representative
elements of ${\pi}_q({\Gamma}_2
\cup\bigcup_{j=1}^ae^q_j,{\Gamma}_2)$, (with $q=2$, $a=3$).\par
\noindent The fundamental group ${\pi}_1({\Gamma}^{(2)};e_0)$ acts
on ${\pi}_q({\Gamma}_2
\cup\bigcup_{j=1}^ae^q_j,{\Gamma}_2)$. If the loop $s$ represent
$[\alpha]\in {\pi}_1({\Gamma}^{(2)};e_0)$, then
$\beta \in  {\pi}_q({\Gamma}_2
\cup\bigcup_{j=1}^ae^q_j,{\Gamma}_2)$ is sent into
$[\alpha]\beta \in {\pi}_q({\Gamma}_2
\cup\bigcup_{j=1}^ae^q_j,{\Gamma}_2)$. The geometrical idea is to
pull the image of $J^{q-1}$ along the path $s$ back to the point
$e_0$ with the image of $Q^q$ being dragged in such a way that the
image of $Q^{q-1}$ is always in ${\Gamma}^{(2)}$. In this way,
${\pi}_1({\Gamma}^{(2)},e_0)$ appears as a {\it gauge group}
for the spaces ${\pi}_q({\Gamma}_2
\cup\bigcup_{j=1}^ae^q_j,{\Gamma}_2)$ which describes the attachement
of new cells to ${\Gamma}^{(2)}_{(m)}$.\par
\vskip 1 cm
{\it Fig.10}. A representative elements of the group
${\pi}_{q+1}({\cal M},{\Gamma}_2
\cup\bigcup_{j=1}^ae^q_j)$, (here with $q=2$ and $a=3$), is
constructed by {\it filling} the tetrahedra in such a way that
the resulting three-cell has the surfaces of the tetrahedra
as boundary.\par
\vfill\eject
{\it Fig.11}. A {\it vector} of
${\pi}^{\theta}_2({\Gamma}_2
\cup\bigcup_{j=1}^ae^2_j,{\Gamma}_2)$ can be thought of as a
collection of euclidean ${\bf R}^p$-vectors ${\xi}^{(i)}$
providing a coloring (labelling) of the two cells
representing the elements of ${\pi}_2({\Gamma}_2
\cup\bigcup_{j=1}^ae^2_j,{\Gamma}_2)$, (cells that we keep
on in describing as the boundaries of tetrahedra).\par
\vskip 1 cm
{\it Fig.12}. The vector $[{\theta}_*(\partial)^{-1}{\xi}]_{(l)}$,
associated with the three-cell $e^3_l$, is obtained from the
vectors ${\xi}^{(k)}$ associated with the two-cells $e^2_k$
through the action of the set of $p\times p$ matrices
$[{\theta}_*(w_{kl})]^{-1}$, ($l$ is fixed while $k$ ranges
over the set of vectors ${\xi}^{(k)}$). Each matrix acts on the
corresponding vector, the resulting transformed vectors add up
to $[{\theta}_*(\partial)^{-1}{\xi}]_{(l)}$.\par
\vskip 1 cm
{\it Fig.13}. In the reciprocal complex associated with
$({\cal M}, {\Gamma}^{(2)})$, the three-cells representing
the classes of ${\pi}^{\theta}_{3}({\cal M}, {\Gamma}_2
\cup\bigcup_{j=1}^ae^2_j)$ are described by the vertices
of a graph ${\Lambda}$ the edges of which corresponds to the
two cells of ${\Gamma}_2\cup\bigcup_{j=1}^ae^2_j$.
Thus, with the vector ${\nu} \in
{\pi}^{\theta}_{3}({\cal M}, {\Gamma}_2
\cup\bigcup_{j=1}^ae^2_j)$, describing a colouring of the
three-cells of ${\pi}^{\theta}_{3}({\cal M}, {\Gamma}_2
\cup\bigcup_{j=1}^ae^2_j)$, we can associate a magnetic vector
model evolving on the graph ${\Lambda}$ and characterized
by an energy $H(\xi)$ given by (\ref{energia}).\par
\vskip 1 cm
{\it Fig.14}. The two-skeleton $({\Gamma}^{(2)}_{(m)}, e_0)$
is homotopically equivalent to a bouquet of circles,
(originating from $e_0$), and disks.\par
\vfill\eject
{\it Fig.15}. The  $CW$-pair $({\cal N}, {\Gamma}^{(2)})$
is equivalent, in the simple homotopy sense, to a bouquet of
circles, disks, and spheres attached at $e_0$. The spheres
bound, non-trivially, $N^{(3)}_{(m)}$ three-cells. This
bouquet carries all the (simple) homotopical information
of the original pair $({\cal N}, {\Gamma}^{(2)})$, while
non-trivial metrical information is encoded into the
numbers $N^{(3)}_{(m)}$ and $N^{(1)}_{(m)}$ which can be
read off from the composition of the bouquet.\par
\vskip 1 cm
{\it Fig.16}. According to the Gaussian distribution of the
fields ${\xi}$, the correlations between the colourings
of the three-cells of $({\cal M}, {\Gamma}^{(2)})$
decay exponentially.\par
\vskip 1 cm
{\it Fig.17}. In the grand-canonical approach, each partition
function is weighted by a fugacity term proportional to the
number of generators of ${\pi}_1({\Gamma}^{(2)}_{(m)},e_0)$.
Upon using the reciprocal complex, this picture corresponds
to a vector model whose average number of particles  is controlled by
the given fugacity.\par
\vskip 1 cm
{\it Fig.18}. A pictorial representation of the relations
between the Whitehead group $W({\bf R})$, the general
linear group $GL({\bf R})$, and the commutator subgroup
generated by the elements of $E({\bf R})$.\par
\vskip 1 cm
{\it Fig.19}. The Gaussian averaging, described by
(\ref{normalized}), maps a real-valued function
$f$ defined over $GL(q, {\bf R})$ into a real-valued
function ${\hat f}_{\cal G}$ defined over $W({\bf R})$.\par
\end{document}